\documentclass[11pt,a4paper]{elsarticle}
\usepackage[colorlinks]{hyperref}
\usepackage{multirow,pifont,lscape}

\usepackage[dvipsnames]{xcolor}

\AtBeginDocument{%
\usepackage{color}
\definecolor{mgray}{cmyk}{0,0,0,.8}
\hypersetup{
    bookmarks=true,         
    unicode=false,          
    pdftoolbar=true,        
    pdfmenubar=true,        
    pdffitwindow=true,      
    pdfnewwindow=true,      
    colorlinks=true,       
    linkcolor=mgray,          
    citecolor=blue,        
    filecolor=magenta,      
    urlcolor=blue          
}
}

\usepackage[utf8]{inputenc}
\usepackage{eso-pic}
\usepackage{amsmath,amsfonts,amsthm,graphicx,ulem,tocvsec2}
\usepackage{titlesec}
\usepackage{sectsty,colortbl}
\usepackage{pict2e}

\allsectionsfont{\normalfont\sffamily\bfseries}

\journal{\hspace{-10em}\colorbox{white}{\phantom{Preprint for submission to Nuclear Physics B}}}

\makeatletter
\def\paragraph{\secdef{\els@aparagraph}{\els@bparagraph}}
\def\els@aparagraph[#1]#2{\elsparagraph[#1]{#2.}}
\def\els@bparagraph#1{\elsparagraph*{#1.}}

\usepackage{wrapfig}

\usepackage{multirow}

\usepackage{mathrsfs}
\usepackage{empheq}

\newcommand{\bal}{\begin{align}}
\newcommand{\eal}{\end{align}}
\newcommand{\ba}{\begin{align}}
\newcommand{\ea}{\end{align}}
\newcommand{\beq}{\begin{equation}}
\newcommand{\eeq}{\end{equation}}
\newcommand\beqa{\begin{eqnarray}}
\newcommand\eeqa{\end{eqnarray}}
\newcommand\bea{\begin{array}}
\newcommand\eea{\end{array}}

\newcommand{\ii}{{\bf i}}

\newcommand{\algu}{\mathfrak{u}}
\newcommand{\su}{\mathfrak{su}}
\renewcommand{\sl}{\mathfrak{sl}}
\newcommand{\psu}{\mathfrak{psu}}
\newcommand{\pu}{\mathfrak{pu}}
\newcommand{\gl}{\mathfrak{gl}}

    \newcommand{\COMMENT}[1]{}
    
    \newcommand{\neqa}{\nonumber\end{eqnarray}}
    
\def\a{{\alpha}}

\def\[{\left[}
\def\]{\right]}

\def\l{\lambda}
\def\e{\epsilon}

\def\a{\alpha}
\def\b{\beta}

\def\<{\langle}
\def\>{\rangle}
\def\sG{\,\slash\!\!\!\! \hat G}

\def\sH{\slash\!\!\!\! H}

\def\i2{\frac{i}{2}}

\def\<{\langle}
\def\>{\rangle}

\def\sG{\,\slash\!\!\!\! \hat G}

\def\i2{\frac{i}{2}}

\DeclareMathOperator{\Tr}{Tr}
\DeclareMathOperator{\Pf}{Pf}

\def\1h{\hat 1}
\def\2h{{\hat 2}}
\def\3h{{\hat 3}}
\def\4h{{\hat 4}}

\def\be{\begin{eqnarray}}
\def\ee{\end{eqnarray}}
\def\no{\nonumber}

    \def\CO{{\cal O}}

    \def\<{\left\langle\,}
    \def\>{\, \right\rangle}
    \def\[{\left[}
    \def\]{\right]}

\newcommand{\ps}{{\bf p}}

   \def\gl{{\mathfrak{gl}}}
   \def\su{{\mathfrak{su}}}
   \def\sl{{\mathfrak{sl}}}

\def\Pt{{\tilde \bP}}

\def\i{{\mathsf{i}}}

\newcommand{\bP}{{\bf P}}

\newcommand{\hl}{\hat{\lambda}}
\newcommand{\hn}{\hat{\nu}}
\newcommand{\emp}{\emptyset}

\colorlet{comp}{blue}

\newcommand{\ZZ}{\mathcal{Z}}
\newcommand{\ZZb}{\bar{\mathcal{Z}}}
\newcommand{\XX}{\mathcal{X}}
\newcommand{\XXb}{\bar{\mathcal{X}}}
\newcommand{\YY}{\mathcal{Y}}
\newcommand{\YYb}{\bar{\mathcal{Y}}}
\newcommand{\DD}{\mathcal{D}}

\newcommand{\dQ}{\mathbb{Q}}

\newcommand{\aaa}{\mathbf{a}}
\newcommand{\bbb}{\mathbf{b}}
\newcommand{\fff}{\mathbf{f}}

\def\Ma{\texttt{Mathematica}}
\def\zhuk{Zhukowsky}

\def\Per{\mathcal{P}}

\def\cA{{\textsf{A}}}

\def\cc{{\textsf{c}}}
\def\cd{{\textsf{d}}}

\def\ccc{{c}}

\def\ii{\mathbbm{i}}
\def\iotwo{{\textstyle \frac{\ii}{2}}}

\def\beauty{\lrcorner\hspace{-0.59mm}\ulcorner}
\def\beautysub{{\lrcorner\hspace{-0.43mm}\ulcorner}}

\def\mi{\text{-}}
\def\ns{\mathbf{n}}

\def\nbname{\texttt{QSCsolver.nb}} 
\def\dbname{\texttt{QSCdata.nb}} 

\usepackage{framed}
\definecolor{shadecolor}{rgb}{0.9,0.9,0.9} 

\usepackage{amssymb}
\usepackage{subfig}
\usepackage{overpic,rotating,bbold,bbm}

\usepackage{varioref}

\usepackage{diagbox}

\usepackage[font=footnotesize,labelfont=bf]{caption}

\begin{document}

\begin{frontmatter}

\title{\sffamily\LARGE{\bf The full spectrum of  AdS$_5$/CFT$_4$ II:} \\ 
Weak coupling expansion via the quantum spectral curve}

\author[1,2,3]{Christian Marboe}
\ead{christian.marboe@su.se}
\author[2,1,4]{Dmytro Volin}

\address[1]{Nordita, Stockholm University \& KTH Royal Institute of Technology, \\ Roslagstullsbacken 23, SE-106 91 Stockholm, Sweden \vspace{2mm}
}
\address[2]{School of Mathematics, Trinity College Dublin, \\ College Green, Dublin 2, Ireland \vspace{2mm}
}
\address[3]{Institut f\"{u}r Mathematik \& Institut f\"{u}r Physik, Humboldt-Universit\"{a}t zu Berlin,\\ Zum Gro\ss en Windkanal 6, 12489 Berlin, Germany \vspace{2mm}
}

\address[4]{Department of Physics and Astronomy, Uppsala University, \\ Box 516, SE-751 20 Uppsala, Sweden \\[9em]
}
\ead{dmytro.volin@physics.uu.se}

\begin{abstract}
We continue the effort to optimise and generalise the solution of the spectral problem of AdS$_5$/CFT$_4$ in the planar limit via integrability. We present a simple strategy to solve the quantum spectral curve perturbatively for general states by focusing on the $\bP\mu$-system. A \Ma\ notebook with an implementation of this algorithm is provided, as well as an extensive database  with a user-friendly interface containing more than 8.000 solutions of the QSC. When investigating the solution space, we observe a curious phenomenon: existence of solutions for which the Q-system degenerates in the limit $g\to 0$. These degeneracies are lifted at higher orders in perturbation theory. The degenerating solutions have auxiliary Bethe roots merging with branch points at weak coupling.

\end{abstract}

%
%


%
%

\end{frontmatter}

\AddToShipoutPictureBG*{%
  \AtPageUpperLeft{%
    \hspace{0.9\paperwidth}%
    \raisebox{-5\baselineskip}{%
      \makebox[0pt][r]{\texttt{NORDITA 2018-127}}}
    \raisebox{-6\baselineskip}{%
      \makebox[0pt][r]{\texttt{ UUITP-61/18\,\,}}}
    \raisebox{-7\baselineskip}{%
      \makebox[0pt][r]{\texttt{ TCDMATH 18-20\,\,\,\,}}}
}}%

\AddToShipoutPictureBG*{%
  \AtPageUpperLeft{%
    \hspace{0.7\paperwidth}%
    \raisebox{-31.2\baselineskip}{%
      \makebox[0pt][r]{

\begin{picture}(190,190)


\put(65,35){\color{black!10}\circle*{18}}
\multiput(0,60)(0.5,0){200}{\color{black!6}\line(1,1){30}}

\linethickness{1mm}
\put(65,35){\color{black!40}\line(0,1){25}}
\put(65,75){\color{black!20}\line(0,-1){15}}

\put(71,41){\circle*{3}}
\put(77,47){\circle*{3}}
\put(59,29){\circle*{3}}
\put(53,23){\circle*{3}}
\put(65,75){\circle*{3}}
\put(65,35){\circle*{3}}

\thicklines
\put(0,20){\line(1,0){100}}
\put(30,50){\line(1,0){100}}
\put(0,20){\line(1,1){30}}
\put(100,20){\line(1,1){30}}

\put(0,60){\line(1,0){100}}
\put(30,90){\line(1,0){100}}
\put(0,60){\line(1,1){30}}
\put(100,60){\line(1,1){30}}

\put(220,-240){\begin{turn}{30} \color{gray}\huge \bf\texttt{Now with} \end{turn}}
\put(230,-260){\begin{turn}{30} \color{gray}\huge \bf\texttt{11 loops} \end{turn}}

\end{picture}

      }}
}}%

\newpage
\thispagestyle{empty}

\begingroup
\hypersetup{linkcolor=black}
\setcounter{tocdepth}{2}
\tableofcontents
\endgroup

\newpage

\section{Introduction}
In this paper we continue our studies \cite{Marboe:2017dmb} of the spectrum of the AdS$_5$/CFT$_4$ correspondence in the planar limit. 
The goal is to provide a general perturbative solution method to the Quantum Spectral Curve (QSC) \cite{Gromov:2013pga,Gromov:2014caa} that is both conceptually simple and powerful in practice.

We will take the point of view of $\mathcal{N}=4$ supersymmetric Yang-Mills theory (SYM), where the spectral problem refers to the eigenvalue problem of the dilatation operator on the space of single-trace operators of fundamental fields,
\be
\mathbb{D} \,\mathcal{O}(x)=\Delta \, \mathcal{O}(x)\,,\ \ \ \ \mathcal{O}(x)=\Tr\left[\ZZ\DD\Psi\hdots\right]\,.
\ee
The dilatation operator, as well as its eigenvalues and -states, depends on the coupling. We work in the planar limit and use a coupling constant defined by $g\equiv \frac{\sqrt{\l}}{4\pi}$, where $\lambda = \frac{g_{\text{YM}}^2}{N}$ is the 't Hooft coupling. The conformal dimension $\Delta$ can be split into two parts: the classical dimension $\Delta_0\equiv\Delta|_{g=0}$ and the anomalous dimension $\gamma(g)$. In this paper, we want to find the spectrum of anomalous dimensions as perturbative expansions around $g=0$,
\be
\Delta(g) &=& \Delta_0 + 
\sum_{j=1}^{\infty} g^{2j} \,\gamma_j \,.
\ee

The QSC supersedes earlier formulations of the Thermodynamic Bethe Ansatz (TBA) equations \cite{Gromov:2009tv,Bombardelli:2009ns,Gromov:2009bc,Arutyunov:2009ur} and phrases the spectral problem in terms of a relatively simple Riemann-Hilbert problem. Its power has been demonstrated in many applications \cite{Gromov:2014bva,Alfimov:2014bwa,Marboe:2014gma,Marboe:2014sya,Gromov:2015wca,Gromov:2015vua,Hegedus:2016eop,Marboe:2016igj,Alfimov:2018cms,Cavaglia:2018lxi}, also including twisted SYM \cite{Kazakov:2015efa,Gromov:2015dfa,Gromov:2016rrp,Gromov:2017cja} and ABJM theory \cite{Cavaglia:2014exa,Bombardelli:2017vhk,Gromov:2014eha,Anselmetti:2015mda,Cavaglia:2016ide,Bombardelli:2018bqz,Lee:2017mhh,Lee:2018jvn}.

We build on our results in \cite{Marboe:2017dmb,Marboe:2016yyn} where we described an efficient method to determine the leading perturbative solution to the QSC. 
One can think of that method as the jump-start, and the algorithm of this paper as the main engine, of our  perturbative solver.



\subsubsection*{Structure of the paper}
In section \ref{sec:QSC}, we review the aspects of the QSC relevant for our approach. Section \ref{sec:QSCweak} contains an analysis of the QSC functions in the weak coupling limit, and lays the foundation for the perturbative algorithm to solve the $\bP\mu$-system for general states presented in section \ref{sec:pert}. Section \ref{sec:deg} discusses a new phenomenon: the appearance of degenerate solutions to the leading Q-system, which require special attention. Section \ref{sec:res} summarises the results obtained with the \Ma\ implementation of the solution algorithm which can be found in the ancillary files at arxiv.org. An introduction to this \Ma\ notebook, \texttt{\nbname}, as well as to a database of our results, \texttt{\dbname}, is given in section \ref{sec:code}. 

\subsubsection*{Definitions and conventions}
All QSC functions are multi-valued functions of the spectral parameter, $u$. We use the standard notation
\be
f^{[n]}(u)\equiv f\left(u+{\textstyle\frac{\ii}{2}}n\right)\,, \quad\quad f^\pm\equiv f^{[\pm1]} \,.
\ee
The analytic continuation of a function $f(u)$ through a cut on the real axis is denoted by $\tilde{f}(u)$.
Perturbative orders in the QSC functions are denoted by $f^{(n)}$, e.g.
\be
f(u,g)=f^{(0)}(u)+g f^{(1)}(u)+g^2 f^{(2)}(u)+...
\ee
In most cases, only even powers will be present, but we will encounter exceptions where the QSC functions are expanded in both even and odd powers of $g$.

The \zhuk\ variable $x$ is defined by
\be\label{zhukdef}
x+\frac{1}{x} = \frac{u}{g}
\ee
and has the important property $\tilde{x}=\frac{1}{x}$. We always choose the branch where $|x(u)|>1$.

The algebra of functions and transcendental numbers in the perturbative QSC functions were discussed thoroughly in \cite{Leurent:2013mr,Marboe:2014gma}. The main functions are Hurwitz $\eta$-functions,
\be \label{etadefap}
\eta_{k_1,k_2,\hdots,k_n}(u) \equiv \sum_{0\le j_1<\hdots< j_n}^\infty \frac{1}{(u+\ii j_1)^{k_1}\cdots(u+\ii j_n)^{k_n}}\,,
\ee
and multiple zeta-values,
\begin{eqnarray} \label{zdef}
\zeta_{k_1,\hdots,k_n}=\sum_{0<j_1<\hdots<j_n<\infty}\frac{1}{j_1^{k_1}\cdots j_n^{k_n}}.
\end{eqnarray}
These functions naturally appear when solving difference equations. As usual, we denote the discrete analog of integration by $\Psi$, i.e.
\be
f(u)-f(u+\ii)=g(u) \quad\Leftrightarrow \quad f(u) = \Psi\left(g(u)\right)+ \Per \,,
\ee
where $\Per$ is an $\ii$-periodic function.

\subsubsection*{Labelling of multiplets}

Single-trace operators are organised in irreducible representations of the superconformal symmetry, $\psu(2,2|4)$. All operators within a multiplet have the same anomalous dimension, and each multiplet corresponds to a solution of the QSC with particular boundary conditions determined by its quantum numbers.

\begin{figure}[b!]
\centering
\begin{picture}(320,210)
\color{gray}
\linethickness{.1mm}

\multiput(15,75)(0,20){5}{\line(1,0){290}}
\multiput(150,175)(0,20){2}{\line(1,0){40}}
\multiput(110,15)(0,20){3}{\line(1,0){40}}

\multiput(30,75)(20,0){4}{\line(0,1){80}}
\multiput(210,75)(20,0){5}{\line(0,1){80}}

\put(110,0){\line(0,1){150}}
\put(130,0){\line(0,1){150}}
\put(150,0){\line(0,1){210}}
\put(170,75){\line(0,1){135}}
\put(190,75){\line(0,1){135}}

\color{black}
\linethickness{0.7mm}

\put(150,195){\line(1,0){20}}
\put(150,176){\line(1,0){40}}
\put(30,155){\line(1,0){180}}
\put(30,135){\line(1,0){200}}
\put(50,115){\line(1,0){220}}
\put(90,95){\line(1,0){200}}
\put(110,75){\line(1,0){180}}
\put(110,55){\line(1,0){40}}
\put(110,35){\line(1,0){40}}
\put(130,15){\line(1,0){20}}

\put(30,135){\line(0,1){20}}
\put(50,115){\line(0,1){40}}
\put(70,115){\line(0,1){40}}
\put(90,95){\line(0,1){60}}
\put(110,35){\line(0,1){120}}
\put(130,15){\line(0,1){140}}
\put(150,15){\line(0,1){180}}
\put(170,75){\line(0,1){120}}
\put(190,75){\line(0,1){100}}
\put(210,75){\line(0,1){80}}
\put(230,75){\line(0,1){60}}
\put(250,75){\line(0,1){40}}
\put(270,75){\line(0,1){40}}
\put(290,75){\line(0,1){20}}




\color{Brown!50}
\linethickness{0.2mm}
\put(30,145){\vector(1,0){120}}\put(150,145){\vector(-1,0){120}}
\put(50,125){\vector(1,0){100}}\put(150,125){\vector(-1,0){100}}
\put(90,105){\vector(1,0){60}}\put(150,105){\vector(-1,0){60}}
\put(110,85){\vector(1,0){40}}\put(150,85){\vector(-1,0){40}}

\color{NavyBlue!60}
\put(120,35){\vector(0,1){40}}\put(120,75){\vector(0,-1){40}}
\put(140,15){\vector(0,1){60}}\put(140,75){\vector(0,-1){60}}

\color{OliveGreen!60}
\put(150,145){\vector(1,0){60}}\put(210,145){\vector(-1,0){60}}
\put(150,125){\vector(1,0){80}}\put(230,125){\vector(-1,0){80}}
\put(150,105){\vector(1,0){120}}\put(270,105){\vector(-1,0){120}}
\put(150,85){\vector(1,0){140}}\put(290,85){\vector(-1,0){140}}

\color{Violet!60}
\put(160,155){\vector(0,1){40}}\put(160,195){\vector(0,-1){40}}
\put(180,155){\vector(0,1){20}}\put(180,175){\vector(0,-1){20}}

\color{black}
\put(180,145){\color{white}\circle*{14}}\put(174,143){$n_{\fff_4}$}
\put(180,125){\color{white}\circle*{14}}\put(174,123){$n_{\fff_3}$}
\put(180,105){\color{white}\circle*{14}}\put(174,103){$n_{\fff_2}$}
\put(180,85){\color{white}\circle*{14}}\put(174,83){$n_{\fff_1}$}

\put(123,145){\color{white}\circle*{8}}\put(130,144){\color{white}\circle*{8}}\put(137,145){\color{white}\circle*{8}}\put(119,142){\small$L {\text-} n_{\fff_4}$}
\put(123,125){\color{white}\circle*{8}}\put(130,124){\color{white}\circle*{8}}\put(137,125){\color{white}\circle*{8}}\put(119,122){\small$L {\text-} n_{\fff_3}$}
\put(123,105){\color{white}\circle*{8}}\put(130,104){\color{white}\circle*{8}}\put(137,105){\color{white}\circle*{8}}\put(119,102){\small$L {\text-} n_{\fff_2}$}
\put(123,85){\color{white}\circle*{8}}\put(130,84){\color{white}\circle*{8}}\put(137,85){\color{white}\circle*{8}}\put(119,82){\small$L {\text-} n_{\fff_1}$}

\put(118,55){\color{white}\circle*{11}}\put(122,55){\color{white}\circle*{11}}\put(113,53){$n_{\bbb_1}$}
\put(138,55){\color{white}\circle*{11}}\put(142,55){\color{white}\circle*{11}}\put(133,53){$n_{\bbb_2}$}

\put(158,165){\color{white}\circle*{9}}\put(162,165){\color{white}\circle*{9}}\put(153,163){$n_{\aaa_1}$}
\put(178,165){\color{white}\circle*{9}}\put(182,165){\color{white}\circle*{9}}\put(173,163){$n_{\aaa_2}$}

\end{picture}
\caption{Young diagram corresponding to the oscillator numbers $n_\aaa$, $n_\fff$, $n_\bbb$ with respect to the $12\hat{1}\hat{2}\hat{3}\hat{4}34$ grading.}
\label{fig:YDdef}
\end{figure}
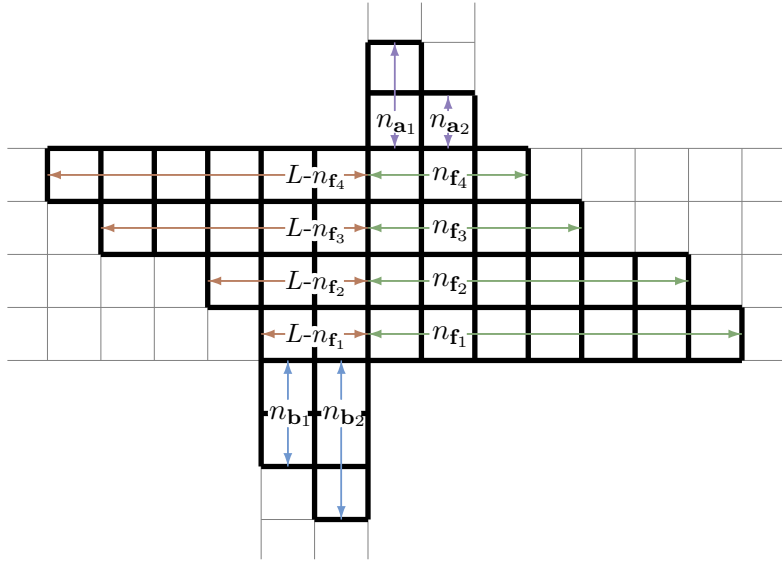

Following the practice of \cite{Marboe:2017dmb}, we label multiplets by the oscillator content used to construct their highest-weight state (HWS) in the ``compact beauty" grading (denoted $12\hat{1}\hat{2}\hat{3}\hat{4}34$ or $2222$ in \cite{Marboe:2017dmb}) at zero coupling. This is the operator within the multiplet with the lowest conformal dimension. We use the notation
\be
\ns=[n_{\bbb_1},n_{\bbb_2}|n_{\fff_1},n_{\fff_2},n_{\fff_3},n_{\fff_4}|n_{\aaa_1},n_{\aaa_2}]
\ee
for these eight numbers, which specify the multiplet and its spin chain interpretation completely at zero coupling. Recall that the oscillators can be used to build the $\mathcal{N}=4$ SYM fields according to
\begin{table}[h]
\centering
\begin{tabular}{c|c|c|c}
scalar&fermion&field strength& cov. derivative\\\hline
\!\!\!\!$\begin{matrix}\Phi_{ab}\equiv \fff_a^\dagger\fff_b^\dagger|0\rangle\\
\scriptscriptstyle \Phi_{12}\equiv\ZZ,\; \Phi_{13}\equiv\XX,\; \Phi_{14}\equiv\YY, \\\scriptscriptstyle \Phi_{23}\equiv\YYb,\; \Phi_{24}\equiv\XXb,\; \Phi_{34}\equiv\ZZb
\end{matrix}$\!\!
&\!\!
$\begin{matrix} \Psi_{a\alpha}\equiv\fff_a^\dagger\aaa_\alpha^\dagger|0\rangle \\[2mm] \bar{\Psi}_{a\dot\a}\equiv \epsilon_{abcd}\fff_b^\dagger\fff_c^\dagger\fff_d^\dagger \bbb_{\dot\a}^\dagger |0\rangle \end{matrix}$
\!\!&\!\!
$\begin{matrix}\mathcal{F}_{\a\b}\equiv\aaa_\a^\dagger\aaa_\b^\dagger|0\rangle \\[2mm] \mathcal{F}_{\dot\a\dot\b}\equiv\bbb_{\dot\a}^\dagger\bbb_{\dot\b}^\dagger\fff_1^\dagger\fff_2^\dagger\fff_3^\dagger\fff_4^\dagger|0\rangle \end{matrix}$
\!\!&
$\DD_{\alpha\dot\alpha}\equiv\aaa_\alpha^\dagger\bbb_{\dot\alpha}^\dagger$
\end{tabular}
\end{table}

\noindent and the length, equal to the number of component fields in the single-trace operator at zero coupling, is given by 
\be
L={\textstyle\frac{1}{2}}\left(n_\fff+n_\aaa-n_\bbb\right).
\ee
Non-compact Young diagrams \cite{Gunaydin:2017lhg}, see figure \ref{fig:YDdef}, provide a nice pictorial way to characterise the multiplets, as well as intuition about how to decompose tensor products into irreps and how to solve rational Q-systems \cite{Marboe:2017dmb,Marboe:2016yyn}.

At finite coupling, the oscillator construction is no longer an adequate description of the representations. All short multiplets, except for the BMN vacua, organise into long ones (see the Konishi example below) containing operators of different length. Furthermore, the eigenstates can be linear combinations of composite operators of different length, but identical $\psu(2,2|4)$ quantum numbers.

In relation to the large $u$ asymptotics of the QSC functions, the relevant quantum numbers are the shifted weights $\hl_a$ and $\hn_i$, which in terms of the oscillator numbers are given by
\begin{subequations}\label{hlhn}
\be \label{hl0}
\hl_a&=& n^{\beautysub}_{\fff_a}+\{2,1,0,-1\}_a+\Lambda \quad \equiv \quad \l^0_a + \Lambda \\
\hn_i|_{g=0}&=& \{-L^\beautysub-n^{\beautysub}_{\bbb_{\dot{\alpha}}},n^{\beautysub}_{\aaa_\a}  \}_i + \{-1,-2,1,0\}_i -\Lambda \quad \equiv \quad \nu_i^0 -\Lambda \,. \label{hn0}
\ee
\end{subequations}
where $\Lambda$ is an arbitrary constant, and the symbol $\beauty$ refers to the grading $12\hat{1}\hat{2}\hat{3}\hat{4}34$. The differences $\hl_a-\hl_{a+1}$ and $\hn_i-\hn_{i+1}$ provide six invariant $\psu(2,2|4)$ quantum numbers that characterise the finite-coupling multiplets.

Finally, recall that the weights $\hl$, related to $S^5$, remain fixed at any value of the coupling $g$, while $\hn$ are related to AdS$_5$ charges, including the conformal dimension $\Delta=\Delta_0+\gamma$, so they will run with the coupling,
\be
\hn_i &=& \hn_i |_{g=0} + \frac{\gamma}{2} \{- 1,- 1,1,1\}_i \,.
\ee


\begin{shaded} 
\noindent {\bf Example} 

\noindent The Konishi multiplet has a highest weight state of the form 
\be\Tr[\ZZ\ZZb]+\Tr[\XX\XXb]+\Tr[\YY\YYb]\no\ee
in the $12\hat{1}\hat{2}\hat{3}\hat{4}34$ grading. In the oscillator construction of the SYM fields, we have e.g. $\ZZ\equiv \fff_1\fff_2|0\rangle$ and $\ZZb\equiv\fff_3\fff_4|0\rangle$, so we denote the Konishi multiplet by 
\be\ns=[0,0|1,1,1,1|0,0]\,.\no\ee
The corresponding weights, governing the large $u$ asymptotics of the QSC functions are 
\be \hl_a=\{3,2,1,0\}_a+\Lambda \quad\quad \hn_i|_{g=0}=\{-3,-4,1,0\}_i-\Lambda\,.\no\ee

The Young diagrams for the four short zero-coupling multiplets that join into the long Konishi multiplet at finite coupling are

\setlength{\unitlength}{0.32mm}
\begin{picture}(430,90)
\put(5,70){$1...4$}
\put(115,70){$\hat1...4$}
\put(225,70){$1...\hat4$}
\put(335,70){$\hat1...\hat4$}

\color{gray}
\linethickness{.2mm}

\multiput(-5,20)(0,10){5}{\line(1,0){90}}
\put(20,0){\line(1,0){20}}
\put(20,10){\line(1,0){20}}
\put(40,70){\line(1,0){20}}
\put(40,80){\line(1,0){20}}
\put(0,20){\line(0,1){40}}
\put(10,20){\line(0,1){40}}
\put(20,-5){\line(0,1){65}}
\put(30,-5){\line(0,1){65}}
\put(40,-5){\line(0,1){90}}
\put(50,20){\line(0,1){65}}
\put(60,20){\line(0,1){65}}
\put(70,20){\line(0,1){40}}
\put(80,20){\line(0,1){40}}

\multiput(105,20)(0,10){5}{\line(1,0){90}}
\put(130,0){\line(1,0){20}}
\put(130,10){\line(1,0){20}}
\put(150,70){\line(1,0){20}}
\put(150,80){\line(1,0){20}}
\put(110,20){\line(0,1){40}}
\put(120,20){\line(0,1){40}}
\put(130,-5){\line(0,1){65}}
\put(140,-5){\line(0,1){65}}
\put(150,-5){\line(0,1){90}}
\put(160,20){\line(0,1){65}}
\put(170,20){\line(0,1){65}}
\put(180,20){\line(0,1){40}}
\put(190,20){\line(0,1){40}}

\multiput(215,20)(0,10){5}{\line(1,0){90}}
\put(240,0){\line(1,0){20}}
\put(240,10){\line(1,0){20}}
\put(260,70){\line(1,0){20}}
\put(260,80){\line(1,0){20}}
\put(220,20){\line(0,1){40}}
\put(230,20){\line(0,1){40}}
\put(240,-5){\line(0,1){65}}
\put(250,-5){\line(0,1){65}}
\put(260,-5){\line(0,1){90}}
\put(270,20){\line(0,1){65}}
\put(280,20){\line(0,1){65}}
\put(290,20){\line(0,1){40}}
\put(300,20){\line(0,1){40}}

\multiput(325,20)(0,10){5}{\line(1,0){90}}
\put(350,0){\line(1,0){20}}
\put(350,10){\line(1,0){20}}
\put(370,70){\line(1,0){20}}
\put(370,80){\line(1,0){20}}
\put(330,20){\line(0,1){40}}
\put(340,20){\line(0,1){40}}
\put(350,-5){\line(0,1){65}}
\put(360,-5){\line(0,1){65}}
\put(370,-5){\line(0,1){90}}
\put(380,20){\line(0,1){65}}
\put(390,20){\line(0,1){65}}
\put(400,20){\line(0,1){40}}
\put(410,20){\line(0,1){40}}

\color{black}
\linethickness{0.7mm}

\put(30,20){\line(1,0){20}}
\put(30,30){\line(1,0){20}}
\put(30,40){\line(1,0){20}}
\put(30,50){\line(1,0){20}}
\put(30,60){\line(1,0){20}}

\put(30,20){\line(0,1){40}}
\put(40,20){\line(0,1){40}}
\put(50,20){\line(0,1){40}}

\put(130,60){\line(1,0){30}}
\put(130,50){\line(1,0){30}}
\put(130,40){\line(1,0){30}}
\put(130,30){\line(1,0){50}}
\put(150,20){\line(1,0){30}}

\put(130,30){\line(0,1){30}}
\put(140,30){\line(0,1){30}}
\put(150,20){\line(0,1){40}}
\put(160,20){\line(0,1){40}}
\put(170,20){\line(0,1){10}}
\put(180,20){\line(0,1){10}}

\put(230,60){\line(1,0){30}}
\put(230,50){\line(1,0){50}}
\put(250,40){\line(1,0){30}}
\put(250,30){\line(1,0){30}}
\put(250,20){\line(1,0){30}}

\put(230,50){\line(0,1){10}}
\put(240,50){\line(0,1){10}}
\put(250,20){\line(0,1){40}}
\put(260,20){\line(0,1){40}}
\put(270,20){\line(0,1){30}}
\put(280,20){\line(0,1){30}}

\put(330,60){\line(1,0){40}}
\put(330,50){\line(1,0){60}}
\put(350,40){\line(1,0){40}}
\put(350,30){\line(1,0){60}}
\put(370,20){\line(1,0){40}}

\put(330,50){\line(0,1){10}}
\put(340,50){\line(0,1){10}}
\put(350,30){\line(0,1){30}}
\put(360,30){\line(0,1){30}}
\put(370,20){\line(0,1){40}}
\put(380,20){\line(0,1){30}}
\put(390,20){\line(0,1){30}}
\put(400,20){\line(0,1){10}}
\put(410,20){\line(0,1){10}}

\put(40,40){\circle*{5}}
\put(150,40){\circle*{5}}
\put(260,40){\circle*{5}}
\put(370,40){\circle*{5}}
\end{picture}
\vspace{0.1mm}

\noindent and the leftmost diagram corresponds to the states that remain of highest weight in 1...4 gradings at finite coupling, so we simply represent the multiplet by this diagram.

\end{shaded}

\vspace{4mm}
A given type of multiplet, specified by a set of oscillator numbers $\ns$, appears with a certain multiplicity in the spectrum. The multiplet content with $\Delta_0\le 8$ (10.535 multiplets in total) is provided in Appendix A.3 of \cite{Marboe:2017dmb}, and the content up to $\Delta_0\le 9$ (84.793 multiplets in total) can be found in the notebook \dbname\ in the ancillary files of this paper at arxiv.org. This information was derived from character theory (see also \cite{Bianchi:2003wx,Beisert:2003te}). 



\section{QSC essentials}\label{sec:QSC}
In this section, we review the features of the QSC that are relevant for our purposes. Our main focus is the self-contained $\bP\mu$-system, and we will only use the larger Q-system in the determination of the leading solutions.

\subsection{$\bP\mu$-system} \label{sec:Pmu}
The $\bP\mu$-system contains two types of multi-valued functions, $\bP$ and $\mu$, of the spectral parameter, $u$, labeled by indices $a,b,c,...\in\{1,2,3,4\}$. 
The crucial information in the QSC is the precise specification of the analytic structure and asymptotic behaviour of these functions. In this paper, we will always describe the QSC functions by choosing {\it short} branch cuts placed between $\pm2g+\ii \mathbb{Z}$.

\subsubsection*{$\bP_a$ and $\bP^a$}
\begin{figure}[h]
\centering
\begin{picture}(200,120)

\put(200,80){\large $\bP(u)$}
\put(200,20){\large $\Pt(u)$}

\color{black}
\thicklines
\put(0,0){\line(1,0){150}}
\put(50,50){\line(1,0){150}}
\put(0,0){\line(1,1){50}}
\put(150,0){\line(1,1){50}}

\color{blue!10}
\linethickness{0.75mm}
\put(91,60){\line(0,1){16}}
\put(93,60){\line(0,1){18}}
\put(95,60){\line(0,1){20}}
\put(97,60){\line(0,1){22}}
\put(99,60){\line(0,1){24}}
\put(101,60){\line(0,1){26}}
\put(103,60){\line(0,1){28}}
\put(105,60){\line(0,1){30}}
\put(107,60){\line(0,1){32}}
\put(109,60){\line(0,1){34}}
\color{blue!40}
\put(91,60){\line(0,-1){44}}
\put(93,60){\line(0,-1){42}}
\put(95,60){\line(0,-1){40}}
\put(97,60){\line(0,-1){38}}
\put(99,60){\line(0,-1){36}}
\put(101,60){\line(0,-1){34}}
\put(103,60){\line(0,-1){32}}
\put(105,60){\line(0,-1){30}}
\put(107,60){\line(0,-1){28}}
\put(109,60){\line(0,-1){26}}

\color{blue}
\linethickness{0.4mm}
\qbezier(90,75)(90,75)(110,95)

\qbezier(30,15)(30,15)(50,35)
\qbezier(60,15)(60,15)(80,35)
\qbezier(90,15)(90,15)(110,35)
\qbezier(120,15)(120,15)(140,35)
\qbezier(150,15)(150,15)(170,35)

\color{black}
\put(90,75){\circle*{3}}\put(110,95){\circle*{3}}

\put(30,15){\circle*{3}}
\put(60,15){\circle*{3}}
\put(90,15){\circle*{3}}
\put(120,15){\circle*{3}}
\put(150,15){\circle*{3}}

\put(50,35){\circle*{3}}
\put(80,35){\circle*{3}}
\put(110,35){\circle*{3}}
\put(140,35){\circle*{3}}
\put(170,35){\circle*{3}}

\thicklines
\put(0,60){\line(1,0){150}}
\put(50,110){\line(1,0){150}}
\put(0,60){\line(1,1){50}}
\put(150,60){\line(1,1){50}}

\put(106,100){\scriptsize$2g$}
\put(82,66){\scriptsize$\mi 2g$}

\put(106,40){\scriptsize$2g$}
\put(82,6){\scriptsize$\mi 2g$}

\put(129,40){\scriptsize$2g\!+\!\ii$}
\put(105,6){\scriptsize$\mi 2g\!+\!\ii$}

\put(69,40){\scriptsize$2g\!-\!\ii$}
\put(45,6){\scriptsize$\mi 2g\!-\!\ii$}

\end{picture}
\caption{Cut-structure for $\bP$. On the first sheet there is a short branch cut between $\pm 2g$, while there is an infinite ladder at $\pm2g+\ii\mathbb{Z}$ on the second sheet.}
\label{fig:Past}
\end{figure}
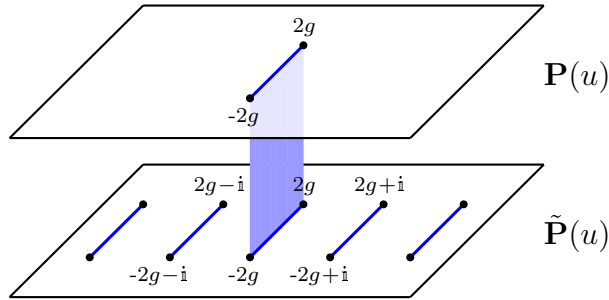

The eight functions $\bP_a$ and $\bP^a$ have a single branch cut between $\pm2g$ on their first Riemann sheet, on which we denote their value $\bP(u)$, while they have an infinite ladder of squareroot-type cuts at $\pm2g+\ii \mathbb{Z}$ on their second sheet, where their value is denoted $\Pt(u)$. See figure \ref{fig:Past}. 

\subsubsection*{$\mu_{ab}$ and $\mu^{ab}$}
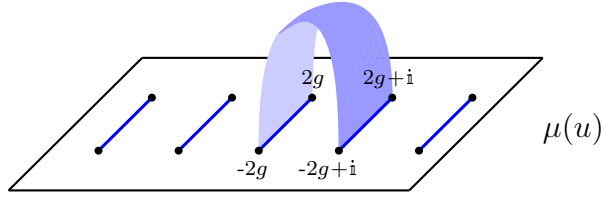
\begin{figure}[t]
\centering
\begin{picture}(200,80)

\put(200,20){\large $\mu(u)$}

\color{black}
\thicklines
\put(0,0){\line(1,0){150}}
\put(50,50){\line(1,0){150}}
\put(0,0){\line(1,1){50}}
\put(150,0){\line(1,1){50}}

\linethickness{0.79mm}
\color{blue!20}
\qbezier(91,16)(91,80)(120,60)
\qbezier(93,18)(93,80)(120,60)
\qbezier(95,20)(95,80)(120,60)
\qbezier(97,22)(97,80)(120,60)
\qbezier(99,24)(99,80)(120,60)
\qbezier(101,26)(101,80)(120,60)
\qbezier(103,28)(103,80)(120,60)
\qbezier(105,30)(105,80)(120,60)
\qbezier(107,32)(107,80)(120,60)
\qbezier(109,34)(109,80)(120,60)
\color{blue!40}
\qbezier(121,16)(121,80)(97,60)
\qbezier(123,18)(123,81)(97,60)
\qbezier(125,20)(125,82)(97,60)
\qbezier(127,22)(127,83)(97,60)
\qbezier(129,24)(129,84)(97,60)
\qbezier(131,26)(131,85)(97,60)
\qbezier(133,28)(133,86)(97,60)
\qbezier(135,30)(135,87)(97,60)
\qbezier(137,32)(137,88)(97,60)
\qbezier(139,34)(139,89)(97,60)
\color{white}
\linethickness{0.3mm}
\qbezier(93,55)(101,65)(111,70)
\qbezier(93,56)(101,66)(111,70)
\qbezier(93,57)(101,66)(111,70)
\qbezier(93,58)(101,66)(111,70)

\linethickness{0.4mm}\color{white}
\qbezier(92,15)(92,15)(112,35)
\qbezier(122,15)(122,15)(142,35)
\color{blue}
\qbezier(30,15)(30,15)(50,35)
\qbezier(60,15)(60,15)(80,35)
\qbezier(90,15)(90,15)(110,35)
\qbezier(120,15)(120,15)(140,35)
\qbezier(150,15)(150,15)(170,35)

\color{black}

\put(30,15){\circle*{3}}
\put(60,15){\circle*{3}}
\put(90,15){\circle*{3}}
\put(120,15){\circle*{3}}
\put(150,15){\circle*{3}}

\put(50,35){\circle*{3}}
\put(80,35){\circle*{3}}
\put(110,35){\circle*{3}}
\put(140,35){\circle*{3}}
\put(170,35){\circle*{3}}

\put(106,40){\scriptsize$2g$}
\put(82,6){\scriptsize$\mi 2g$}

\put(129,40){\scriptsize$2g\!+\!\ii$}
\put(105,6){\scriptsize$\mi 2g\!+\!\ii$}


\end{picture}
\caption{Cut-structure for $\mu$. On their defining sheet, $\mu$ have an infinite ladder of branch points at $\pm 2g+\ii \mathbb{Z}$. These functions have the very particular property $\tilde{\mu}=\mu^{[2]}$.}
\label{fig:muast}
\end{figure}

The functions $\mu$ carry two antisymmetric indices, $\mu_{ab}=-\mu_{ba}$, and thus six of these functions are inequivalent generically. The functions $\mu^{ab}$ satisfy $\mu_{ab}\mu^{bc}=\delta_a^c$ and can be written in terms of $\mu_{ab}$ through\footnote{Note that in the literature \cite{Gromov:2014caa} it is custom to set $\Pf(\mu)=1$. This corresponds to a particular way to fix the symmetries of the QSC. However, we choose to keep $\Pf(\mu)$ as a free parameter that will be automatically fixed by our chosen way to fix the symmetry.}
\be\label{muPf}
\mu^{ab}=-\frac{1}{2}\epsilon^{abcd}\mu_{cd}\frac{1}{\Pf(\mu)}\,,
\ee
where the Pfaffian is given by
\be\label{Pfaf}
\Pf(\mu) = \mu_{12}\mu_{34}-\mu_{13}\mu_{24}+\mu_{14}\mu_{23} = \frac{1}{8}\epsilon^{abcd}\mu_{ab}\mu_{cd}\,.
\ee
These functions have an infinite ladder of squareroot-type branch points at $\pm 2g+\ii \mathbb{Z}$ on all sheets. An important property is that the analytic continuation through the cut on the real axis on their first sheet, denoted $\tilde{\mu}(u)$, is equivalent to shifting the value on the first sheet by $\ii$, i.e.
\be \label{mumu}
\tilde{\mu}(u)=\mu(u+\ii)\,,
\ee
as depicted in figure \ref{fig:muast}.

By using the below-discussed relation \eqref{mudiff} and the property \eqref{PP0}
, one can show that $\Pf(\mu)=\Pf(\mu^{[2]})=\Pf(\tilde\mu)$, i.e. it is $\ii$-periodic and has no branch points. As it has powerlike asymptotics, it must then be a constant, and this constant can be controlled by the choice of normalisations $\cA_a$ and $\cA^a$ in \eqref{AABB} below.

\subsubsection*{Relations}
The functions $\bP$ and $\mu$ are related by a Riemann-Hilbert problem. They satisfy the relations
\be \label{muPtP}
\mu_{ab}-\tilde\mu_{ab}=\Pt_a\bP_b- \bP_a\Pt_b\,,
\ee
and
\begin{subequations}\label{Ptmu}
\be
\Pt_a\!\!&\!\!=\!\!&\!\!\mu_{ab}\bP^b\\ 
\Pt^a\!\!&\!\!=\!\!&\!\!{\mu^{ab}}\bP_b \,. 
\ee
\end{subequations}
Furthermore, the functions $\bP$ satisfy
\be\label{PP0}
\bP_a\bP^a=0\,.
\ee

By combining \eqref{mumu}, \eqref{muPtP} and \eqref{Ptmu}, we see that the functions $\mu_{ab}$ are solutions to the six coupled first-order difference equations
\be \label{mudiff}
\boxed{
\mu_{ab}-\mu_{ab}^{[2]}=\bP_b\bP^d\mu_{ad}^{[1\pm 1]}-\bP_a\bP^d\mu_{bd}^{[1\pm 1]}}
\ee
where we used the property $\mu_{ab}\bP^b=\mu_{ab}^{[2]}\bP^b$. These equations play a central role in our approach.



\subsubsection*{Asymptotics}

At large $u$, the functions $\bP$ behave as
\be \label{PPas}
\bP_a\simeq \cA_a u^{-\hl_a}
\quad\quad\quad\quad
\bP^a\simeq \cA^a u^{\hl_a-1} \,,
\ee
and the constant prefactors $\cA$ can be seen to satisfy the relations \cite{Gromov:2014caa}
\be\label{AABB}
\cA_a\cA^a &=&  \ii\frac{\prod_j ({\color{black}\hl_a}+{\color{black}\hn_j})}{\prod_{b\neq a} ({\color{black}\hl_a}-{\color{black}\hl_b})}\,,
\ee
where the reader should note that there is no sum over the index $a$ on the left hand side, in contrast to the conventions used elsewhere in the paper.

For solutions to the QSC that correspond to the single-trace operators of $\mathcal{N}=4$ SYM, the functions $\mu$ likewise have powerlike asymptotics at large $u$. This is an important constraint that singles out these ``physical" solutions. 

\subsection{Q-system} \label{sec:Qsys}
To understand the complete set of solutions to the difference equations \eqref{mudiff}, it is convenient to introduce a $\gl(4|4)$ Q-system. This system involves 256 functions
\be
Q_{A|I}=Q_{a_1a_2\hdots|i_1i_2\hdots}
\ee
with between zero and four antisymmetric indices of each of the two types $a_k,i_k\in\{1,2,3,4\}$, related by the QQ-relations
\begin{subequations}
\label{QQ}
\be
Q_{A|I}Q_{Aab|I}&=&Q^+_{Aa|I}Q^-_{Ab|I}-Q^-_{Aa|I}Q^+_{Ab|I}\label{QQ1}\\
Q_{A|I}Q_{A|Iij}&=&Q^+_{A|Ii}Q^-_{A|Ij}-Q^-_{A|Ii}Q^+_{A|Ij}\label{QQ2}\\
Q_{Aa|I}Q_{A|Ii}&=&Q^+_{Aa|Ii}Q^-_{A|I}-Q^-_{Aa|Ii}Q^+_{A|I} \label{QQ3}\,.
\ee
\end{subequations}
We can additionally define a Hodge dual Q-system through
\be \label{QHodge}
Q^{A|I}\equiv (-1)^{|\bar{A}||I|}\epsilon^{\bar{A}A}\epsilon^{\bar{I}I}Q_{\bar{A}|\bar{I}}\,,
\ee
which then satisfy QQ-relations identical to \eqref{QQ}. 	$|A|$ denotes the number of indices in the multi-index $A$, $\bar{A}$ is the complement of $A$, and $\epsilon^{1234}=1$. No sum over the indices $\bar{A}$ and $\bar{I}$ is implied in \eqref{QHodge}.

\subsubsection*{Identification of $\bP\mu$-system}
The functions $\bP$ are identified as
\be
\bP_a = Q_{a|\emp}\,,\quad \bP^a = Q^{a|\emp}\,.
\ee
A consequence of the QQ-relations is that $Q_{ab|ij}$ satisfy the equations
\be \label{QabijP}
Q_{ab|ij}^- - Q_{ab|ij}^+ = -\bP_a\bP^c Q_{bc|ij}^\pm + \bP_b\bP^c Q_{ac|ij}^\pm \,.
\ee
We then see that the functions $Q_{ab|ij}^-$ are the complete set of $6\times6$ solutions to \eqref{mudiff}. The functions $\mu_{ab}$ are thus linear combinations of $Q_{ab|ij}^-$ with $\ii$-periodic coefficients $\omega^{ij}$,
\be \label{muom}
\mu_{ab}=\frac{1}{2}\omega^{ij}Q_{ab|ij}^-\,.
\ee

\subsection{Symmetries} \label{sec:QSCsym}
As reviewed in section 3.6 of \cite{Marboe:2017dmb}, the Q-system has a large amount of symmetry. Only two types of symmetry transformations preserve the analytic properties and powerlike asymptotic behaviour of the functions. First, the $H$-symmetry which rotates $\bP$ and $\mu$,
\be \label{Hsym}
\bP_a\to H_a^{\;b} \bP_b\,,\quad\quad \mu_{ab}\to H_a^{\;c}H_b^{\;d}\mu_{cd}\,,
\ee
where $H$ is a constant matrix, which is restricted to be lower-triangular if we want to maintain the distinct asymptotics \eqref{PPas}. 
Second, the recalings of $\bP$ by powers of the \zhuk-variable $x$,
\be \label{xrescaling}
\bP_a \to x^\Lambda \bP_a \quad\quad
\bP^a \to x^{-\Lambda} \bP^a\,,
\ee
which leave $\mu$ invariant.

An advantage of considering the $\bP\mu$-system compared to the full Q-system is that the $H$-symmetry on the indices $i,j,...$ is avoided. This symmetry can be controlled by demanding pure asymptotic behaviour \cite{Gromov:2014caa,Gromov:2015vua}, but in practice this requires power expansion of $\eta$-functions at $u=\infty$, which is a computational burden that we decided to avoid.


\section{$\bP\mu$-system at weak coupling}\label{sec:QSCweak}
In this section, we analyse the properties of the QSC functions in the limit $g\to0$. This information will enable us to systematically construct perturbative corrections to the functions in section \ref{sec:pert}.

\begin{figure}[t]
\centering

\setlength{\unitlength}{0.9pt}

\begin{picture}(300,100)

\put(139,45){\huge $\rightarrow$}
\put(90,90){$u$}
\put(290,90){$u$}

\linethickness{0.6mm}
\put(35,10){\color{blue}\line(1,0){30}}
\put(35,30){\color{blue}\line(1,0){30}}
\put(35,50){\color{blue}\line(1,0){30}}
\put(35,70){\color{blue}\line(1,0){30}}
\put(35,90){\color{blue}\line(1,0){30}}

\put(35,10){\circle*{4}}
\put(65,10){\circle*{4}}
\put(35,30){\circle*{4}}
\put(65,30){\circle*{4}}
\put(35,50){\circle*{4}}
\put(65,50){\circle*{4}}
\put(35,70){\circle*{4}}
\put(65,70){\circle*{4}}
\put(35,90){\circle*{4}}
\put(65,90){\circle*{4}}

\put(250,10){\circle*{4}}
\put(250,30){\circle*{4}}
\put(250,50){\circle*{4}}
\put(250,70){\circle*{4}}
\put(250,90){\circle*{4}}

\thicklines
\put(0,0){\line(1,0){100}}
\put(00,100){\line(1,0){100}}
\put(00,0){\line(0,1){100}}
\put(100,0){\line(0,1){100}}

\put(200,0){\line(1,0){100}}
\put(200,100){\line(1,0){100}}
\put(200,0){\line(0,1){100}}
\put(300,0){\line(0,1){100}}

\end{picture}
\caption{As $g\to0$ the branch points at $\ii \mathbb{Z} \pm 2g$ collide at $\ii \mathbb{Z}$. The branch cuts disappear, and leave behind possible singularities.}
\label{fig:colcut}
\end{figure}
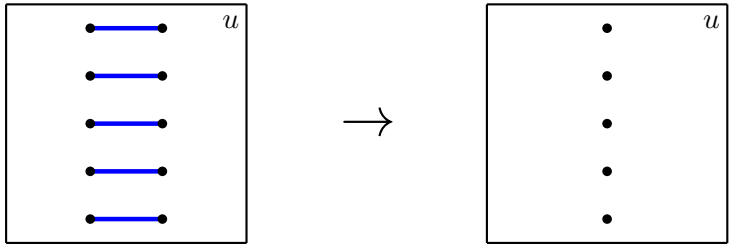

Considering the analytic structure of the QSC functions, we see that, as depicted in figure \ref{fig:colcut}, the branch points at $\ii \mathbb{Z} -2g$ and $\ii \mathbb{Z} +2g$ collide at $\ii \mathbb{Z}$ when $g\to0$. The branch cuts vanish, and the functions develop the possibility of having poles at these points instead.

Our basic assumption is that for solutions corresponding to multiplets of single-trace operators, any function in the QSC can be written as a power expansion in $g$, i.e.
\be \label{perts}
f(u,g)=g^n\sum_{j=0}^\infty f^{(j)}(u)g^{(j)}=g^n\left(f^{(0)}(u)+ g\,f^{(1)}(u)+g^2\,f^{(2)}+\hdots\right)\,.
\ee
Note that some functions may come with overall factors of $g$, and that such prefactors can usually be modified through the symmetry transformations mentioned in section \ref{sec:QSCsym}.

For solutions of the QSC corresponding to single-trace operators, the conformal dimension follows the pattern \eqref{perts} with only even powers being nonzero,
\be
\Delta(g) = \Delta_0 + \sum_{j=1}^{\infty} g^{2j}\gamma_j\,,
\ee
where the classical dimension $\Delta_0$ has the value
\be
\Delta_0 = n_{\aaa}+\frac{n_{\fff}}{2}\,.
\ee
This means that at the leading order, all functions have integer power-like asymptotics at $u\to\infty$, 
like the Q-functions of a rational spin chain.

\subsection{Structure of $\mu$}\label{sec:muprop}
Let us first analyse the properties of the functions $\mu$ at weak coupling.

\subsubsection*{Behaviour at the origin}
Due to the property $\tilde{\mu}=\mu^{[2]}$, the combinations 
\be\label{regmu}
\mu+\mu^{[2]} \quad\quad \text{and} \quad\quad \frac{\mu-\mu^{[2]}}{\sqrt{u^2-4g}} 
\ee
have no branch points on the real axis.
As singularities at weak coupling only arise from colliding branch points, these combinations should be regular at $u=0$ at any perturbative order. This turns out to be a useful constraint.

\subsubsection*{Polynomiality at the leading order} 
To satisfy both regularity constraints \eqref{regmu}, $\mu$ can neither have poles at  $u=0$ or $u=\ii$ at the leading order. Furthermore, $\mu$ satisfies relations of the type \eqref{mudiff}, which schematically have the form
\be
\mu-\mu^{[2]}=(\mu\bP)\bP-\bP(\mu\bP) = (\mu^{[2]}\bP)\bP-\bP(\mu^{[2]}\bP) \,,
\ee
and we can use these relations to make the recursive replacements
\begin{subequations}
\be
\mu^{[n-2]}&=&\mu^{[n]}-\left((\mu\bP^{[-2]})\bP^{[-2]}-\bP^{[-2]}(\mu\bP^{[-2]})\right)^{[n]}
\label{sub1}
\\
\mu^{[n+2]}&=&\mu^{[n]}-\left((\mu\bP)\bP-\bP(\mu\bP)\right)^{[n]}
\,.\label{sub2}
\ee
\end{subequations}
Using \eqref{sub1} to replace $\mu^{[2]}$ in the combination $\mu+\mu^{[2]}$, and recalling that $\bP$ can only be singular at $u=0$, we see that $\mu^{[4]}$ has to be regular at $u=0$, i.e. $\mu$ is regular at $u=2\ii$. This argument can be applied recursively to argue that $\mu$ has no poles at $\ii\mathbb{Z}_+$. Furthermore, the replacement \eqref{sub2} is used in the same way to argue that $\mu$ has no poles at $\ii\mathbb{Z}_-$. In conclusion, $\mu$ are completely regular functions at the leading order with power-like asymptotics. Thus they are polynomials, and hence their asymptotics are not only power-like but also integer.

Through \eqref{Ptmu}, this observation also implies that, at the leading order, $\Pt$ can only have poles at $u=0$ and have integer power-like asymptotics.

\subsubsection*{The zero-momentum condition}
One can also write
\be
\mu-\mu^{[2]}=\sqrt{u^2-4g^2}\times \frac{\mu-\mu^{[2]}}{\sqrt{u^2-4g^2}}\,.
\ee
As the second term on the r.h.s. has no cut $[-2g,2g]$ on the real axis and hence also no poles there by the regularity assumption, one gets  $\mu-\mu^{[2]}=u\times{\rm Reg}$ at the leading order of the weak coupling expansion. The same conclusion can be made about any derivative of $\mu$, and therefore one has
\be \label{muzmc}
\lim_{u\to0} \frac{\mu^{(0)}(u+\ii)}{\mu^{(0)}(u)} = 1\,,
\ee
which we will refer to as the {\it zero-momentum condition}. 

\subsubsection*{Maximal order of poles in the perturbative expansion}
We assume that the perturbative expansion of $\mu$ has the form
\be
g^n \mu= \mu^{(0)}+g\,\mu^{(1)}+g^2\,\mu^{(2)}+\hdots\,,
\ee
where $n$ is some number. The regularity of $\mu+\mu^{[2]}$ on the real axis tells us that the poles of $\mu$ and $\mu^{[2]}$ at $u=0$ must be identical up to a sign. For $\frac{\mu-\mu^{[2]}}{\sqrt{u^2-4g^2}}$ to also be regular, the maximal order of the poles at $u=0$ is restricted. Perturbatively, this quantity looks like
\be 
g^n\frac{\mu-\mu^{[2]}}{\sqrt{u^2-4g^2}} &=& \frac{\mu^{(0)}-\mu^{(0)[2]}}{u} + g \frac{\mu^{(1)}-\mu^{(1)[2]}}{u} \\
&& + g^2 \left( \frac{\mu^{(2)}-\mu^{(2)[2]}}{u} +2 \frac{\mu^{(0)}-\mu^{(0)[2]}}{u^3} \right) \no\\ 
&& + g^3 \left( \frac{\mu^{(3)}-\mu^{(3)[2]}}{u} +2 \frac{\mu^{(1)}-\mu^{(1)[2]}}{u^3} \right) \no\\ 
&& + g^4\left(\frac{\mu^{(4)}-\mu^{(4)[2]}}{u} +2 \frac{\mu^{(2)}-\mu^{(2)[2]}}{u^3}+6\frac{\mu^{(0)}-\mu^{(0)[2]}}{u^5}\right) + \CO(g^5) \no \,.
\ee
As $\mu^{(0)}-\mu^{(0)[2]}\sim u$, the maximal pole in $\mu^{(2)}$ is $u^{-1}$. Then the maximal pole in $\mu^{(4)}$ is $u^{-3}$, and recursively we then see that $\mu^{(2n)}$ and $\mu^{(2n+1)}$ have the maximal pole $u^{-2n+1}$ at $u=0$. The same holds for $\mu^{[2]}$ and then, by the recursive argument as above, the same restriction applies for all poles of $\mu$ at $u=\ii\mathbb{Z}$.

\subsection{Structure of $\bP$} \label{sec:Pstructure}
An important property that makes it possible to construct perturbative corrections to the QSC functions is the simple structure of the functions $\bP$. As discussed in section \ref{sec:Pmu}, these functions have integer power-like asymptotics at any value of $g$, and they have only a single short branch cut on the first Riemann sheet. In this section, we explain how to write a precise ansatz for these functions that contains only a finite number of free parameters at each perturbative order.

\subsubsection*{\zhuk\ parametrisation}
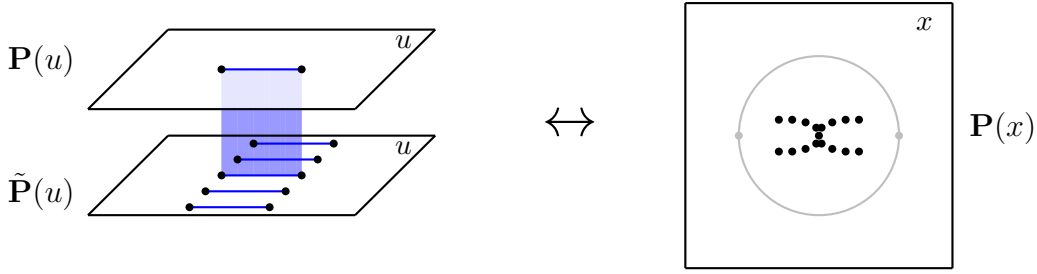
\begin{figure}[t]
\centering
\begin{picture}(330,95)

\put(-30,75){\large $\bP(u)$}
\put(-30,25){\large $\Pt(u)$}
\put(170,50){\huge $\leftrightarrow$}
\put(115,83){$u$}
\put(115,43){$u$}
\put(310,90){$x$}
\put(330,50){\large $\bP(x)$}

\color{blue!10}
\linethickness{0.75mm}
\put(51,60){\line(0,1){15}}
\put(53,60){\line(0,1){15}}
\put(55,60){\line(0,1){15}}
\put(57,60){\line(0,1){15}}
\put(59,60){\line(0,1){15}}
\put(61,60){\line(0,1){15}}
\put(63,60){\line(0,1){15}}
\put(65,60){\line(0,1){15}}
\put(67,60){\line(0,1){15}}
\put(69,60){\line(0,1){15}}
\put(71,60){\line(0,1){15}}
\put(73,60){\line(0,1){15}}
\put(75,60){\line(0,1){15}}
\put(77,60){\line(0,1){15}}
\put(79,60){\line(0,1){15}}
\color{blue!40}
\put(51,60){\line(0,-1){25}}
\put(53,60){\line(0,-1){25}}
\put(55,60){\line(0,-1){25}}
\put(57,60){\line(0,-1){25}}
\put(59,60){\line(0,-1){25}}
\put(61,60){\line(0,-1){25}}
\put(63,60){\line(0,-1){25}}
\put(65,60){\line(0,-1){25}}
\put(67,60){\line(0,-1){25}}
\put(69,60){\line(0,-1){25}}
\put(71,60){\line(0,-1){25}}
\put(73,60){\line(0,-1){25}}
\put(75,60){\line(0,-1){25}}
\put(77,60){\line(0,-1){25}}
\put(79,60){\line(0,-1){25}}

\color{black}
\thicklines
\put(50,35){\color{blue}\line(1,0){30}}
\put(50,75){\color{blue}\line(1,0){30}}
\put(56,41){\color{blue}\line(1,0){30}}
\put(62,47){\color{blue}\line(1,0){30}}
\put(44,29){\color{blue}\line(1,0){30}}
\put(38,23){\color{blue}\line(1,0){30}}

\thinlines

\put(56,41){\circle*{3}}
\put(86,41){\circle*{3}}
\put(62,47){\circle*{3}}
\put(92,47){\circle*{3}}
\put(44,29){\circle*{3}}
\put(74,29){\circle*{3}}
\put(38,23){\circle*{3}}
\put(68,23){\circle*{3}}

\put(50,75){\circle*{3}}
\put(80,75){\circle*{3}}
\put(50,35){\circle*{3}}
\put(80,35){\circle*{3}}

\thicklines
\put(0,20){\line(1,0){100}}
\put(30,50){\line(1,0){100}}
\put(0,20){\line(1,1){30}}
\put(100,20){\line(1,1){30}}

\put(0,60){\line(1,0){100}}
\put(30,90){\line(1,0){100}}
\put(0,60){\line(1,1){30}}
\put(100,60){\line(1,1){30}}

{
\color{gray!50}
\put(270,50){\circle{60}}
\put(240,50){\circle*{3}}
\put(300,50){\circle*{3}}
}

\thicklines
\put(220,0){\line(1,0){100}}
\put(220,100){\line(1,0){100}}
\put(220,0){\line(0,1){100}}
\put(320,0){\line(0,1){100}}

\put(270,50){\circle*{3}}

\put(271,53){\circle*{3}}
\put(269,53){\circle*{3}}
\put(271,47){\circle*{3}}
\put(269,47){\circle*{3}}

\put(275,55){\circle*{3}}
\put(265,55){\circle*{3}}
\put(275,45){\circle*{3}}
\put(265,45){\circle*{3}}

\put(280,56){\circle*{3}}
\put(260,56){\circle*{3}}
\put(280,44){\circle*{3}}
\put(260,44){\circle*{3}}

\put(285,56){\circle*{3}}
\put(255,56){\circle*{3}}
\put(285,44){\circle*{3}}
\put(255,44){\circle*{3}}

\end{picture}
\caption{For the functions $\bP$, the first two Riemann sheets in $u$ (left) get mapped to a single Riemann sheet in $x$ (right). The {\color{gray}grey} circle on the $x$-sheet is the unit circle, and the branch points on the real axis in the $u$-sheets are mapped to the points $x=\pm1$.}
\label{fig:zhukP}
\end{figure}

Make a change of variables from the spectral parameter $u$ to the \zhuk\ variable $x(u)$ \eqref{zhukdef}. This maps the first two Riemann sheets in $u$ to a single Riemann sheet in $x$. The first sheet, where we denote the value $\bP(u)$, is mapped outside the unit circle, $|x|>1$, while the second sheet, denoted $\Pt(u)$, is mapped to the inside of the unit circle, $|x|<1$, see figure \ref{fig:zhukP}. The branch cut on the real axis is resolved and it is mapped to the unit circle in the $x$-plane.

$\bP(x)$ can be written as a power expansion around $x=\infty$. Since $x$ behaves like $x\sim \frac{u}{g}$ at large $x$, the functions $\bP(x)$ have integer power-like asymptotics, $\bP_a(x)\sim x^{-\hl_a}$ and $\bP^a(x)\sim x^{\hl_a-1}$. The power expansion of $\bP(x)$ around $x=\infty$ then has the form
\be \label{pansx}
\bP_a(x)=\sum_{k=\hl_a}^{\infty} \frac{C_{a,k}}{x^k}
\quad\quad\quad
\quad\quad\quad
\bP^a(x)=\sum_{k=-\hl_a+1}^{\infty} \frac{C^{a,k}}{x^k}\;,
\ee
where $C$ are coefficients that depend on $g$. This expansion converges until the first branch points are reached in the $x$-plane, i.e. for $|x|>|x(\ii+2g)|$. This means that the expansion is convergent everywhere on the first $u$-sheet, $\bP(u)$, and in a finite region around the branch cut on the real axis on the second $u$-sheet, $\Pt(u)$.

Crossing the unit circle in the $x$-plane corresponds to crossing the short branch cut in $x(u)$. As we restrict $x(u)$ to be the solution of \eqref{zhukdef} with $|x|>1$, this corresponds to the replacement $x\to\frac{1}{x}$ in \eqref{pansx}.

\subsubsection*{Explicit ansatz at weak coupling}

Let us analyse the $g$-dependence of the coefficients $C$ in the ansatz \eqref{pansx} more carefully. Assuming that the asymptotic behaviour \eqref{PPas} is valid at $g=0$, it follows from the constraints \eqref{AABB} that the products $\cA_a \cA^a$ are regular in $g$. These products are generically $\CO(g^0)$, except in the case of shortening:
\begin{subequations} \label{shortAA}
\be
n_{\fff_1}^\beautysub=L-1 & \Rightarrow & \cA_1\cA^1=\CO(g^2) \label{S1} \\
n_{\fff_4}^\beautysub=1 & \Rightarrow & \cA_4\cA^4=\CO(g^2) \label{S2} \,.
\ee
\end{subequations}
 
This means that we can always choose to fix the freedom in $\cA$ such that all $\bP_a(u)$ and $\bP^a(u)$ are regular in $g$ with the possibility of some of them being $\CO(g^2)$ due to shortening. 

We can then write a more precise version of \eqref{pansx}:
\begin{subequations}\label{pansi}
\be 
\bP_a(x)&=&
\frac{\cA_a}{(gx)^{\l_a^0+\Lambda}}\sum_{k=0}^{\infty} \frac{\ccc_{a,k}}{(gx)^k} \\
\bP^a(x)&=&
\frac{\cA^a}{(gx)^{-\l_a^0+1-\Lambda}}\sum_{k=0}^{\infty} \frac{\ccc^{a,k}}{(gx)^k}\,,
\ee
\end{subequations}
where $\l_a^0$ is invariant under the symmetries described in section \ref{sec:QSCsym} and was defined in \eqref{hl0}, while $\Lambda$ corresponds to the symmetry \eqref{xrescaling}. 
The constants $\ccc$ are here strictly regular in $g$, and we assume that they can be written as expansions in $g$,
\be
\ccc=\sum_{j=0}^\infty \ccc^{(j)} g^{j}\,.
\ee
The first coefficient should match the choice of normalisations $\cA$, i.e.
\be
\ccc_{a,0}=\ccc^{a,0}=1\,.
\ee

Restricting to the branch $|x|>1$, the expansion that describes the value on the second sheet $\Pt(u)$ is obtained through the replacement $x\to\frac{1}{x}$ in \eqref{pansi}. This gives
\begin{subequations}\label{ptansi}
\be 
\Pt_a(x)&=& \cA_a
{\left(\frac{x}{g}\right)^{\l_a^0+\Lambda}}\sum_{k=0}^{\infty} {\ccc_{a,k}}{\left(\frac{x}{g}\right)^k} \\
\Pt^a(x)&=& \cA^a
{\left(\frac{x}{g}\right)^{-\l_a^0+1-\Lambda}}\sum_{k=0}^{\infty} {\ccc^{a,k}}{\left(\frac{x}{g}\right)^k}\,.
\ee
\end{subequations}
The expansion is convergent in a finite region around $u=0$.

\subsubsection*{Scaling considerations}
As discussed in section \ref{sec:muprop}, $\mu_{ab}$ are polynomials in $u$ at the leading order of maximal degree $-\l_a^0-\l_b^0-\nu_1^0-\nu_2^0$, corresponding to the strongest possible asymptotics among the functions $Q_{ab|ij}$. Since $\Pt(u)$ are given by \eqref{Ptmu}, the ansatz \eqref{ptansi} must coincide with the exact expression for $\Pt^{(0)}\!(u)$ at the leading order in $g$. In other words, the sums in \eqref{ptansi} are truncated at the leading order. The leading contributions $\Pt^{(0)}(u)$ have integer power-like asymptotics of maximal degree
\begin{subequations}\label{Ptmax}
\be
\Pt^{(0)}_a(u) &\sim& u^{-\l_a^0-\nu_1^0-\nu_2^0-1+\Lambda} \\
\Pt^a_{(0)}(u) &\sim& u^{\l_a^0+\nu_3^0+\nu_4^0-\Lambda} \,.
\ee
\end{subequations}
Denote the term in the sums \eqref{ptansi} responsible for the leading power in $u$ of $\Pt^{(0)}$ by $k=M_a$ and $k=M^a$, respectively. These numbers are limited by \eqref{Ptmax}. 
Use $N\ge0$ to denote the scaling in $g$ of these coefficients:
\be
c_{a,M_a}\sim g^{N_a} \quad\quad\quad c^{a,M^a}\sim g^{N^a}\,.
\ee
The behaviour $x\sim\frac{u}{g}+\CO(g)$ and the bounds \eqref{Ptmax} put constraints on how the coefficients $\ccc$ in the ansatz \eqref{ptansi} scale in $g$. As we assume $c_{a,0}^{(0)}=c^{a,0}_{(0)}=1$, we see that $N_a\le 2M_{a}$ and $N^a\le 2M^a$, since otherwise $c_{a,M_a}$ and $c^{a,M^a}$ would not contribute to $\Pt^{(0)}$. We can also conclude that
\be
\ccc_{a,k}^{(j)}=0 \quad \text{for} \quad \left\{ \begin{matrix} 
j\le 2(k-M_a)-N_a & \wedge & 
M_a-\frac{N_a}{2}<k\le M_a \\ 
j\le 2(k-M_a)-N_a +1 & \wedge& 
k >M_a 
\end{matrix} \right. \,.
\ee

\subsubsection*{Parity of $N$}
There are now two possible scenarios: $N$ being odd or even. The parity of $N$ has important consequences for the ansatz. See table \ref{tab:cs} for an overview of the scaling of the coefficients $c$ in the two scenarios.

\begin{table}[t]

\def\SubZero{\cellcolor{gray!10}}

\def\LO{\cellcolor{Dandelion!100}}
\def\NLO{\cellcolor{ProcessBlue!100}}
\def\NNLO{\cellcolor{Blue!60}}
\def\NNNLO{\cellcolor{Blue!50}}
\def\NNNNLO{\cellcolor{Blue!40}}
\def\NNNNNLO{\cellcolor{Blue!30}}

\def\suA{\cellcolor{Magenta!60}}
\def\suB{\cellcolor{Magenta!50}}
\def\suC{\cellcolor{Magenta!40}}
\def\suD{\cellcolor{Magenta!30}}
\def\suE{\cellcolor{Magenta!20}}
\def\suF{\cellcolor{Magenta!10}}

\def\subA{\cellcolor{Emerald!70}}
\def\subB{\cellcolor{Emerald!60}}
\def\subC{\cellcolor{Emerald!50}}
\def\subD{\cellcolor{Emerald!40}}
\def\subE{\cellcolor{Emerald!30}}
\def\subF{\cellcolor{Emerald!20}}

\def\sA{\cellcolor{RawSienna!80}}
\def\sB{\cellcolor{RawSienna!70}}
\def\sC{\cellcolor{RawSienna!60}}
\def\sD{\cellcolor{RawSienna!50}}
\def\sE{\cellcolor{RawSienna!40}}
\def\sF{\cellcolor{RawSienna!30}}
\def\sG{\cellcolor{RawSienna!20}}
\def\sH{\cellcolor{RawSienna!10}}

\centering
{\small\begin{tabular}{|c||c|c|c|c|c|c|c|c|c|c|c||c|c|c|c|c|c|c|c|c|c|c|} \hline

$c^{(j)}_{\bullet,k}$ &\multicolumn{11}{c||}{even $N$}& \multicolumn{10}{c|}{odd $N$} \\\hline

\backslashbox{$j$}{$k$} & 
\!\!\rotatebox{90}{0}\!\! & 
\!\rotatebox{90}{1}\! & 
\!\rotatebox{90}{2}\! & 
\!\!$\cdots$\!\!\! &
\!\rotatebox{90}{$\!\scriptstyle M-\frac{N}{2}-1$}\! &
\!\rotatebox{90}{$\!\scriptstyle M-\frac{N}{2}$}\! &
\!\rotatebox{90}{$\!\scriptstyle M-\frac{N}{2}+1$}\! &
\!\!$\cdots$\!\!\! &
\!\rotatebox{90}{$\!\scriptstyle M$}\! &
\!\rotatebox{90}{$\!\scriptstyle M+1$}\! &
\!\!$\cdots$\!\!\! &
\!\rotatebox{90}{0}\! & 
\!\rotatebox{90}{1}\! & 
\!\rotatebox{90}{2}\! & 
\!\!$\cdots$\!\!\! &
\!\rotatebox{90}{$\!\scriptstyle M-\frac{N}{2}-\frac{1}{2}$}\! &
\!\rotatebox{90}{$\!\scriptstyle M-\frac{N}{2}+\frac{1}{2}$\,}\! &
\!\!$\cdots$\!\!\! &
\!\rotatebox{90}{$\!\scriptstyle M$}\! &
\!\rotatebox{90}{$\!\scriptstyle M+1$}\! &
\!\!$\cdots$\!\!\! 
\\\hline\hline

0 & 
\!\subE{\color{gray}1}\!&
\subC  & 
\subA  &
\!\!$\cdots$\!\!\! &
\NNLO&
\LO&
\multicolumn{5}{c||}{\SubZero} &
\subE{\color{gray}1}&
\subC  & 
\subA  &
\!\!$\cdots$\!\!\! &
\NLO&
\multicolumn{5}{c|}{{\color{gray!10}\!$\vdots$\!} \SubZero}
\\ \cline{1-7}\cline{13-18}

1 & 
\SubZero\multirow{8}{*}{0}   &
\subD  &
 \subB   &
\!\!$\cdots$\!\!\! & \NNNLO &
\NLO&
\multicolumn{5}{c||}{\SubZero} &
\SubZero\multirow{8}{*}{0}   &
\subD  &
\subB   &
\!\!$\cdots$\!\!\! & \NNLO &
\LO&
\multicolumn{4}{c|}{{\color{gray!10}\!$\vdots$\!}\SubZero}
\\\cline{1-1}\cline{3-8}\cline{14-18}

2 & \SubZero &
\subE   &
\subC    &
\!\!$\cdots$\!\!\! &
\NNNNLO&
\NNLO&
\LO&
\multicolumn{4}{c||}{\SubZero}&
\SubZero &
\subE   &
\subC    &
\!\!$\cdots$\!\!\! &
\NNNLO&
\NLO&
\multicolumn{4}{c|}{{\color{gray!10}\!$\vdots$\!}\SubZero}
\\ \cline{1-1}\cline{3-8}\cline{14-19}

3 & 
\SubZero  &
\subF  &
\subD  & 
\!\!$\cdots$\!\!\! &
\NNNNNLO&
\NNNLO&
\NLO&
\multicolumn{4}{c||}{\SubZero}&
\SubZero  &
\subF  &
\subD  & 
\!\!$\cdots$\!\!\! &
\NNNNLO&
\NNLO&
\!$\ddots$\!&
\multicolumn{3}{c|}{\SubZero} 
\\ \cline{1-1}\cline{3-9}\cline{14-19}

\!$\vdots$\! & \SubZero  & \!$\vdots$\! & \!$\vdots$\! & \!$\ddots$\! & \!$\vdots$\! & \!$\vdots$\! & \!$\vdots$\! & \!$\ddots$\! & \multicolumn{3}{c||}{\SubZero} &
\SubZero  & \!$\vdots$\! & \!$\vdots$\! & \!$\ddots$\! & \!$\vdots$\! & \!$\vdots$\! & \!$\ddots$\! & \multicolumn{3}{c|}{\SubZero}
\\ \cline{1-1}\cline{3-10}\cline{14-20}


$N$ & \SubZero  &
\suC  &
\suA    &
\!\!$\cdots$\!\!\! &
\sE&
\sC&
\sA&
\!$\vdots$\!&
\LO&
\multicolumn{2}{c||}{\SubZero}&
 \SubZero  &
\suC  &
\suA    &
\!\!$\cdots$\!\!\! &
\sC&
\sA&
\!$\vdots$\!&
\LO&
\multicolumn{2}{c|}{\SubZero}
\\ \cline{1-1}\cline{3-10}\cline{14-20}

$\!\!N\!+\!1\!\!$ & 
\SubZero  &  
\suD&
\suB  & 
\!\!$\cdots$\!\!\! &
\sF&
\sD&
\sB&
\!$\vdots$\!&
\NLO&
\multicolumn{2}{c||}{\SubZero}&
\SubZero  &  
\suD&
\suB  & 
\!\!$\cdots$\!\!\! &
\sD&
\sB&
\!$\vdots$\!&
\NLO&
\multicolumn{2}{c|}{\SubZero} 
\\ \cline{1-1}\cline{3-10}\cline{14-20}

$\!\!N\!+\!2\!\!$ & 
\SubZero  &
\suE  & 
\suC & 
\!\!$\cdots$\!\!\! &
\sG&
\sE&
\sC&
\!$\vdots$\!&
\NNLO&
\multicolumn{2}{c||}{\SubZero}&
\SubZero  &
\suE  & 
\suC & 
\!\!$\cdots$\!\!\! &
\sE&
\sC&
\!$\vdots$\!&
\NNLO&
\multicolumn{2}{c|}{\SubZero}
\\ \cline{1-1}\cline{3-10}\cline{14-21}

$\!\!N\!+\!3\!\!$ & 
\SubZero  & 
\suF & 
\suD & 
\!\!$\cdots$\!\!\! &
\sH&
\sF&
\sD&
\!$\vdots$\!&
\NNNLO&
\NLO&
\SubZero&
\SubZero  & 
\suF & 
\suD & 
\!\!$\cdots$\!\!\! &
\sF&
\sD&
\!$\vdots$\!&
\NNNLO&
\NLO&
\SubZero
\\ \cline{1-1}\cline{3-12}\cline{14-22}

\!$\vdots$\! & 
\SubZero  & 
\!$\vdots$\! & 
\!$\vdots$\!  & 
\!\!$\cdots$\!\!\! & 
\!$\vdots$\! & 
\!$\vdots$\! & 
\!$\vdots$\! & 
\!$\ddots$\! & 
\!$\vdots$\! & 
\!$\vdots$\! & 
\!$\ddots$\! &
\SubZero  & 
\!$\vdots$\! & 
\!$\vdots$\!  & 
\!\!$\cdots$\!\!\! & 
\!$\vdots$\! & 
\!$\vdots$\! & 
\!$\ddots$\! & 
\!$\vdots$\! & 
\!$\vdots$\! & 
\!$\ddots$\! \\\hline

\end{tabular}}
\caption{Overview of the nonzero constants $\ccc_{a,k}^{(j)}$ for even and odd $N$. Rows contribute to $\bP(u)$ at same order in $g$. Boxes with the same colour contribute to $\Pt(u)$ at same order in $g$, with the leading order marked in {\color{Dandelion}\bf yellow}, the subleading order marked in {\color{ProcessBlue}\bf blue}, etc. As discussed below, only even orders in $j$ turn out to be present for even $N$.}
\label{tab:cs}
\end{table}

\vspace{4mm}
\noindent {\bf If $N$ is even}, the leading order of $\bP$ and $\Pt$ have the form
\begin{subequations}\be
\bP_a^{(0)}(u)&=&
\cA_a u^{-\l_a^0-\Lambda-M_a+\frac{N_a}{2}} \left(
u^{M_a-\frac{N_a}{2}}+\sum_{k=1}^{M_a-\frac{N_a}{2}} \ccc_{a,k}^{(0)}\; u^{M_a-\frac{N_a}{2}-k} \right)
\\
\Pt_a^{(0)}(u)&=&
\cA_a\left(\frac{u}{g^2}\right)^{\l_a^0+\Lambda+M_a-\frac{N_a}{2}} 
\left(
\sum_{k=0}^{\frac{N_a}{2}} 
\ccc_{a,M_a-\frac{N_a}{2}+k}^{(2k)}\; u^k 
\right)\,,
\ee
\end{subequations}
and similarly for $\bP^a_{(0)}$ and $\Pt^a_{(0)}$. Notice that, if $N$ is even, then $\bP^{(0)}$ and $\Pt^{(0)}$ have the coefficient of the lowest power, $\ccc^{(0)}_{M-\frac{N}{2}}$, in common (though this coefficient can in principle be zero). Their product behaves as
\be
\bP_a^{(0)}\Pt_a^{(0)}&\!\!\!=\!\!\!&\cA_a^2 g^{-2(\l_a^0+\Lambda+M_a-\frac{N_a}{2})} \times \\ 
&&\left( \left(c_{a,M_a-\frac{N_a}{2}}^{(0)}\right)^2 + u \, \ccc_{a,M_a-\frac{N_a}{2}}^{(0)}\left( \ccc_{a,M_a-\frac{N_a}{2}-1}^{(0)} + \ccc_{a,M_a-\frac{N_a}{2}+1}^{(2)} \right) + \CO(u^2) \right)\,.\quad \no
\ee
Consequently, the product $\bP_a^{(0)}\Pt_a^{(0)}$ starts from a constant term, unless $\ccc_{a,M_a-\frac{N_a}{2}}^{(0)}=0$, in which case the product starts from a quadratic term or higher.

\vspace{4mm}
\noindent {\bf If $N$ is odd}, the leading order of $\bP$ and $\Pt$ instead have the form
\begin{subequations}
\be
\bP_a^{(0)}(u)&\!\!=\!\!&
\cA_a u^{-\l_a^0-\Lambda-M_a+\frac{N_a}{2}+\frac{1}{2}} \left(\!
u^{M_a-\frac{N_a}{2}-\frac{1}{2}}+\sum_{k=1}^{M_a-\frac{N_a}{2}-\frac{1}{2}} \ccc_{a,k}^{(0)}\; u^{M_a-\frac{N_a}{2}-\frac{1}{2}-k} \!\right)\quad\quad
\\
\Pt_a^{(0)}(u)&\!\!=\!\!&
\cA_a\left(\frac{u}{g^2}\right)^{\l_a^0+\Lambda+M_a-\frac{N_a}{2}+\frac{1}{2}}   g
\left(
\sum_{k=0}^{\frac{N_a-1}{2}} 
\ccc_{a,M_a-\frac{N_a}{2}+\frac{1}{2}+k}^{(2k+1)}\; u^k \right)\,.
\ee
\end{subequations}
So for odd $N$, $\bP^{(0)}$ and $\Pt^{(0)}$ do not contain overlapping coefficients. This is only the case at subleading orders. Their product behaves like
\be
\bP_a^{(0)}\Pt_a^{(0)}&=&\cA_a^2 g^{-2(\l_a^0+\Lambda+M_a-\frac{N_a}{2})}
\left( u \, \ccc^{(0)}_{a,M_a-\frac{N_a}{2}-\frac{1}{2}} \ccc^{(1)}_{a,M_a-\frac{N_a}{2}+\frac{1}{2}} +\CO(u^2) \right)\,,\quad
\ee
i.e. it starts with a linear term.

\subsubsection*{Comparison to the leading Q-system}

The crucial numbers to understand are the differences $M_a-\frac{N_a}{2}$. They determine the degree of the polynomial parts of $\bP^{(0)}$. To understand what these numbers are, we can compare with solutions of the leading Q-system, which was analysed in detail in \cite{Marboe:2017dmb}. The structure is $Q_{a|\emp}=\bP_a^{(0)}=u^{-L^\beautysub-\Lambda} \ps_a$ and $Q^{a|\emp}=\bP^a_{(0)}=u^\Lambda \ps^a$, where $\ps$ are polynomials of degree
\begin{subequations}\label{degP}
\be
\text{degree}(\ps_a)
&=&L^{\beautysub}-\lambda_a^0\\
\text{degree}(\ps^a)
&=&\lambda_a^0-1\,.
\ee
\end{subequations}
Here we made the choice to work in the grading $12\hat{1}\hat{2}\hat{3}\hat{4}34$. This implies the choice $\bP_1\sim \CO(g^2)$ and $\bP^1 \sim \CO(g^0)$ in the case of the shortening $\l_1^0 = L^\beautysub + 1$, and similarly $\bP_4\sim \CO(g^0)$ and $\bP^4 \sim \CO(g^2)$ in the case of the shortening $\l_4^0 = 0$. 

\subsubsection*{Classification of solutions}
In section \ref{sec:pert}, we describe an algorithm to explicitly construct the perturbative $\bP\mu$-system from the leading Q-system. This algorithm uses the explicit structure for the ansatz of $\bP$, but it is of course possible to start from a rather general ansatz that includes enough terms to be on the safe side. The solutions fall into two classes that we denote
{
\begin{center}
\begin{tabular}{cc}
{\bf Typical} solutions: & $N$ even \\ 
{\bf Degenerating} solutions:& $N_\bullet$ and/or $N^\bullet$ odd 
\end{tabular}
\end{center}
}

\noindent For the typical solutions, we observe that all QSC functions have expansions in $g^2$, i.e. only the even orders $f^{(2n)}$ in \eqref{perts} will be nonzero. For the degenerating solutions, the QSC functions are expanded in $g$.

\subsubsection*{Final ansatz - typical solutions}
We have made an extensive analysis of the spectrum with $\Delta_0\le 9$, and the vast majority of the states correspond to typical solutions. For these states, based on the comparison \eqref{degP}, we can identify
\begin{subequations}
\be
M_a-\frac{N_a}{2}&=&L^\beautysub - \l_a^0 + \delta_{a,1}\,\delta_{\l_1^0, L^\beautysub + 1} \\
M^a-\frac{N^a}{2}&=& \l_a^0 -1 + \delta_{a,4}\,\delta_{\l_4^0,0}\,.
\ee
\end{subequations}
We can then rephrase the ansatz \eqref{pansi} as
\be\label{PANS}
\boxed{
\begin{matrix} &\bP_a &=&  \displaystyle (g x)^{-L^\beautysub-\Lambda} \left(\sum_{k=0}^{L^\beautysub-\l_a^0} \cd_{a,k}\left({g}{x}\right)^k+\sum_{k=1}^\infty \cc_{a,k}\left(\frac{g}{x}\right)^k\right) & \\[20pt]
&\bP^a &=& \displaystyle  (g x)^{\Lambda} \left(\sum_{k=0}^{\l_a^0-1} \cd^{a,k}\left({g}{x}\right)^k + \sum_{k=1}^\infty \cc^{a,k}\left(\frac{g}{x}\right)^k \right)\quad\quad\quad\quad&
\end{matrix}}
\ee
where we recall that the weights $\l_a^0$ are defined as
\be
\l_a^0&=&n_{\fff_a}^\beautysub + \{2,1,0,-1\}_a \,,
\ee
and where all coefficients $\cc$ and $\cd$ have a regular expansion in $g^2$,
\be\label{cexp}
\cc=\sum_{j=0}^\infty \cc^{(2j)}g^{2j}
,\quad\quad
\cd=\sum_{j=0}^\infty \cd^{(2j)}g^{2j}.
\ee
The leading coefficient in $gx$ should be identified with the corresponding prefactor $\cA$, which is constrained by \eqref{AABB}, i.e $\cd_{a,L^\beautysub-\l_a^0}=\cA_a$ and $\cd^{a,\l_a^0-1}=\cA^a$. In the case of shortening, $\cc_{1,1}=\cA_1 g^{-2}$ and/or $\cc^{4,1}=\cA^4 g^{-2}$.

\subsubsection*{Final ansatz - degenerating solutions}
The degenerating solutions are rare in the lowest-lying spectrum and only start appearing for $\Delta_0 \ge 7$. We will return to these states in section \ref{sec:deg}, where we explain how they are related to degenerate solutions of the leading Q-system. For these solutions, we introduce new coefficients with the behaviour
\be\label{cexpD}
\cc=\sum_{j=0}^\infty \cc^{(j)}g^{j}
,\quad\quad
\cd=\sum_{j=0}^\infty \cd^{(j)}g^{j}.
\ee
As mentioned, these cases are characterised by the behaviour $\bP_a^{(0)}\Pt_a^{(0)}\sim u + \CO(u^2)$ and/or $\bP^a_{(0)}\Pt^a_{(0)}\sim u + \CO(u^2)$. 
We can write these functions as
\begin{subequations}
\be
\bP_a^{(0)}&=&u^{-L^\beautysub-\Lambda} \ps_a \\
\bP^a_{(0)}&=&u^\Lambda \ps^a \\
\Pt_a^{(0)}&=&u^{L^\beautysub+\Lambda} \tilde\ps_a \\
\Pt^a_{(0)}&=&u^{-\Lambda} \tilde\ps^a
\ee
\end{subequations}
separating out polynomial factors $\ps$ and $\tilde\ps$. It is important whether the zero at $u=0$ in the product $\ps\tilde\ps$ stems from $\ps$ or $\tilde\ps$. 
In the case $\bP_a^{(0)}\Pt_a^{(0)}\sim u + \CO(u^2)$, we can write the ansatz for $\bP_a$ as
\begin{subequations}\label{PANSD}
\be\label{PANSDL}
\boxed{\bP_a =
(g x)^{-L^\beautysub-\Lambda-\delta_{0,\tilde\ps_a(0)}} 
\left(
\sum_{k=1}^{L^\beautysub-\l_a^0+\delta_{0,\tilde\ps_a(0)}} \cd_{a,k}\left({g}{x}\right)^k
+
g\sum_{k=0}^\infty \cc_{a,k}\left(\frac{g}{x}\right)^k
\right)\,.
}\ee
Similarly, in the case $\bP^a_{(0)}\Pt^a_{(0)}\sim u + \CO(u^2)$, the ansatz for $\bP^a$ becomes
\be\label{PANSDU}
\boxed{
\bP^a = (g x)^{\Lambda-\delta_{0,\tilde\ps^a(0)}} 
\left(\sum_{k=1}^{\l_a^0-1+\delta_{0,\tilde\ps^a(0)}} \cd^{a,k}\left({g}{x}\right)^k 
+ g \sum_{k=0}^\infty \cc^{a,k}\left(\frac{g}{x}\right)^k 
\right)\,.
}
\ee\end{subequations}
If the phenomenon $\bP\Pt\sim u+\CO(u^2)$ only happens for $\bP_a$, then the functions $\bP^a$ (and vice versa) have the structure \eqref{PANS}, but with coefficients expanded in powers of $g$ as in \eqref{cexpD}.


\section{Perturbative solution of the $\bP\mu$-system} \label{sec:pert}

A perturbative algorithm appropriate for the $\sl(2)$ subsector based solely on the $\bP\mu$-system was given in \cite{Marboe:2014gma}, while a more general algorithm solving the full Q-system was first outlined in \cite{Gromov:2015vua}. 
In this section we present a conceptually simple algorithm to solve the $\bP\mu$-system for general states, which is in some sense a hybrid of the two algorithms \cite{Marboe:2014gma,Gromov:2015vua}. We continue where we stopped in \cite{Marboe:2017dmb}: from the leading solutions of the QSC. 

Using only the $\bP\mu$-system has clear advantages. It consists of only 14 functions, eight $\bP$'s and six $\mu$'s. The indices $i$, $j$, etc. are absent, which means that this part of the $H$-symmetry is absent as well. Finally, the linearisation of products of $\eta$-functions is computationally demanding, and such products are only encountered in a very mild way in the presented algorithm.

The algorithm is general, but in the examples of this section we assume that we are dealing with typical solutions, i.e. the QSC functions are expanded in $g^2$. We will treat the degenerating solutions which are expanded in $g$ more explicitly in section \ref{sec:deg}.


\subsection{Exploiting the ansatz for $\bP$} \label{sec:usingansatz}
The ansatz \eqref{PANS} is the key to constructing the perturbative corrections to the QSC functions, since it gives us a concrete ansatz for the functions $\bP$ in terms of a finite number of unknown coefficients at each loop order. This ansatz will be the building block from which we can construct the functions $\mu$ and $\Pt$. Furthermore, it converges in a finite region on the second sheet, so it imposes consistency constraints on the found expressions for $\Pt$. We now describe how to use this ansatz in practice.

\subsubsection*{Fixing symmetries}
To avoid free parameters in the expressions, one should fix the symmetries described in section \ref{sec:QSCsym}.

The choice is not crucial. In our implementation of the algorithm, we choose to fix the symmetry \eqref{xrescaling} by setting $\Lambda=0$. We fix the $H$-symmetry \eqref{Hsym} by setting the coefficients $\cA_a$ to
\be
\cA_a &=& -\frac{ ({\hl_a}+{\hn_1})({\hl_a}+{\hn_2})}{\prod_{b > a} \ii (\hl_a-\hl_b)}\,,
\ee
which fixes $\cA^a$ through \eqref{AABB}, and by cancelling subleading powers in $\bP_a$ by subtracting the leading power of other $\bP_a$ with weaker asymptotics. Concretely, we impose
\be
\cd_{a,L^\beautysub-\lambda_b^0} = 0 \quad\quad \text{for}\quad a>b
\ee
in the ansatz \eqref{PANS}. In the case of the shortening $L^\beautysub -\lambda_1^0 = -1$, we set $\cc_{a,1}=0$ for $a>1$.
Note that this choice does not treat $\bP_a$ and $\bP^a$ symmetrically and therefore does not manifest the so-called left/right symmetry when present.

\subsubsection*{Reintroducing $u$ and expanding in $g$}

We would like to express the ansatz \eqref{PANS} in terms of $u$ instead of $x$ and expand it in $g$. 
On the sheet with $|x|>1$, the expansion of the \zhuk\ variable looks like
\be
x(u,g) = \frac{u}{g}-\frac{g}{u}-\frac{g^3}{u^3}-\frac{2g^5}{u^5} + \CO(g^7)\,.
\ee
Expanding $x$ in this way, we see that only a finite number of terms contribute to $\bP(u)$ at a given order in $g$. This is not the case for $\Pt(u)$, but we do not need the infinite set of coefficients $\cc_{a,k}^{(j)}$. Only those that appear in $\bP_{a}^{(n)}$, where $n$ is the maximal perturbative order that we wish to construct, are relevant.

The coefficient $\cc_{a,k}^{(j)}$ will appear in $\bP_a^{(j+2k)}$, so if we wish to go to $\CO(g^{n})$, then we should control the coefficients $\cc_{a,k}^{(j)}$ for which $2k+j\le n$ along the way. However, there is another feature that means that we need to control additional coefficients at each order. The singular and constant terms in the ansatz for $\Pt_a^{(n)}$ are given in terms of the coefficient $\cd_{a,0}^{(n)}$ which also appears in $\bP_{a}^{(n)}$ and otherwise only coefficients $\cc$ and $\cd$ from subleading orders. This means that these terms impose consistency constraints on the functions $\Pt$ constructed through the algorithm. A term $\cc_{a,k}^{(j)}g^{j}\left(g x\right)^k$ will give rise to a $\CO(u^0)$ term at $\CO(g^{j+k})$ for even $k$. If we want full knowledge of the singular and constant terms at $\CO(g^{n})$, then we thus need to know the coefficients $\cc_{a,k}^{(j)}$ with $j+k\le n$. For example, if we want to construct the perturbative correction at $\CO(g^{20})$, 
then we should determine $\cc_{a,k\le 20}^{(0)}$, $\cc_{a,k\le 18}^{(2)}$, $\cc_{a,k\le 16}^{(4)}$, etc. along the way.

\begin{shaded} 
\noindent {\bf Example} 

\noindent As an example of the above procedure, consider the Konishi multiplet. With the symmetries fixed in the above way, the expansion of $\bP^2$ looks like
\be \label{PP2ex}
\bP^2 = -6\ii u + \cd^{2,0}_{(0)} + g^2 \left( -\frac{5}{2}\ii u \gamma_1 + \cd^{2,0}_{(2)} + \frac{6\ii + \cc^{2,1}_{(0)}}{u} \right)+ \CO(g^4)
\ee
while the expanded ansatz for $\Pt^2$ has the form
\be \label{PPt2ex}
\Pt^2=
\cd^{2,0}_{(0)} + \sum_{k=1}^\infty \cc^{2,k}_{(0)}\, u^k 
+ g^2 \left( \cd^{2,0}_{(2)}  - \frac{6\ii}{u}  + \sum_{k=1}^\infty \cc^{2,k}_{(2)}\, u^k - \sum_{k=1}^\infty k\, \cc^{2,k}_{(0)}\, u^{k-2}\right) +\CO(g^4)\,.
\ee
Notice how the two expansions have coefficients in common, and that some appear at different orders in $g$.

Let us take a look at $\bP^4$:
\be\label{PP4ex}
\bP^4 &=&g^2 \frac{\ii \gamma_1}{12u} + \CO(g^4)\,.
\ee 
For multiplets with the shortening $n_{\fff_4}^\beautysub=1$, which is the case for the Konishi multiplet, this function vanishes at the leading order. Similarly, $\bP_1=\CO(g^2)$ when $n_{\fff_1}^\beautysub=L-1$. The important point is that in these cases $\cA_1$ and/or $\cA^4$ will contain $\gamma_1$ in their leading term. This coefficient will also be present in $\Pt$, but instead at the leading order:
\be\label{PPt4ex}
\Pt^4 &=& \frac{\ii}{12} \gamma_1 \, u  +  \sum_{k=2}^\infty \cc^{4,k}_{(0)}\, u^k + \CO(g^2)\,.
\ee
This means that, for multiplets subject to shortening, we will be able to determine $\gamma_n$ at the order $g^{2(n-1)}$, i.e. from the $n$'th contribution to the $\bP\mu$-system, while for long multiplets we need to determine the $\bP\mu$-system at order $g^{2n}$ to fix $\gamma_n$.

\end{shaded}

\subsection{Leading solution}
In \cite{Marboe:2017dmb}, we described an efficient algorithm to determine the leading Q-system for general states in the spectrum. We will take this as our starting point. In fact, we only need the 4+4+36 functions
\be
Q_{a|\emp}^{(0)}\,,\quad Q_{abc|1234}^{(0)}\,,\quad Q_{ab|ij}^{(0)}\,,
\ee
the first two types being directly related to $\bP_a^{(0)}$ and $\bP^a_{(0)}$. As explained in section \ref{sec:Qsys}, the 36 functions $Q_{ab|ij}^{(0)}$ are the complete set of solutions to the equation \eqref{mudiff} at the leading order, and $\mu_{ab}^{(0)}$ is a linear combination of these functions, cf. \eqref{muom}. As discussed in section \ref{sec:muprop}, the functions $\mu^{(0)}$ are polynomials. As a consequence, the periodic functions $\omega^{ij}$ must be constants at the leading order. Futhermore, for long multiplets, we observe that all $Q_{ab|ij}^{(0)}$ are non-rational functions except for $ij=12$, in which case $\omega^{12}$ should be the only nonzero coefficient at the leading order. In practice, this turns out to be true for short multiplets as well, so we can simply set\footnote{It is of course possible to use a more general ansatz. It will be constrained when matching the found expressions for $\Pt$ to the ansatz.}
\be\label{leadmu}
\mu_{ab}^{(0)} = \omega^{12}_{(0)} Q_{ab|12}^{(0)-}\,.
\ee
The functions $\mu^{ab}_{(0)}$ are given through \eqref{muPf} with the introduction of the constant $\Pf(\mu)^{(0)}$.

With $\mu^{(0)}$ at hand, $\Pt^{(0)}$ is constructed directly from \eqref{Ptmu}. To fix the parameters $\Pf(\mu)^{(0)}$ and the appearing coefficients $\cc$, $\cd$ (for short multiplets also $\gamma_1$), we can now match the obtained expressions for $\bP^{(0)}$ and $\Pt^{(0)}$ with their ansatz \eqref{PANS}. This constraint is sufficient to fix all these parameters.

\begin{shaded} 
\noindent {\bf Example}

\noindent For the Konishi multiplet, the procedure described in \cite{Marboe:2017dmb}, with our chosen way of fixing symmetries, yields
\be\label{konPs}
\bP_a^{(0)} = Q_{a|\emp}^{(0)} = \left\{0,\frac{1}{u^2},\frac{6\ii}{u},-12\right\}
\,, \quad \quad 
\bP^a_{(0)} = Q^{a|\emp}_{(0)}= \left\{-6-12u^2,-6\ii u,1,0\right\}\,,
\ee 
and $Q_{ab|ij}^{(0)}$ are e.g. (the complete set is given in appendix \ref{ap:Qabij})
\be
Q_{12|12}^{(0)} &=& \frac{u^2}{2880}-\frac{1}{34560}\\
Q_{13|12}^{(0)} &=& \frac{\ii u^3}{720}+\frac{\ii u}{2880} \no\\
Q_{12|13}^{(0)} &=& 5 \ii u + \left(5u^2 - \frac{5}{12} \right) \eta_2^+\no \,.
\ee
By introducing two parameters $\omega^{12}_{(0)}$ and $\Pf(\mu)^{(0)}$, we obtain $\mu_{ab}^{(0)}$ and $\mu^{ab}_{(0)}$ through \eqref{leadmu} and \eqref{muPf}. We then construct $\Pt^{(0)}$ from \eqref{Ptmu} and obtain e.g.
\be\label{PPttrue}
\Pt^4_{(0)}&=& \frac{\ii \,\omega^{12}_{(0)}}{1440\Pf(\mu)^{(0)}} \, u \,. 
\ee 
When matched with the ansatz \eqref{PPt4ex}, we get the constraint $\frac{\ii}{12} \gamma_1 = \frac{\ii \,\omega^{12}_{(0)}}{1440\Pf(\mu)^{(0)}}$, and by combining the constraints for all $\Pt$, we can fix all the free parameters to be
\be
\omega_{(0)}^{12}=1440\,, \quad \Pf(\mu)^{(0)}=1\,,\quad \gamma_1=12\,.
\ee

\end{shaded}

\subsection{Coupled difference equations on $\mu$}

At a given loop order $n$ in perturbation theory, the only information we start out with is the ansatz for $\bP$ in terms of a finite number of unknown coefficients $\cd$. To find the functions $\mu$, we consider the six coupled linear difference equations \eqref{mudiff}. At a given order in perturbation theory, they look like
\be \label{mukeyn}
\mu_{ab}^{(n)} - \mu_{ab}^{(n)[2]} = -\bP_a^{(0)}\bP^c_{(0)} \mu_{bc}^{(n)[2]} + \bP_b^{(0)}\bP^c_{(0)} \mu_{ac}^{(n)[2]} + U_{ab}^{(n)} \,.
\ee
Apart from $\bP^{(n)}$, the source term $U_{ab}^{(n)}$ contains only completely fixed subleading orders of $\bP$ and $\mu$.
The source vanishes at the leading order, i.e. $U_{ab}^{(0)}=0$. Let us label the six sets of solutions to the homogeneous equation by 
\be
f_{ab|k}\equiv Q_{ab|\{12,13,14,23,24,34\}_k}^{(0)-} \,,
\quad
f^{ab|k}\equiv Q^{ab|\{12,13,14,23,24,34\}_k -}_{(0)} \,,
\quad
k=1,\hdots,6\,.
\ee

The solution to the inhomogeneous equations \eqref{mukeyn} is given in terms of the solutions to the homogenous equations and the source by the strikingly simple expression
\be\label{musol}\boxed{
\mu_{ab}^{(n)}= \frac{1}{2} 
f_{ab|k} \Psi \left( f^{cd|k} U_{cd}^{(n)} \right)\,.\!\!\!\!\!
}\ee
This is the only point in our algorithm where a non-linearity in $\eta$-functions arises, but it is only a mild non-linearity, because two of the three factors are leading-order Q-functions containing only simple $\eta$-functions.

\subsubsection*{Proof of \eqref{musol}}
The proof of \eqref{musol} goes along the same lines as the one for the construction of $Q_{a|i}^{(n)}$ given in \cite{Gromov:2015vua}. We start out by writing $\mu_{ab}$ as
\be
\mu_{ab} = (A^k + B^k)f_{ab|k}\,,
\ee
where $A^k$ are constants that pick out the correct linear combination at the leading order (so in practice only $A^1$ is nonzero), and $B^k(u,g)=\CO(g)$.

Plugging this ansatz into \eqref{mukeyn} yields
\be
B^{k(n)} f_{ab|k} - B^{k(n)[2]} f_{ab|k}^{[2]} = -\bP_a^{(0)} \bP^c_{(0)} B^{k(n)[2]} f_{bc|k}^{[2]} + \bP_b^{(0)} \bP^c_{(0)} B^{k(n)[2]} f_{ac|k}^{[2]} + U_{ab}^{(n)}
\ee
Using the homogeneous equation to replace $f_{ab|k}^{[2]}$ on the left hand side, we are left with
\be
\left(B^{k(n)}-B^{k(n)[2]}\right)f_{ab|k} = U_{ab}^{(n)}\,.
\ee
Now, contracting both sides with $f^{ab|l}$ and using the property\footnote{This is equivalent to the property $Q_{ab|ij}Q^{ab|kl}=2\left(\delta_i^k\delta_j^l-\delta_i^l\delta_j^k\right)$, see e.g. \cite{Gromov:2014caa}.}
$f_{ab|k}f^{ab|l}=2\delta_k^l$, we get
\be
B^{k(n)}-B^{k(n)[2]}=\frac{1}{2} f^{ab|k} U_{ab}^{(n)}\,,
\ee
which has the solution
\be
B^{k(n)}=\frac{1}{2}\Psi\left( f^{ab|k} U_{ab}^{(n)} \right)\,,
\ee
and this is exactly the statement \eqref{musol}.

\subsubsection*{The $\Psi$-operation and $\ii$-periodic functions}

The $\Psi$-operation was thoroughly described in e.g. \cite{Marboe:2014gma}. As the finite-difference analog of integration, it gives rise to an $\ii$-periodic ambiguity. 
The $\ii$-periodic functions have at most constant asymptotics at $u\to\infty$. They are only allowed to have poles at $u=\ii \mathbb{Z}$. A general ansatz for such $\ii$-periodic functions is given by 
\be
\Phi^k = \phi^{k,0} + \sum_{n=1}^\infty \phi^{k,n} \Per_n \,,
\ee
where $\phi$ are constants and $\Per_n$ are the $\ii$-periodic functions
\be \label{Pper}
\mathcal{P}_n(u)=\sum_{j=-\infty}^\infty \frac{1}{(u+\ii j)^n} \,.
\ee
As discussed in section \ref{sec:muprop}, the sum must truncate at each perturbative order, since the maximal order of the poles in $\mu$ are bounded.

\subsubsection*{Regularity constraints}
Another property discussed in section \eqref{sec:muprop} was the fact that the combinations $\mu+\mu^{[2]}$ and $\frac{\mu-\mu^{[2]}}{\sqrt{u^2-4g^2}}$ should be regular at $u=0$. We can expand the obtained expressions for $\mu$ at $u=0$ and demand that all poles vanish. This fixes the coefficients $\phi^{k,n}$ with $n\ge1$ and the $\phi^{k,0}$ multiplying $f_{ab|k}$ that contain poles at $u=0$.

\begin{shaded}
\noindent {\bf Example}

\noindent For the Konishi multiplet, the ansatz for $\bP^{(2)}$ contains three coefficients that are not known from the leading order: $\cd^{1,0}_{(2)}$, $\cd^{1,1}_{(2)}$ and $\cd^{2,0}_{(2)}$. For example, the source term $U_{12}^{(2)}$ looks like
\be
U_{12}^{(2)}&=& -6\ii u -6 -\frac{2\ii}{u} + \cd^{2,0}_{(2)} \left( \frac{1}{2} + \frac{\ii}{2u} - \frac{1}{6u^2} \right)
\ee
Using \eqref{musol}, we get e.g.
\be
\mu_{12}^{(2)}&=&\left(u^2-\ii u -\frac{1}{3}\right) \left(-6\eta_1 +\eta_2 \frac{\cd_{(2)}^{2,0}+10\phi^{2,0}}{2}+\frac{\phi^{1,0}+\phi^{1,1}\mathcal{P}_1}{2880}\right) \\
&&+\cd_{(2)}^{2,0}\left(\frac{5}{8}u^2-\frac{3\ii}{8}u\right)+\phi^{2,0}\left(5\ii u+\frac{5}{2}\right)-2u^2+2\ii u\no
\ee
Expanding at $u=0$, we get e.g.
\be
\mu_{12}^{(2)}+\mu_{12}^{(2)[2]}=\frac{-\frac{\cd_{(2)}^{2,0}}{6}-\frac{5\phi^{2,0}}{3}}{u^2}+\frac{2\ii - \frac{\ii\cd_{(2)}^{2,0}}{2} - 5\ii \phi^{2,0} - \frac{\phi^{1,1}}{4320} }{u}+\CO(u^0)
\ee
and imposing all regularity constraints fixes most of the free parameters. For example, only one free parameter is left in $\mu_{12}^{(2)}$,
\be
\mu_{12}^{(2)}=\left(u^2-\ii u -\frac{1}{3}\right) \left(-6\eta_1 + 3\ii \mathcal{P}_1 +\frac{\phi^{1,0}}{2880} \right) - 2u^2+2\ii u\,.
\ee
\end{shaded}

\begin{figure}[t]
\centering
\begin{picture}(400,220)

\linethickness{1mm}
\put(127,163){\footnotesize\rotatebox{38}{$\,n\!\to\! n\!\!+\!\!1$}}

\linethickness{0.8mm}

\color{gray!40}
\qbezier(325,196)(280,220)(220,215)
\color{gray!70}
\put(290,215){\footnotesize\rotatebox{-10}{$n\!=\!0$}}

\color{Brown!40}
\qbezier(325,196)(280,190)(290,152)

\qbezier(88,210)(150,190)(170,198)

\qbezier(60,180)(60,180)(90,140)

\color{black}
\linethickness{1.5mm}
\put(270,181.2){\vector(2,-1){0}}
\put(182,0.3){\vector(-1,0){0}}
\put(145,190){\vector(3,2){0}}

\color{Brown!40}
\linethickness{0.8mm}
\put(59.5,180.5){\vector(-3,4){0}}
\put(84,148){\vector(3,-4){0}}
\put(125,199.8){\vector(5,-1){0}}
\put(290,178.3){\vector(-1,-1){0}}

\color{gray!40}
\put(290,209.6){\vector(-4,1){0}}

\color{black}
\linethickness{1mm}
\qbezier(200,200)(300,200)(300,100)
\qbezier(300,100)(300,0)(200,0)
\qbezier(200,0)(100,0)(100,100)
\qbezier(100,100)(100,200)(200,200)

\linethickness{0.6mm}

\put(35,180){\line(1,0){53}}
\put(35,215){\line(1,0){53}}
\put(35,180){\line(0,1){35}}
\put(88,180){\line(0,1){35}}
\put(40,200){\Large Ansatz}
\put(46,185){\Large for $\bP$}



\linethickness{0.6mm}

\put(47,173){\color{white}\circle*{18}}
\put(47,173){\color{Dandelion}\circle{18}}
\put(40,169){$\cc^{(\!n\!)}$}

\put(80,173){\color{white}\circle*{18}}
\put(80,173){\color{Dandelion}\circle{18}}
\put(73,169){$\gamma^{(\!n\!)}$}

\linethickness{0.6mm}

\put(325,178){\line(1,0){65}}
\put(325,215){\line(1,0){65}}
\put(325,178){\line(0,1){37}}
\put(390,178){\line(0,1){37}}
\put(330,200){\Large Leading}
\put(330,185){\Large Q-system}

\linethickness{0.6mm}
\put(200,195){\color{white}\circle*{60}}
\put(200,195){\color{NavyBlue}\circle{60}}
\put(187,189){\huge $\bP^{(n)}$}

\put(200,160){\color{white}\circle*{20}}
\put(200,160){\color{Dandelion}\circle{20}}
\put(193,156){$\cd^{(\!n\!)}$}

\put(300,115){\color{white}\circle*{78}}
\put(300,115){\color{NavyBlue}\circle{78}}
\put(289,125){\huge $\mu^{(n)}$}
\put(267,103){\footnotesize$=\!Q\left(\Psi[QU]\! +\! \mathcal{P}\right)$}
\put(255,115){\color{white}\circle*{19}}
\put(255,115){\color{Dandelion}\circle{19}}
\put(250,112){$\phi^{i}$}

\put(270,25){\color{white}\circle*{80}}
\put(270,25){\color{Red}\circle{80}}
\put(241,33){\large \bf Regularity}
\put(234,14){\small $\mu\!+\!\mu^{[2]}$, $\frac{\mu-\mu^{[2]}}{\sqrt{u^2-4g^2}}$}
\put(250,-3){\scriptsize at $u=0$}
\put(240,60){\color{white}\circle*{23}}
\put(240,60){\color{Green}\circle{23}}
\put(231,56){\small$\phi^{i\ge1}$}

\put(130,25){\color{white}\circle*{70}}
\put(130,25){\color{NavyBlue}\circle{70}}
\put(116,29){\huge $\Pt^{(n)}$}
\put(110,8){\small$=\mu\bP|^{(n)}$}

\put(100,115){\color{white}\circle*{72}}
\put(100,115){\color{Red}\circle{72}}
\put(74,122){\large \bf Matching}
\put(68,105){$\Pt|_{u\simeq 0} = \Pt_{\text{ans}}$}

\put(128,145){\color{white}\circle*{20}}
\put(128,145){\color{Green}\circle{20}}
\put(121,141){$\cd^{(\!n\!)}$}

\put(142,126){\color{white}\circle*{19}}
\put(142,126){\color{Green}\circle{19}}
\put(137,123){$\phi^{0}$}

\put(142,100){\color{white}\circle*{19}}
\put(142,100){\color{Green}\circle{19}}
\put(135,96){$\cc^{(\!n\!)}$}

\put(123,80){\color{white}\circle*{19}}
\put(123,80){\color{Green}\circle{19}}
\put(116,76){$\gamma^{(\!n\!)}$}

\end{picture}
\vspace{5mm}
\caption{Overview of the perturbative algorithm. Coefficients encircled in {\color{Dandelion}yellow} means that they are introduced at this step, while a {\color{Green}green} encircling means that they are fixed at the given step. For long multiplets, $\gamma^{(n)}\equiv\gamma_{\frac{n}{2}+1}\to\gamma_{\frac{n}{2}}$, as there is a delay of one order in the determination of the anomalous dimension. One can optionally impose $\bP_a\bP^a=0$ after the first step to fix a few of the coefficients $\cd$.}
\label{fig:alg}
\end{figure}
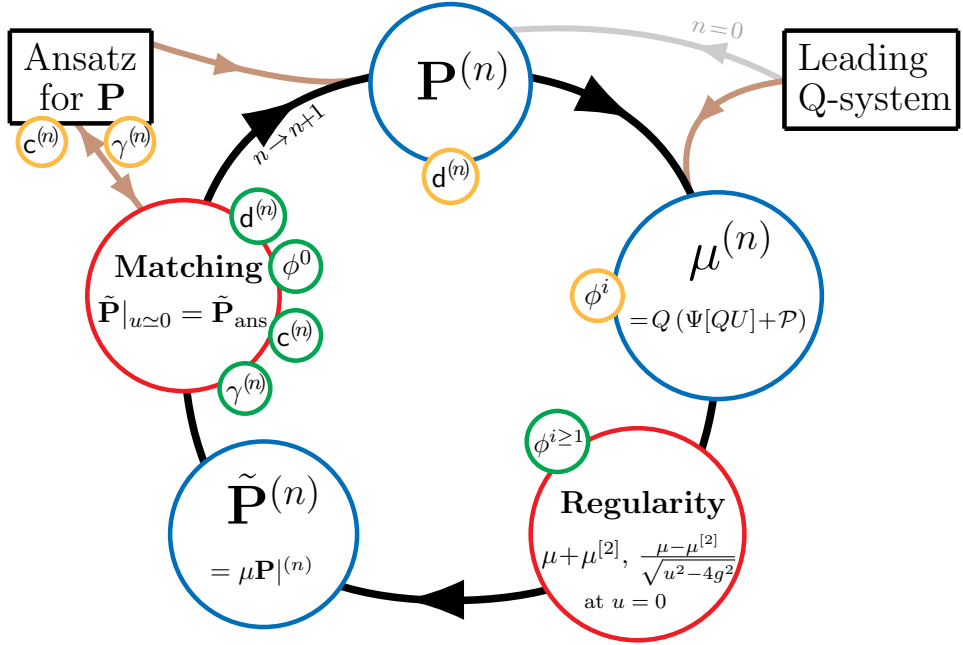

\subsection{Matching $\Pt$ to ansatz}
We proceed, as we did at the leading order, by constructing $\Pt^{(n)}$ through \eqref{Ptmu}. These functions now need to be power expanded at $u=0$ and matched with the ansatz \eqref{PANS}. This fixes the left-over parameters. An overview of the algorithm is given in figure \ref{fig:alg}.

In practice, power expansion is one of the time-consuming steps in the algorithm. In general, the non-trivial constraints in the matching come from the singular parts of $\Pt$. One should therefore be careful to fix all $\cc_{a,k}^{(j)}$ that appear in singular terms at the maximal perturbative order that one wants to reach, as described above in section \ref{sec:usingansatz}.

\begin{shaded}
\noindent {\bf Example}

\noindent For the Konishi solution, inserting the expressions for $\mu^{(2)}$ found in the previous step in \eqref{Ptmu}, we get e.g.
\be
\Pt_1^{(2)} &=& -12 u^3 \eta_1 + 6 u^3 \mathcal{P}_1 + u^3\left(4\ii -\frac{\ii \phi^{1,0}}{1440}\right) +6u^2+3\ii u  
\ee
Expanding at $u=0$ and matching all $\Pt$ to their ans\"{a}tze fixes all remaining degrees of freedom, e.g.
\be
\phi^{1,0}=8640(1+2\zeta_1)\,,\quad\quad \gamma_2=-48\,.
\ee
Note that, for both $\eta$-functions and $\zeta$-values, we follow the regularisation procedure described in \cite{Leurent:2013mr}.

\end{shaded}

\section{Degenerating solutions}
\label{sec:deg}

The aim of our work is to investigate a much larger part of the spectrum than done previously. When one gets far enough, an interesting phenomenon occurs: at the leading order, certain multiplets with different oscillator content at zero coupling, i.e. with different $\pu(2,2|4)\oplus\algu(1)$ charges, but with coinciding $\psu(2,2|4)$ quantum numbers, have identical solutions to the $\gl(4|4)$ Q-system up to gauge transformations. The apparent difference in the number of Bethe roots at different nodes in the Q-system is neatly accounted for by roots placed at $u=\frac{\ii}{2}n$, $n\in \mathbb{Z}$, and differences in the length of the states.\footnote{The phenomenon is reminiscent of the mechanism that allows solutions corresponding to short $\psu(2,2|4)$ multiplets with different Young diagrams to join into one solution. In the example, 
the four multiplets would hypothetically have the ability to join into a long $\psu(3,3|6)$ representation.}

Recall the notion of distinguished Q-functions $\dQ_{a,s}$ from \cite{Marboe:2017dmb}, and the key role they play in the solution algorithm of the leading Q-system. To illustrate the degenerating solutions, we here give two examples of how these functions can overlap on the $\gl(4|4)$ part of the Young diagram Q-system even though the Q-systems seemingly have different boundary conditions.

\begin{shaded}
\noindent {\bf Example: 4-fold degeneracy}

\noindent We observe the first example of a degeneracy at $\Delta_0=7$, where four multiplets correspond to identical solutions of the leading $\gl(4|4)$ Q-system up to symmetry transformations. These solutions appear for the oscillator numbers
\be
\quad [1,2|2,2,2,2|2,1]\,,\quad [1,2|3,3,3,3|1,0]\,,\quad [0,1|2,2,2,2|2,1]\,,\quad [0,1|3,3,3,3|1,0] \,,\no
\ee
corresponding to length 5, 6, 6 and 7, respectively. 
All four multiplets have $\psu(2,2|4)$ quantum numbers
\be
\hl_1-\hl_2=\hl_2-\hl_3=\hl_3-\hl_4=1\,,\quad \hn_1-\hn_2=\hn_3-\hn_4=2\,,\quad \hn_2-\hn_3=-11\,.
\ee
The Young diagrams and number of Bethe roots in the $\gl(4|4)$ Q-system for the four multiplets are

\begin{picture}(390,130)

\color{black}
\linethickness{0.7mm}

\put(30,120){\line(1,0){15}}
\put(30,105){\line(1,0){30}}
\put(0,90){\line(1,0){60}}
\put(0,75){\line(1,0){60}}
\put(0,60){\line(1,0){60}}
\put(0,45){\line(1,0){60}}
\put(0,30){\line(1,0){60}}
\put(0,15){\line(1,0){30}}
\put(15,0){\line(1,0){15}}

\put(0,15){\line(0,1){75}}
\put(15,0){\line(0,1){90}}
\put(30,0){\line(0,1){120}}
\put(45,30){\line(0,1){90}}
\put(60,30){\line(0,1){75}}

\put(120,105){\line(1,0){15}}
\put(90,90){\line(1,0){75}}
\put(90,75){\line(1,0){75}}
\put(90,60){\line(1,0){75}}
\put(90,45){\line(1,0){75}}
\put(90,30){\line(1,0){75}}
\put(90,15){\line(1,0){30}}
\put(105,0){\line(1,0){15}}

\put(90,15){\line(0,1){75}}
\put(105,0){\line(0,1){90}}
\put(120,0){\line(0,1){105}}
\put(135,30){\line(0,1){75}}
\put(150,30){\line(0,1){60}}
\put(165,30){\line(0,1){60}}

\put(240,120){\line(1,0){15}}
\put(240,105){\line(1,0){30}}
\put(195,90){\line(1,0){75}}
\put(195,75){\line(1,0){75}}
\put(195,60){\line(1,0){75}}
\put(195,45){\line(1,0){75}}
\put(195,30){\line(1,0){75}}
\put(225,15){\line(1,0){15}}

\put(195,30){\line(0,1){60}}
\put(210,30){\line(0,1){60}}
\put(225,15){\line(0,1){75}}
\put(240,15){\line(0,1){105}}
\put(255,30){\line(0,1){90}}
\put(270,30){\line(0,1){75}}

\put(345,105){\line(1,0){15}}
\put(300,90){\line(1,0){90}}
\put(300,75){\line(1,0){90}}
\put(300,60){\line(1,0){90}}
\put(300,45){\line(1,0){90}}
\put(300,30){\line(1,0){90}}
\put(330,15){\line(1,0){15}}

\put(300,30){\line(0,1){60}}
\put(315,30){\line(0,1){60}}
\put(330,15){\line(0,1){75}}
\put(345,15){\line(0,1){90}}
\put(360,30){\line(0,1){75}}
\put(375,30){\line(0,1){60}}
\put(390,30){\line(0,1){60}}

\color{blue}
\footnotesize

\put(60,90){\color{shadecolor}\circle*{8}}\put(58,87){0}
\put(45,90){\color{shadecolor}\circle*{8}}\put(43,87){1}
\put(30,90){\color{shadecolor}\circle*{8}}\put(28,87){3}
\put(15,90){\color{shadecolor}\circle*{8}}\put(13,87){5}
\put(0,90){\color{shadecolor}\circle*{8}}\put(-2,87){0}

\put(60,75){\color{shadecolor}\circle*{8}}\put(58,72){0}
\put(45,75){\color{shadecolor}\circle*{8}}\put(43,72){2}
\put(30,75){\color{shadecolor}\circle*{8}}\put(28,72){5}
\put(15,75){\color{shadecolor}\circle*{8}}\put(13,72){4}
\put(0,75){\color{shadecolor}\circle*{8}}\put(-2,72){0}

\put(60,60){\color{shadecolor}\circle*{8}}\put(58,57){0}
\put(45,60){\color{shadecolor}\circle*{8}}\put(43,57){3}
\put(30,60){\color{shadecolor}\circle*{8}}\put(28,57){7}
\put(15,60){\color{shadecolor}\circle*{8}}\put(13,57){3}
\put(0,60){\color{shadecolor}\circle*{8}}\put(-2,57){0}

\put(60,45){\color{shadecolor}\circle*{8}}\put(58,42){0}
\put(45,45){\color{shadecolor}\circle*{8}}\put(43,42){4}
\put(30,45){\color{shadecolor}\circle*{8}}\put(28,42){5}
\put(15,45){\color{shadecolor}\circle*{8}}\put(13,42){2}
\put(0,45){\color{shadecolor}\circle*{8}}\put(-2,42){0}

\put(60,30){\color{shadecolor}\circle*{8}}\put(58,27){0}
\put(45,30){\color{shadecolor}\circle*{8}}\put(43,27){5}
\put(30,30){\color{shadecolor}\circle*{8}}\put(28,27){3}
\put(15,30){\color{shadecolor}\circle*{8}}\put(13,27){1}
\put(0,30){\color{shadecolor}\circle*{8}}\put(-2,27){0}

\put(150,90){\color{shadecolor}\circle*{8}}\put(148,87){0}
\put(135,90){\color{shadecolor}\circle*{8}}\put(133,87){0}
\put(120,90){\color{shadecolor}\circle*{8}}\put(118,87){1}
\put(105,90){\color{shadecolor}\circle*{8}}\put(103,87){5}
\put(90,90){\color{shadecolor}\circle*{8}}\put(88,87){0}

\put(150,75){\color{shadecolor}\circle*{8}}\put(148,72){1}
\put(135,75){\color{shadecolor}\circle*{8}}\put(133,72){2}
\put(120,75){\color{shadecolor}\circle*{8}}\put(118,72){4}
\put(105,75){\color{shadecolor}\circle*{8}}\put(103,72){4}
\put(90,75){\color{shadecolor}\circle*{8}}\put(88,72){0}

\put(150,60){\color{shadecolor}\circle*{8}}\put(148,57){2}
\put(135,60){\color{shadecolor}\circle*{8}}\put(133,57){4}
\put(120,60){\color{shadecolor}\circle*{8}}\put(118,57){7}
\put(105,60){\color{shadecolor}\circle*{8}}\put(103,57){3}
\put(90,60){\color{shadecolor}\circle*{8}}\put(88,57){0}

\put(150,45){\color{shadecolor}\circle*{8}}\put(148,42){3}
\put(135,45){\color{shadecolor}\circle*{8}}\put(133,42){6}
\put(120,45){\color{shadecolor}\circle*{8}}\put(118,42){5}
\put(105,45){\color{shadecolor}\circle*{8}}\put(103,42){2}
\put(90,45){\color{shadecolor}\circle*{8}}\put(88,42){0}

\put(150,30){\color{shadecolor}\circle*{8}}\put(148,27){4}
\put(135,30){\color{shadecolor}\circle*{8}}\put(133,27){8}
\put(120,30){\color{shadecolor}\circle*{8}}\put(118,27){3}
\put(105,30){\color{shadecolor}\circle*{8}}\put(103,27){1}
\put(90,30){\color{shadecolor}\circle*{8}}\put(88,27){0}

\put(270,90){\color{shadecolor}\circle*{8}}\put(268,87){0}
\put(255,90){\color{shadecolor}\circle*{8}}\put(253,87){1}
\put(240,90){\color{shadecolor}\circle*{8}}\put(238,87){3}
\put(225,90){\color{shadecolor}\circle*{8}}\put(223,87){8}
\put(210,90){\color{shadecolor}\circle*{8}}\put(208,87){4}

\put(270,75){\color{shadecolor}\circle*{8}}\put(268,72){0}
\put(255,75){\color{shadecolor}\circle*{8}}\put(253,72){2}
\put(240,75){\color{shadecolor}\circle*{8}}\put(238,72){5}
\put(225,75){\color{shadecolor}\circle*{8}}\put(223,72){6}
\put(210,75){\color{shadecolor}\circle*{8}}\put(208,72){3}

\put(270,60){\color{shadecolor}\circle*{8}}\put(268,57){0}
\put(255,60){\color{shadecolor}\circle*{8}}\put(253,57){3}
\put(240,60){\color{shadecolor}\circle*{8}}\put(238,57){7}
\put(225,60){\color{shadecolor}\circle*{8}}\put(223,57){4}
\put(210,60){\color{shadecolor}\circle*{8}}\put(208,57){2}

\put(270,45){\color{shadecolor}\circle*{8}}\put(268,42){0}
\put(255,45){\color{shadecolor}\circle*{8}}\put(253,42){4}
\put(240,45){\color{shadecolor}\circle*{8}}\put(238,42){4}
\put(225,45){\color{shadecolor}\circle*{8}}\put(223,42){2}
\put(210,45){\color{shadecolor}\circle*{8}}\put(208,42){1}

\put(270,30){\color{shadecolor}\circle*{8}}\put(268,27){0}
\put(255,30){\color{shadecolor}\circle*{8}}\put(253,27){5}
\put(240,30){\color{shadecolor}\circle*{8}}\put(238,27){1}
\put(225,30){\color{shadecolor}\circle*{8}}\put(223,27){0}
\put(210,30){\color{shadecolor}\circle*{8}}\put(208,27){0}

\put(375,90){\color{shadecolor}\circle*{8}}\put(373,87){0}
\put(360,90){\color{shadecolor}\circle*{8}}\put(358,87){0}
\put(345,90){\color{shadecolor}\circle*{8}}\put(343,87){1}
\put(330,90){\color{shadecolor}\circle*{8}}\put(328,87){8}
\put(315,90){\color{shadecolor}\circle*{8}}\put(313,87){4}

\put(375,75){\color{shadecolor}\circle*{8}}\put(373,72){1}
\put(360,75){\color{shadecolor}\circle*{8}}\put(358,72){2}
\put(345,75){\color{shadecolor}\circle*{8}}\put(343,72){4}
\put(330,75){\color{shadecolor}\circle*{8}}\put(328,72){6}
\put(315,75){\color{shadecolor}\circle*{8}}\put(313,72){3}

\put(375,60){\color{shadecolor}\circle*{8}}\put(373,57){2}
\put(360,60){\color{shadecolor}\circle*{8}}\put(358,57){4}
\put(345,60){\color{shadecolor}\circle*{8}}\put(343,57){7}
\put(330,60){\color{shadecolor}\circle*{8}}\put(328,57){4}
\put(315,60){\color{shadecolor}\circle*{8}}\put(313,57){2}

\put(375,45){\color{shadecolor}\circle*{8}}\put(373,42){3}
\put(360,45){\color{shadecolor}\circle*{8}}\put(358,42){6}
\put(345,45){\color{shadecolor}\circle*{8}}\put(343,42){4}
\put(330,45){\color{shadecolor}\circle*{8}}\put(328,42){2}
\put(315,45){\color{shadecolor}\circle*{8}}\put(313,42){1}

\put(375,30){\color{shadecolor}\circle*{8}}\put(373,27){4}
\put(360,30){\color{shadecolor}\circle*{8}}\put(358,27){8}
\put(345,30){\color{shadecolor}\circle*{8}}\put(343,27){1}
\put(330,30){\color{shadecolor}\circle*{8}}\put(328,27){0}
\put(315,30){\color{shadecolor}\circle*{8}}\put(313,27){0}
\end{picture}

\vspace{5mm}


\noindent Taking the point of view of the multiplet with $\ns=[1,2|2,2,2,2|2,1]$, the distinguished Q-functions for the degenerate solutions have the form
\be
\dQ_{a,s}
&=&{\small\begin{array}{c|c|c|c|c}
\dQ_{4,0}&\dQ_{4,1}&\dQ_{4,2}&\dQ_{4,3}&\dQ_{4,4} \\\hline
\dQ_{3,0}&\dQ_{3,1}&\dQ_{3,2}&\dQ_{3,3}&\dQ_{3,4} \\\hline
\dQ_{2,0}&\dQ_{2,1}&\dQ_{2,2}&\dQ_{2,3}&\dQ_{2,4} \\\hline
\dQ_{1,0}&\dQ_{1,1}&\dQ_{1,2}&\dQ_{1,3}&\dQ_{1,4} \\\hline
\dQ_{0,0}&\dQ_{0,1}&\dQ_{0,2}&\dQ_{0,3}&\dQ_{0,4}
\end{array}}
\\[3mm]
&\propto&
\small\begin{array}{c|c|c|c|c}
\frac{1}{\color{gray}(u^{[3]}u^+u^-u^{[-3]})^4}&\frac{u^5+\frac{u}{3}}{\color{gray}(u^{[2]}uu^{[-2]})^4}&{\color{purple}u^+u^-}u&{\color{purple}u}&1 
\\[2mm] \hline
\frac{1}{\color{gray}(u^{[2]}uu^{[-2]})^4}&\frac{u^4-\frac{u^2}{2}+\frac{19}{240}}{\color{gray}(u^+u^-)^4}&{\color{purple}u}\left(u^4+\frac{13}{35}\right)&u^2-\frac{1}{4}&1 
\\[2mm] \hline
\frac{1}{\color{gray}(u^+u^-)^4}&\frac{u^3-\frac{u}{2}}{\color{gray}u^4}&u^7-\frac{u^5}{4}+\frac{61u^3}{112}+\frac{75u}{448}&u^3-\frac{u}{2}&1
\\[2mm] \hline
\frac{1}{\color{gray}u^4}&u^2-\frac{1}{4}&{\color{gray}u^4}\cdot {\color{purple}u}\left(u^4+\frac{13}{35}\right)&u^4-\frac{u^2}{2}+\frac{19}{240}&1 
\\[2mm] \hline
1&{\color{gray}u^4}\cdot {\color{purple}u}& {\color{gray}(u^+u^-)^4} \cdot {\color{purple}u^+u^-}u&u^5+\frac{u}{3}&1
%
%
\end{array}\,.\no
\ee
By absorbing factors of $u+\frac{\ii}{2}n$, in particular those marked in {\color{purple}purple}, into the length-dependent factors marked in {\color{gray}grey}, and by making symmetry transformations of the kind \eqref{xrescaling}, one can see that this is also a solution to the leading Q-system for the three other multiplets shown above.
\end{shaded}

\begin{shaded}
\noindent {\bf Example: 2-fold degeneracy}

\noindent Another example of a degenerating solutions appears for two multiplets with $\Delta_0=\frac{15}{2}$ and oscillator numbers
{\begin{center}
\begin{picture}(240,80)

\color{black}
\linethickness{0.7mm}

\put(50,0){\line(1,0){10}}
\put(50,10){\line(1,0){10}}
\put(40,20){\line(1,0){50}}
\put(30,30){\line(1,0){60}}
\put(30,40){\line(1,0){50}}
\put(30,50){\line(1,0){50}}
\put(30,60){\line(1,0){50}}
\put(60,70){\line(1,0){20}}
\put(60,80){\line(1,0){10}}
\put(30,30){\line(0,1){30}}
\put(40,20){\line(0,1){40}}
\put(50,00){\line(0,1){60}}
\put(60,00){\line(0,1){80}}
\put(70,20){\line(0,1){60}}
\put(80,20){\line(0,1){50}}
\put(90,20){\line(0,1){10}}

\put(240,0){\line(1,0){10}}
\put(240,10){\line(1,0){10}}
\put(230,20){\line(1,0){60}}
\put(220,30){\line(1,0){70}}
\put(220,40){\line(1,0){60}}
\put(220,50){\line(1,0){60}}
\put(220,60){\line(1,0){60}}
\put(250,70){\line(1,0){10}}
\put(220,30){\line(0,1){30}}
\put(230,20){\line(0,1){40}}
\put(240,0){\line(0,1){60}}
\put(250,0){\line(0,1){70}}
\put(260,20){\line(0,1){50}}
\put(270,20){\line(0,1){40}}
\put(280,20){\line(0,1){40}}
\put(290,20){\line(0,1){10}}

\put(-60,37){$[0,2|3,2,2,2|2,1]$}
\put(130,37){$[0,2|4,3,3,3|1,0]$}

\put(60,40){\circle*{5}}
\put(250,40){\circle*{5}}

\end{picture}
\end{center}}
\noindent and similarly for the Hodge rotated cases $[1,2|3,3,3,2|2,0]$ and $[0,1|3,3,3,2|2,0]$.
The distinguished Q-functions for these solutions are
\be
\!\!\dQ_{a,s}
&\!\!\propto\!\!&
\small\begin{array}{c|c|c|c|c}
\frac{u^3+\frac{13u}{28}}{\color{gray}(u^{[3]}u^+u^-u^{[-3]})^5}&\frac{u^7+\frac{47u^5}{28}-\frac{3u^3}{2}-\frac{13u}{28}}{\color{gray}(u^{[2]}uu^{[-2]})^5}&{\color{purple}u^+u^-}u&{\color{purple}u}&1 
\\[2mm] \hline
\frac{u^2+\frac{1}{14}}{\color{gray}(u^{[2]}uu^{[-2]})^5}&\frac{u^5+\frac{9u^3}{14}-\frac{201u}{560}}{\color{gray}(u^+u^-)^5}&{\color{purple}u}\left(u^4-\frac{u^2}{4}+\frac{7}{20}\right)&u^2-\frac{3}{8}&1 
\\[2mm] \hline
\frac{u}{\color{gray}(u^+u^-)^5}&\frac{u^3+\frac{u}{4}}{\color{gray}u^4}&u^7-u^5-\frac{3u^3}{16}+\frac{u}{32}&u^3-\frac{7u}{8}&1
\\[2mm] \hline
\frac{1}{\color{gray}u^5}&u&{\color{gray}u^5}\cdot \left(u^4-\frac{3}{2}u^2+\frac{1}{6}\right)&u^4-\frac{5u^2}{4}+\frac{85}{72}&1 
\\[2mm] \hline
1&{\color{gray}u^5}& {\color{gray}(u^+u^-)^5} \cdot (u^2-1)&u^6-\frac{5u^4}{8}+\frac{72u^2}{24}-\frac{35}{12}&u
\end{array}\,.\no\\
\ee

\end{shaded}

The appearance of degenerating solutions is not a unique exception but apparently a reoccuring phenomenon. The examples of degenerating solutions that we have encountered are listed in table \ref{tab:deg}, but this is not an extensive list, as we are unable to solve the leading Q-system for many states with $\Delta_0\ge 7$.

\begin{table}[h!]
\small
\centering
\def\arraystretch{1.2}
\begin{tabular}{|c|c|c|c|c|} \hline
$\Delta_0$ & Degeneracy & $\ns$ & $\begin{matrix} \text{\# total} \\ \text{solutions} \end{matrix}$ & $\begin{matrix} \text{\# degenerating} \\ \text{solutions found} \end{matrix}$ \\\hline\hline

7 & 4-fold & 
$\begin{matrix} [1,2|2,2,2,2|2,1] \\ [0,1|2,2,2,2|2,1] \\ [1,2|3,3,3,3|1,0] \\ [0,1|3,3,3,3|1,0] \end{matrix}$ & 
$\begin{matrix}  3 \\ 7 \\ 7 \\ {\color{gray}23} \end{matrix}$ & 
1 \\\hline

$\frac{15}{2}$ & 
$\begin{matrix} \text{2-fold} \\ \text{+ conj.} \end{matrix}$ & 
$\begin{matrix} [0,4|3,2,2,2|2,1] \\ [0,4|4,3,3,3|1,0] \end{matrix}$ & 
$\begin{matrix} 7 \\ 11 \end{matrix}$ & 
1 \\\hline

$\frac{15}{2}$ & 
$\begin{matrix} \text{2-fold} \\ \text{+ conj.} \end{matrix}$ & 
$\begin{matrix} [0,0|3,2,2,2|2,1] \\ [0,0|4,3,3,3|1,0] \end{matrix}$ & 
$\begin{matrix} 11 \\ {\color{gray}43} \end{matrix}$ & 
1 \\\hline

$\frac{15}{2}$ & 
$\begin{matrix} \text{2-fold} \\ \text{+ conj.} \end{matrix}$ & 
$\begin{matrix} [1,1|3,2,2,2|2,1] \\ [1,1|4,3,3,3|1,0] \end{matrix}$ & 
$\begin{matrix} 14 \\ {\color{gray}38} \end{matrix}$ & 
2 \\\hline

$\frac{15}{2}$ & 
$\begin{matrix} \text{2-fold} \\ \text{+ conj.} \end{matrix}$ & 
$\begin{matrix} [0,2|3,2,2,2|2,1] \\ [0,2|4,3,3,3|1,0] \end{matrix}$ & 
$\begin{matrix} 31 \\ {\color{gray}77} \end{matrix}$ & 
1 \\\hline

$8$ & 4-fold &
$\begin{matrix} [1,1|4,4,2,2|1,1] \\ [0,0|4,4,2,2|1,1] \\ [1,1|5,5,3,3|0,0] \\ [0,0|5,5,3,3|0,0] \end{matrix}$ & 
$\begin{matrix} 39 \\ {\color{gray}43} \\ {\color{gray}43} \\ {\color{gray}71} \end{matrix}$ & 
4 \\\hline

$\frac{17}{2}$ & 
$\begin{matrix} \text{2-fold} \\ \text{+ conj.} \end{matrix}$ & 
$\begin{matrix} [0,6|3,3,3,2|2,1] \\ [0,6|4,3,3,3|1,0] \end{matrix}$ & 
$\begin{matrix} 3 \\ 3 \end{matrix}$ & 
1 \\\hline

$\frac{17}{2}$ & 
$\begin{matrix} \text{2-fold} \\ \text{+ conj.} \end{matrix}$ & 
$\begin{matrix} [0,2|5,2,2,2|2,1] \\ [0,2|6,3,3,3|1,0] \end{matrix}$ & 
$\begin{matrix} 27 \\ 81 \end{matrix}$ & 
3 \\\hline

$\frac{17}{2}$ & 
$\begin{matrix} \text{2-fold} \\ \text{+ conj.} \end{matrix}$ & 
$\begin{matrix} [0,0|5,2,2,2|2,1] \\ [0,0|6,3,3,3|1,0] \end{matrix}$ & 
$\begin{matrix} 33 \\ {\color{gray}117} \end{matrix}$ & 
3 \\\hline

$9$ & 4-fold &
$\begin{matrix} [1,4|2,2,2,2|4,1] \\ [0,3|2,2,2,2|4,1] \\ [1,4|3,3,3,3|3,0] \\ [0,3|3,3,3,3|3,0] \end{matrix}$ & 
$\begin{matrix} 14 \\ {\color{gray}54} \\ {\color{gray}54} \\ {\color{gray}272} \end{matrix}$ & 
2 \\\hline
\end{tabular}
\caption{Examples of degenerate solutions among the treated cases.
If the total number of solutions is {\color{gray}grey}, it means that we did not manage to find the complete set of solutions to the leading Q-system and hence potentially more degenerate solutions exist.
}
\label{tab:deg}
\end{table}

\subsubsection*{Mechanism of degeneration}
Historically, the first systematic perturbative approach to find the AdS/CFT spectrum via integrability  was  the  Beisert-Staudacher asymptotic Bethe Ansatz (ABA) \cite{Beisert:2005fw} which is valid up to the wrapping orders. In this approach, different states in the spectrum are mapped out by the position of seven types of Bethe roots that are zeros of seven selected Q-functions. This point of view turns out to be useful to understand the phenomenon of degeneration.  We  hence briefly recall how to link the QSC to the ABA formalism, which was already discussed in \cite{Gromov:2014caa}, and then use this link towards explaining the degenerating solutions.

A consequence of the $\bP\mu$-system is the following relation (no summation over $a$):
\begin{equation}
\e_{abcd}(\tilde\bP_a\tilde\bP^d-\bP_a\bP^d)=\mu_{ab}^{[2]}\,\mu_{ac}-\mu_{ab}\,\mu_{ac}^{[2]}\,.
\end{equation}
The term $\bP_a\bP^d$ is suppressed in $g$ compared to $\tilde\bP_a\tilde\bP^d$ and starts to contribute only at wrapping orders. But then, with  $\bP_a\bP^d$ omitted, the above equation looks precisely like a QQ-relation. And indeed, at the first few orders of the weak coupling expansion the following approximations are valid \cite{Gromov:2014caa}
\begin{equation}
\tilde \bP_a\simeq  Q_{a|12}\,\omega^{12}\,,\quad \tilde \bP^d\simeq -\frac 16\e^{dabc}Q_{abc|12}\,\omega^{12}\,,\quad \mu_{ab}^+= Q_{ab|12}\,(\omega^{12})^+\,.
\end{equation}

Then make the identifications $Q_1=\bP_1=Q_{1|0}$, $Q_2=Q_{1|1}$, $Q_{3}=\frac{\tilde\bP_1}{\omega^{12}}\simeq Q_{1|12}$, $Q_{4}=\frac{\mu_{12}^+}{(\omega^{12})^+}\simeq Q_{12|12}$, $Q_5=\frac{\tilde\bP^4}{\omega^{12}}\simeq Q_{123|12}$, $Q_6=Q_{123|123}$, and $Q_7=\bP^4\simeq Q_{123|1234}$. Zeros of $Q_k$, $k=1,\ldots, 7$, on the physical sheet are denoted by $u^{(k)}_i$, $i=1,2,\ldots, K_k$. These are the Bethe roots that solve the asymptotic Bethe equations associated to the ``compact ABA" Dynkin diagram, $\hat{1}12\hat{2}\hat{3}34\hat{4}$. Explicit expressions for these Q-functions are given e.g. in section 5.3  of \cite{Gromov:2014caa}. Consider for instance
\begin{eqnarray}
Q_1\propto \bP_1&\propto& x^{-L}\prod_{i=1}^{K_1}\left(x(u)-x(u_i^{(1)})\right)\prod_{i=1}^{K_3}\left(\frac 1{x(u)}-x(u_i^{(3)})\right)\times \ldots\,,\nonumber\\
Q_3\propto \tilde \bP_1&\propto& x^{L}\prod_{i=1}^{K_1}\left(\frac 1{x(u)}-x(u_i^{(1)})\right)\prod_{i=1}^{K_3}\left(x(u)-x(u_i^{(3)})\right)\times \ldots\,,
\end{eqnarray}
where $\ldots $ stand for the terms that do not have zeros on both physical and next-to-physical sheets. These terms are irrelevant for the current discussion, and they are equal to one at the leading order.

The value of $u_i$ depends on coupling and in most cases it approaches some finite value at $g=0$. Then at the leading order one has $x(u_i)\simeq \frac{u_i}{g}$ and
\begin{eqnarray}
Q_1&\propto& u^{-L}\prod_{i=1}^{K_1}\left(u-u_i^{(1)}\right)\times (1+\CO(g^2))\,,\nonumber\\
Q_3&\propto& u^{L}\prod_{i=1}^{K_3}\left(u-u_i^{(3)}\right)\times (1+\CO(g^2))\,.
\end{eqnarray}
One can simply identify $Q_1^{(0)}=\dQ_{1,0}$, $Q_3^{(0)}=\dQ_{1,2}$ at the leading order and this type of identifications are sufficient to launch the perturbative algorithm as explained in the previous section.

However it may happen that some $u_i$ approach the origin at $g=0$. Then $x(u_i)$ remains a finite quantity at zero coupling which we denote as $x(u_i)=x_*$. In all found examples there was at most one root with this property, so we assume  uniqueness of $x_*$ for simplicity of the discussion. To get a real solution of the QSC one needs $x_*$ real which translates into $u_i$ being real and $|u_i|\geq 2g$. Thus $u_i$ approaches the origin no faster than $u_i\simeq \Lambda g$ with $\Lambda=x_*+\frac 1{x_*}\geq 2$.

If the special root is the one of $Q_1$, one has the following structure at the leading and the linear orders in $g$ 
\begin{subequations}
\label{eq:some1}
\begin{eqnarray}
Q_1&\propto& u^{-L}(u-gx_*)\prod_{i=1}^{K_1-1}\left(u-u_i^{(1)}\right)\times (1+\CO(g^2))\,,\nonumber\\
Q_3&\propto& u^{L}(1-\frac{g}{x^*u})\prod_{i=1}^{K_3}\left(u-u_i^{(3)}\right)\times (1+\CO(g^2))\,.
\end{eqnarray}
If the special root is the one of $Q_3$, one has the following structure
\begin{eqnarray}
Q_1&\propto& u^{-L'}(1-\frac{g}{x_*u})\prod_{i=1}^{K_1'}\left(u-u_i^{(1)}\right)\times (1+\CO(g^2))\,,\nonumber\\
Q_3&\propto& u^{L'}(u-g{x^*})\prod_{i=1}^{K_3'-1}\left(u-u_i^{(3)}\right)\times (1+\CO(g^2))\,.
\end{eqnarray}
\end{subequations}
At the leading order, one cannot distinguish the two cases, if one sets $L'=L-1$, $K_3'=K_3+1$, and $K_1'=K_1-1$, so the knowledge of $\dQ_{a,s}$ would not be sufficient to jump-start the perturbative expansion. This is the observed degeneracy of two different solutions at weak coupling. One has to go to the linear order in $g$ to lift the degeneracy.

To make judgements about the presence of this special zero from the value of $\dQ_{a,s}$, it is not convenient to consider separately $Q_1^{(0)}$ and $Q_3^{(0)}$ as the Bethe root at the origin is disguised by the presence of $u^{\pm L}$ factors. Instead one can consider the product $Q_1^{(0)}Q_3^{(0)}$ where the $u^{\pm L}$ factors cancel. A simple zero of $Q_1^{(0)}Q_3^{(0)}$ at the origin is the property that singles out the degenerating solutions. 

Of course, the same discussion is valid for $Q_5$ and $Q_7$. To summarise, one (or both) of the behaviours of the products of $\bP$ and $\Pt$
\be
&&Q_1^{(0)}Q_3^{(0)} \sim \dQ_{1,0}\dQ_{1,2}\sim  \bP_1^{(0)}\Pt_1^{(0)} \sim  u + \CO(u^2)\,,\nonumber\\
&&Q_5^{(0)}Q_7^{(0)} \sim \dQ_{3,4}\dQ_{3,2}\sim \bP^4_{(0)}\Pt^4_{(0)} \sim  u + \CO(u^2)\,,
\ee
 is  the property that, as discussed in section \ref{sec:Pstructure}, separates the degenerating solutions from the typical ones. From \eqref{eq:some1} we see that the perturbative expansion of such solutions will contain odd powers in $g$ as well and we should thus use the ansatz \eqref{PANSD} for these states and expand the QSC functions in powers of $g$.


\subsubsection*{Perturbative breaking of degeneracy}

At the leading order, the Q-system has completely degenerate solutions, and we lack constraints to fix the free parameters $\omega^{12}_{(0)}$ and $\Pf(\mu)^{(0)}$ that appear in the functions $\mu$ and $\Pt$. We have to proceed to $\CO(g)$, where it turns out that a nonlinear constraint on these parameters will arise. The constants $\omega^{12}_{(0)}$ and $\Pf(\mu)^{(0)}$ are present in $\mu^{(0)}$, but also in $\bP^{(1)}$ through the value assigned to $\cc_{a,0}^{(0)}$ and/or $\cc^{a,0}_{(0)}$ while matching $\Pt$ with the ans\"{a}tze at the leading order. These functions are multiplied when constructing $\mu^{(1)}$ and $\Pt^{(1)}$. 
The nonlinear constraint arises in the matching of $\Pt^{(1)}$ to the ans\"{a}tze. The number of solutions matches the degeneracy, i.e. either two or four in the found examples. Consequently, the QSC functions are distinct for the separate solutions. In the treated cases, we find that the anomalous dimension is the same for both degenerating solutions when the degeneracy is 2-fold, while it splits up into two different values in the case of 4-fold degeneracy.

\begin{shaded}
\noindent {\bf Example}

\noindent For the $\Delta_0=7$ example, with our choice of fixing the symmetries, the nonlinear conditions on $\omega^{12}_{(0)}$ and $\Pf(\mu)^{(0)}$ in the matching of $\Pt^{(1)}$  have the four distinct solutions
\be
\omega^{12}_{(0)} &=& \left\{ -187110 \sqrt{42/127}, 187110 \sqrt{42/127}, -62370 \sqrt{42/17}, 62370 \sqrt{42/17}  \right\} \no\\
\Pf(\mu)^{(0)}&=&\{-1,-1,1,1\}\,.
\ee
We see that for the degenerating solutions, in contrast to the typical ones, the Q-functions can contain only rational coefficients at the leading order, but have more general algebraic numbers appearing after the degeneracy is broken.

Of the four solutions, two solutions take each of two possible values for the anomalous dimension,
{\small\be
\gamma = \left\{ 
\begin{matrix} 
\frac{50}{3}g^2 - \frac{288445}{6858}g^4 + \frac{929745505475}{3982056552}g^6 
- 2134.2942 g^8 + 24531.263 g^{10} - 313147.66 g^{12}
\\[2mm]
\frac{50}{3}g^2 - \frac{46475}{918}g^4 + \frac{3332354125}{9550872}g^6 
- 3433.6267 g^8
+ 39165.126 g^{10}
- 481452.99 g^{12}\,.\quad\quad
\end{matrix}
\right.
\ee}

\end{shaded}


\section{Results and performance} \label{sec:res}
In this section we take a look at the results that can be produced by the perturbative algorithm on a standard laptop. Our \Ma\ implementation of the algorithm is available in the ancillary files at arxiv.org, and we give a brief introduction to this program in section \ref{sec:code}. Here, we discuss what we found achieveable with such relatively basic tools.

\subsubsection*{The ``full" spectrum - how many solutions can we actually find?}

The title of the paper is of course a bit misleading, since there are infinitely many multiplets of single-trace operators. We provided the multiplet content containing operators with $\Delta_0\le 8$ in Appendix A.3 of \cite{Marboe:2017dmb}. This information was derived from character theory (see also \cite{Bianchi:2003wx,Beisert:2003te}). In the database \dbname, we provide an overview of all 84.793 multiplets with $\Delta_0\le 9$, and a gifted representation theorist would definitely be able to use character theory to list the multiplets for a few more values of $\Delta_0$.

\begin{table}[b!]
\centering
\def\arraystretch{1.05}
\begin{tabular}{|c||c|c||c|c|}
\hline
\multirow{2}{*}{$\Delta_0$} & \multicolumn{2}{c||}{\# distinct $\ns$} & \multicolumn{2}{c|}{\# solutions}   
\\ \cline{2-5}
 & total & found & total & found
\\ \hline\hline
2 & $1_{\,\color{black!60}1/0/0}$ & \multirow{7}{*}{all} & $1_{\,\color{black!60}1/0/0}$  & \multirow{7}{*}{all} \\ \cline{1-2}\cline{4-4}
3 & $1_{\,\color{black!60}1/0/0}$ &  & $1_{\,\color{black!60}1/0/0}$  &\\\cline{1-2}\cline{4-4}
4 & $7_{\,\color{black!60}4/2/1}$ &  & $10_{\,\color{black!60}6/2/2}$ & \\\cline{1-2}\cline{4-4}
5 & $13_{\,\color{black!60}7/4/2}$ &  & $27_{\,\color{black!60}13/8/6}$ & \\\cline{1-2}\cline{4-4}
$\frac{11}{2}$ & $12_{\,\color{black!60}6/4/2}$ & & $36_{\,\color{black!60}12/16/8}$ & \\\cline{1-2}\cline{4-4}
6& $39_{\,\color{black!60}11/16/12}$ &  & $144_{\,\color{black!60}33/48/63}$ & \\\cline{1-2}\cline{4-4}
$\frac{13}{2}$ & $36_{\,\color{black!60}10/18/8}$ &  & $276_{\,\color{black!60}40/108/128}$ & \\\hline
7 & $77_{\,\color{black!60}19/30/28}$ & $73_{\,\color{black!60}\text{all}/\text{all}/24}$ & $918_{\,\color{black!60}109/284/525}$  & $726_{\,\color{black!60}\text{all}/\text{all}/333}$ \\\hline
$\frac{15}{2}$ & $84_{\,\color{black!60}18/38/28}$ & $62_{\,\color{black!60}\text{all}/36/8}$ & $2204_{\,\color{black!60}152/660/1392}$ & $864_{\,\color{black!60}\text{all}/564/148}$ \\\hline
$8$& $180_{\,\color{black!60}30/70/80}$ & $116_{\,\color{black!60}\text{all}/62/24}$ & $6918_{\,\color{black!60}348/1624/4946}$ & $1724_{\,\color{black!60}\text{all}/1100/276}$ \\\hline
$\frac{17}{2}$& $186_{\,\color{black!60}32/78/76}$ & $84_{\,\color{black!60}\text{all}/48/4}$ & $18168_{\,\color{black!60}572/3808/13788}$  & $1938_{\,\color{black!60}\text{all}/1216/150}$  \\\hline
$9$& $327_{\,\color{black!60}39 / 120 /168}$ &  $111_{\,\color{black!60}34/62/15}$ & $56090_{\,\color{black!60}1133 / 9128 / 45829}$ & $2296_{\,\color{black!60}664/1348/284}$ \\\hline
\end{tabular}
\caption{Overview of the multiplets with $\Delta_0\le 9$, not counting the protected multiplets of which there is one for each integer $\Delta_0$. The subscripts denote the number of multiplets subject to both shortenings, only one shortening, and no shortenings, respectively. Two hours of computation time was allowed to solve the leading Q-system for each set of oscillator numbers.}
\label{tab:mul}
\end{table}

One thing is to know how many solutions to expect, another thing is to actually solve the leading Q-system. Table \ref{tab:mul} gives an overview of how well we do on this task. For the 495 multiplets with $\Delta_0\le 6.5$, we can solve everything. At $\Delta_0=7$ there are four sets of oscillator numbers for which our code does not manage to solve the leading Q-system. From then on, the fraction of the spectrum for which we can actually find the solutions rapidly decreases. The success of the algorithm to solve the leading Q-system depends on the total number of variables that it needs to solve for, i.e. the number of Bethe roots. Roughly, the code is efficient when the number of variables is less than 10.

\subsubsection*{Performance - how many loops can we do?} 

In principle, the algorithm can run iteratively to any order. In practice, there are of course limitations. 
Roughly, the runtime and memory usage seem to grow factorially with the loop order. Table \ref{tab:kotime} shows the runtime and memory usage for the calculation of the anomalous dimension of the Konishi multiplet up to 11 loop orders. We have managed to perform 11-loop calculations for several multiplets on a standard laptop. By a rough interpolation, the 12-loop calculation of the Konishi anomalous dimension would take less than two weeks, but it should probably be run on a cluster (or a very powerful laptop) due to the memory usage.

\begin{table}[t!]
\centering
\begin{tabular}{c|c|c}
loop & time & memory usage \\ \hline
5 & 14 s & 30 MB 
\\ 
6 & 43 s & 53 MB
\\ 
7 & 2.5 m & 86 MB 
\\ 
8 & 11 m & 134 MB
\\ 
9 & 53 m & 340 MB 
\\ 
10 & 5.5 h & 1.6 GB
\\
11 & 34 h & 8.5 GB
\end{tabular}
\caption{Runtime on a single 2.2 GHz core of a standard laptop for the calculation of the anomalous dimension of the Konishi multiplet up to 11 loops. Note that this calculation used the settings needed to go to 11 loops, and the performance can be improved slightly on lower orders with different settings. The memory usage was measured by the \Ma-function \texttt{MaxMemoryUsed}.}
\label{tab:kotime}
\end{table}

An important ingredient in the algorithm is the substitution of multiple zeta-values to an irreducible basis. For speed purposes, it is preferable to have a precomputed list with reductions of all these numbers. Such lists for multiple zeta-values with transcendentality up to 24 (which is sufficient for 14-loop calculations) can be found in \cite{Blumlein:2009cf}, and can more or less readily be used. The relations up to transcendentality 17 (which is sufficient for 11-loop calculations) take up 200 MB, while those up to transcendentality 19 (enabling 12-loops) take up a few GB\footnote{The ancillary files of this paper only contain the relations up to transcendentality 13, enabling 9-loop calculations. The extended files will be shared upon request. In generating these files, we made use of the extensive dataset related to \cite{Blumlein:2009cf}, see \href{https://www.nikhef.nl/~form/datamine/mzv/mzv.html}{https://www.nikhef.nl/$\sim$form/datamine/mzv/mzv.html}.}. This is another reason why 11-loop calculations are the limit of what can comfortably be done on a standard laptop.

Certainly, our algorithm could be optimised to reduce runtime and memory usage. For example, we do not exploit either parity properties or left/right symmetry when present. 
All in all, our estimate is that a determined programmer would be able to perform a 12-loop calculation on a powerful laptop and a 13-loop calculation on a cluster.

To get a feeling for how the computational effort scales with increasing quantum numbers, we list the normalised runtime and memory usage for the first multiplets containing twist-2 operators  $\DD_{12}^S\ZZ^2$, $\ns=[0,S-2|1,1,1,1|S-2,0]$, and the multiplets containing the first {\it exceptional} $\su(2)$ operators $\ZZ^{L-3}\XX^3$  \cite{Arutyunov:2012tx}, appearing for $\ns=[0,0|L-3,L-4,2,1|0,0]$, in table \ref{tab:tw2ex}. These examples show how the effort grows with increasing quantum numbers, and the general pattern that an increase in length has a larger effect than an increase in a spin-parameter.

For multiplets that remain long at $g=0$, the computations are more demanding. Both because one needs to determine the QSC functions at one order higher to obtain the anomalous dimension at a given order, and because all $Q_{ab|ij}^{(0)}$ are nonzero, making the solution of the difference equations \eqref{mukeyn} more involved.

\begin{table}[t]
\centering
\begin{tabular}{c|c|c}
\multicolumn{3}{c}{\small Twist-2 operators, $\DD_{12}^S\ZZ^2$} \\[2mm] \hline
$S$ & time & memory usage \\ \hline
2 & 1 & 1
\\ 
4 & 1.6 & 1.1
\\ 
6 & 1.9 & 1.6
\\ 
8 & 2.4 & 2.1
\\ 
10 & 3.2 & 2.6
\\ 
12 & 4.2 & 3.3
\\
\end{tabular}
\hspace{1cm}
\begin{tabular}{c|c|c}
\multicolumn{3}{c}{\small Exceptional solutions, $\ZZ^{L-3}\XX^3$} \\[2mm] \hline
$L$ & time & memory usage \\ \hline
6 & 1 & 1
\\ 
8 & 1.8 & 1.7
\\ 
10 & 2.8 & 2.3
\\ 
12 & 4.5 & 3.2
\\ 
14 & 6.4 & 4.3
\\ 
16 & 9.7 & 5.8
\end{tabular}

\caption{Normalised runtime and memory usage for the calculation of the 7-loop anomalous dimension of twist-2 operators and exceptional solutions.}
\label{tab:tw2ex}
\end{table}

\subsubsection*{Selected results}

In \ref{app:res}, we list the new results for two of the most commonly studied multiplets, the Konishi multiplet and the multiplet containing the first exceptional solution in the $\su(2)$ sector. 
For both multiplets, we managed to go to 11 loops, and both results can be expressed solely in terms of the so-called {\it single-valued multiple zeta-values} \cite{Brown:2013gia,Schnetz:2013hqa} given in \eqref{SVMZV}. These numbers are a special subset of multiple zeta-values that appear in perturbative quantum field theory, and the fact that the results contain only these numbers is a good sign that they are not just random numbers.

For our remaining results, the total of 8.043 multiplets with $\Delta_0\le9$ for which our code succesfully found solutions, we invite the reader to have a look at the database \dbname.

\subsubsection*{Working with algebraic numbers}

In general, the solution of the leading Q-system contains complicated algebraic numbers. We did not find it feasible to work with such numbers in their exact form in \Ma. As a consequence we evaluate such number numerically with very high precision. However, one can exploit the fact that if one algebraic number, i.e. a root of some algebraic equation with rational coefficients, appears in a solution, then there will be other solutions for which the other roots of the same equation appear. 
In the simplest example of a squareroot, it is easy to reconstruct the analytic result by adding and subtracting the two results and using a function such as \Ma's \texttt{Rationalize} to reconstruct rational numbers. The same principle can be used for roots of higher degree. A discussion of this was given in \cite{Marboe:2017zdv}.

For multiplets that are not subject to both shortenings, we have not been able to write a code that can safely handle all solutions while keeping the zeta-values as symbols. This is due to instabilities when working with numerical expressions, in particular when solving the large set of equations resulting from the regularity constraints on the functions $\mu$ and in the matching of $\Pt$ to its ansatz. For time efficiency, the solution of these equations must be done iteratively, but the bulky intermediate results sometimes lead to instabilities. To be able to run all states without these problems, we have thus used a version of the code that also evaluates the zeta-values numerically for most of these solutions. We introduce the \Ma\ implementation of our algorithm in the next section.


\section{Introduction to \nbname\ and \dbname} \label{sec:code}

This paper is supplemented by two \Ma\ notebooks on arxiv.org. The notebook \nbname\ contains an implementation of the perturbative algorithm. The notebook \dbname\ contains a database of results and a user-friendly interface to browse through the data. Both notebooks have been writtten in \Ma\ \texttt{11.2}.


\subsection{\nbname}
In \nbname, the described algorithm for solving the $\bP\mu$-system perturbatively is implemented. The algorithm for determining the leading solutions to the Q-system described in \cite{Marboe:2017dmb} is contained in the notebook as well. The notebook imports the file \texttt{z13.txt} containing reductions of all multiple zeta values up to transcendentality 13, which is sufficient to do 9 loop calculations\footnote{Additional files containing relations up to transcendentality 15 (26MB) and 17 (220MB) are available upon request.}.

\subsubsection*{Settings}
The code can be run in three different {\it modes}:

{\centering

\vspace{5mm}

\begin{tabular}{c|c|c}
\texttt{mode} & description & usage \\ \hline\hline
\texttt{exact} & keeps all expressions exact & $\begin{matrix} \text{solutions with rational coefficients} \\ \text{in the leading Q-system} \end{matrix}$ 
\\ \hline
\texttt{semi} & $\begin{matrix} \text{evaluates algebraic} \\ \text{numbers numerically} \end{matrix}$ & $\begin{matrix} \text{solutions subject to} \\ \text{both types of shortening} \end{matrix}$
\\ \hline
\texttt{full} & $\begin{matrix} \text{evaluates algebraic and} \\ \text{transcendental numbers numerically} \end{matrix}$ & all solutions
\\
\end{tabular}

\vspace{5mm}

}

\noindent As discussed, the need for different modes is the inability to systematically simplify expressions containing irrational numbers, and the numerical instabilities introduced by working with numerics.

When using the code in the modes \texttt{semi} and \texttt{full}, the user needs to set the value of \texttt{digits}, that is the number of digits precision in the numerical evaluation of the leading Q-system, which is the seed of the perturbative calculation. For each loop order, the precision drops by 50-300 digits, but it is unproblematic to start from a very high precision, e.g. \texttt{digits=3000}.

Importantly, the user has to specify the value of \texttt{maxord}, which denotes the maximal perturbative order that should be calculated. This variable controls how many constants $\cc_{a,k}^{(j)}$ and $\cc^{a,k}_{(j)}$ are determined when matching $\Pt$ to the ansatz, cf. the discussion in \ref{sec:usingansatz}. It affects the speed of the code, as a lower \texttt{maxord} means that the power expansions performed by the code can be truncated earlier.

\subsubsection*{Syntax}
The individual solutions to the QSC are referred to through a set of oscillator numbers in the format
\be
\{\{n_{\bbb_1},n_{\bbb_2}\},\{n_{\fff_1},n_{\fff_2},n_{\fff_3},n_{\fff_4}\},\{n_{\aaa_1},n_{\aaa_2}\}\}\,,\no
\ee
and a solution number ranging from 1 to the multiplicity of solutions with the given set of quantum numbers.

To calculate the perturbative anomalous dimension for a given solution up to a given order, the function \texttt{find$\gamma$} can be called,
\be
&\texttt{find}\gamma\texttt{[order][osc.no.][sol.no.]} \,.\no
\ee
For example, to get the 7-loop anomalous dimension of the Konishi multiplet, the user should call
\be
&\texttt{find}\gamma\texttt{[7][\{\{0,0\},\{1,1,1,1\},\{0,0\}\}][1]} \,.\no
\ee

The individual QSC-functions at a given perturbative order can be easily be accessed through simple commands. For example, for $\bP_a$ this is done by calling
\be
&\texttt{P[a][order][osc.no.][sol.no.]} \,. \no
\ee
The notebook contains a tutorial on how to use its basic functions.


\subsection{\dbname}
The notebook \dbname\ contains data for all multiplets with $\Delta_0\le 9$ for which we were able to solve the leading Q-system (8.043 out of 84.793). This data includes the solutions of the leading Q-system and the corresponding perturbative contributions to the anomalous dimension. 

\begin{figure}[t!]
\includegraphics[scale=0.55,trim={6mm 1mm 3mm 1mm},clip]{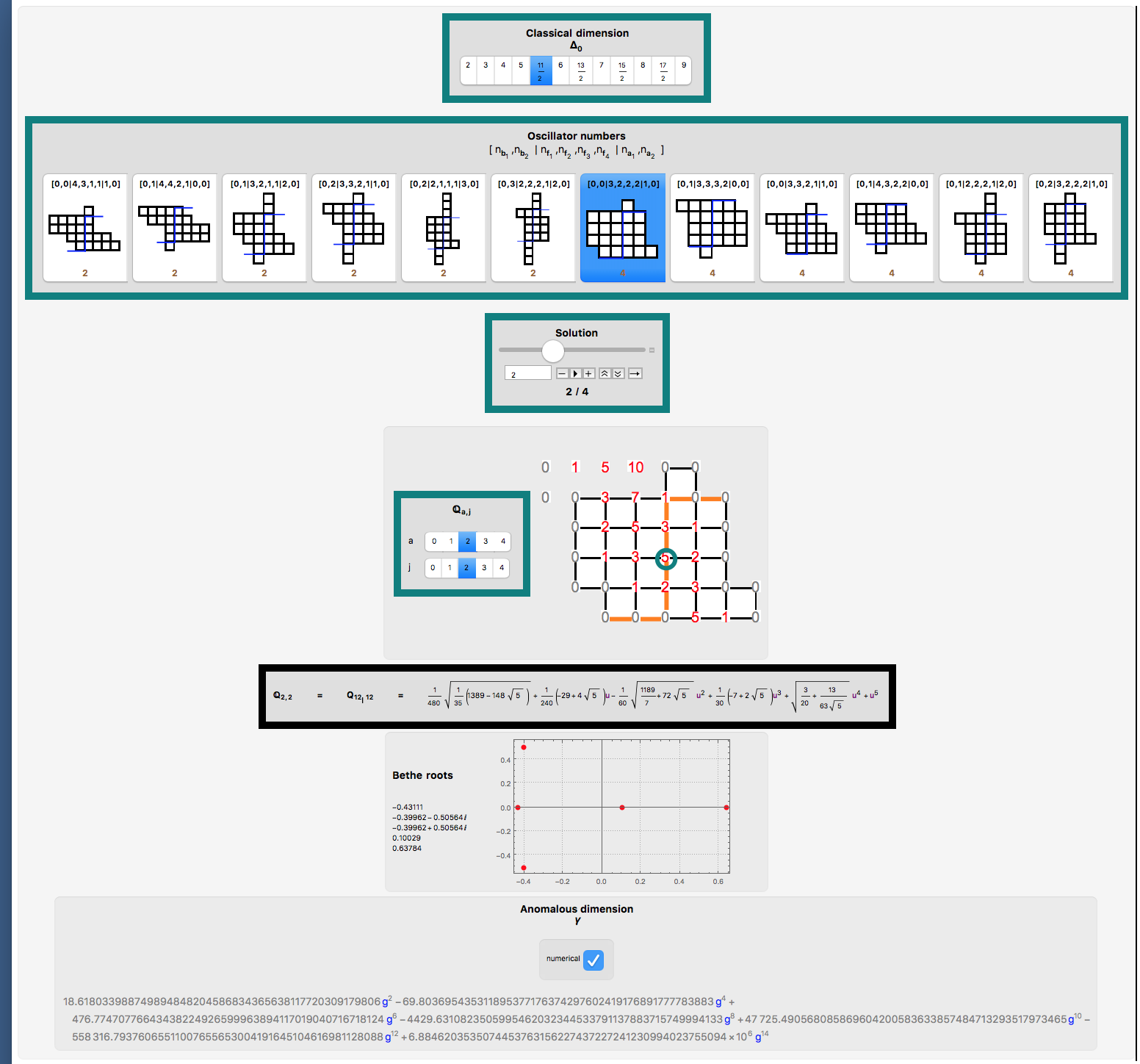}
\caption{A screenshot of the interface in \dbname.}
\label{fig:data}
\end{figure}

\subsubsection*{Interface}
At the top of the notebook, it is possible to open an interface to browse through the spectrum. This is done by first running the \texttt{Initialization cells}, which can be done by pressing the button \boxed{\texttt{Load}}, and then pressing the button \boxed{\texttt{Open interface}}. 

The interface allows the user to choose a classical dimension $\Delta_0$ and then to choose a particular set of oscillator numbers from a complete list. The user can then browse through the individual solutions. One can then choose a particular distinguished Q-function in the leading Q-system, and plot the position of the corresponding Bethe roots. The interface furthermore displays the anomalous dimension of the chosen state to as many order as we have run the code. See figure \ref{fig:data} for a screenshot of the interface.

\subsubsection*{Format of the saved data}
For each possible set of oscillator numbers, the full set of solutions is saved as a list with data for each solution
\be
\texttt{data[}
\texttt{osc.no.}
\texttt{] = \{}\texttt{data}_1,\texttt{data}_2,..., \texttt{data}_n\}
\ee
where each entry is of the form
\be
\texttt{data}_j=\{\{\mathbb{q}_{0,1},\mathbb{q}_{0,2},\mathbb{q}_{1,2},\mathbb{q}_{2,2},\mathbb{q}_{3,2},\mathbb{q}_{4,2},\mathbb{q}_{4,3}\},\{\gamma_1,\gamma_2,...\}\}
\ee
where $\mathbb{q}_{a,j}$ are the polynomial parts of the distinguished Q-functions $\dQ_{a,j}$, and $\gamma_i$ are the perturbative contributions to the anomalous dimension of the corresponding state.


\section{Conclusion}

In this paper, we continued our effort to explicitly compute the perturbative spectrum of planar $\mathcal{N}=4$ SYM. In \cite{Marboe:2017dmb} we introduced a new efficient way of finding general solutions to the QSC at $g=0$, and we built upon this by introducing a likewise efficient and general algorithm to construct perturbative corrections to these solutions.

We focused on the $\bP\mu$-system, consisting of eight $\bP$- and six $\mu$-functiond, instead of the much larger Q-system. We analysed the perturbative structure of the $\bP$- and $\mu$-functions, and used this knowledge to design a simple algorithm to solve this system perturbatively for general multiplets.

While generating the large dataset of QSC solutions and anomalous dimensions provided in \dbname, we encountered a new phenomenon: solutions that degenerate in the limit $g\to0$. We found that such solutions are expanded in $g$ instead of $g^2$ and split up at $\CO(g^1)$ due to non-linearities that normally do not appear.

We hope that the vast amount of data provided with this paper will help others see patterns in the spectrum, as well as aid in the efforts to understand other quantities such as one- and three-point functions.

\vspace{4mm}
\section*{Acknowledgements}
We thank Oliver Schnetz for discussions.
CM would like to thank the Niels Bohr Institute for kind hospitality during the final stages of this work.
The work of CM was supported by the grant {\it Exact Results in Gauge and String Theories} from the Knut and Alice Wallenberg foundation, and by the People Programme (Marie Curie Actions) of the European Union's Seventh Framework Programme FP7/2007-2013/ under REA Grant Agreement No 317089 (GATIS).

\appendix
\addtocontents{toc}{\protect\setcounter{tocdepth}{1}}

\newpage

\section{Selected results} \label{app:res}
For reference, we here list additional data and new results for the most commonly studied multiplets.

\subsection{$Q_{ab|ij}^{(0)}$ for the Konishi multiplet} \label{ap:Qabij}
For the Konishi multiplet, with our chosen way of fixing symmetries, the nonzero $Q_{ab|ij}^{(0)}$ are
{\footnotesize
\be
Q_{12|12}^{(0)} &=& \frac{u^2}{2880}-\frac{1}{34560}\\
Q_{13|12}^{(0)} &=& \frac{\ii u^3}{720}+\frac{\ii u}{2880} \no\\
Q_{14|12}^{(0)} &=&-\frac{u^4}{480}-\frac{u^2}{960}-\frac{1}{7680} \no\\
Q_{23|12}^{(0)} &=& -\frac{u^4}{480}-\frac{u^2}{320}+\frac{1}{23040} \no\\
Q_{24|12}^{(0)} &=&-\frac{\ii u^5}{240}-\frac{7\ii u^3}{1440}-\frac{11\ii u}{11520} \no\\
Q_{34|12}^{(0)} &=& \frac{u^6}{240}-\frac{u^4}{960}-\frac{u^2}{768}-\frac{1}{5120} \no\\
Q_{12|13}^{(0)} &=& 5 \ii u + \left(5u^2 - \frac{5}{12} \right) \eta_2^+\no \\
Q_{13|13}^{(0)} &=& -20 u^2 - \frac{20}{3} + \left(20\ii u^3 +5\ii u\right) \eta_2^+ \no\\
Q_{14|13}^{(0)} &=&-30 \ii u^3 -\frac{35\ii}{2}u - \left(30u^4 + 15u^2 + \frac{15}{8} \right) \eta_2^+ \no\\
Q_{23|13}^{(0)} &=& -30 \ii u^3 -\frac{95\ii}{2}u - \left(30u^4 + 45u^2 - \frac{5}{8} \right) \eta_2^+ \no\\
Q_{24|13}^{(0)} &=& 60 u^4 +75 u^2 - \left(60\ii u^5 + 70\ii u^3 + \frac{55\ii u}{4} \right) \eta_2^+ \no\\
Q_{34|13}^{(0)} &=& 60\ii u^5 -10\ii u^3 - \frac{265\ii u}{4} + \left(60 u^6 -15u^4 -\frac{75u^2}{4} - \frac{45}{16} \right) \eta_2^+ \no\\
Q_{14|14}^{(0)} = Q_{23|23}^{(0)} &=& -1\no\\
Q_{24|14}^{(0)} = Q_{24|23}^{(0)} &=& -6\ii u \no\\
Q_{34|14}^{(0)} &=& 18u^2 - \frac{3}{2} \no\\
Q_{34|23}^{(0)} &=& 18u^2 + \frac{9}{2}\no\\
Q_{24|24}^{(0)} &=& \frac{9u^2}{5}-\frac{3}{20} \no\\
Q_{34|24}^{(0)} &=& \frac{36\ii u^3}{5}+\frac{9\ii u}{5} \no\\
Q_{24|34}^{(0)} &=& 25920\ii u + \left(25920u^2-2160\right)\eta_2^+\no\\
Q_{34|34}^{(0)} &=& -103680 u^2 - 34560 + \left(103680\ii u^3+25920 \ii u\right)\eta_2^+\no\,.
\ee}

\subsection{11-loop anomalous dimension of Konishi multiplet}
We here give the 11-loop contribution for the anomalous dimensiom of the Konishi multiplet, $\ns=[0,0|1,1,1,1|0,0]$. 
See \cite{Marboe:2014gma} for the first 10 orders in the perturbative expansion $\gamma=\sum_{j=1}^\infty g^{2j}\gamma_j$.

{\small\be
\gamma_{11}&=&\textstyle
-242508705792 + 107663966208 \zeta_{3} + 70251466752 \zeta_{3}^2 - 
 12468142080 \zeta_{3}^3 
\no \\&& \textstyle
 + 1463132160 \zeta_{3}^4 - 71663616 \zeta_{3}^5 + 
 180173002752 \zeta_{5} - 16655486976 \zeta_{3} \zeta_{5} 
 \no\\&& \textstyle
 - 24628230144 \zeta_{3}^2 \zeta_{5} - 2895575040 \zeta_{3}^3 \zeta_{5} + 
 19278176256 \zeta_{5}^2 - 9619845120 \zeta_{3} \zeta_{5}^2 
 \no\\&& \textstyle
 + 2504494080 \zeta_{3}^2 \zeta_{5}^2 + \frac{882108048384}{175} \zeta_{5}^3 + 
 45602231040 \zeta_{7} + 14993482752 \zeta_{3} \zeta_{7} 
 \no\\&& \textstyle
 - 12034759680 \zeta_{3}^2 \zeta_{7} + 
 1406730240 \zeta_{3}^3 \zeta_{7} + 30605033088 \zeta_{5} \zeta_{7} + 
 21217637376 \zeta_{3} \zeta_{5} \zeta_{7} 
 \no\\&& \textstyle
 - \frac{1309941061632}{275} \zeta_{5}^2 \zeta_{7} - 
 13215327552 \zeta_{7}^2 - 4059901440 \zeta_{3} \zeta_{7}^2 - 69762034944 \zeta_{9} 
 \no\\&& \textstyle
 + 
 23284599552 \zeta_{3} \zeta_{9} - 3631889664 \zeta_{3}^2 \zeta_{9} - 
 11032374528 \zeta_{5} \zeta_{9} - 6666706944 \zeta_{3} \zeta_{5} \zeta_{9} 
 \no\\&& \textstyle
- 
 23148129024 \zeta_{7} \zeta_{9} - 10024051968 \zeta_{9}^2 - 54555179184 \zeta_{11} 
 + \frac{10048541184}{5} \zeta_{3} \zeta_{11} 
 \no\\&& \textstyle
 - 726029568 \zeta_{3}^2 \zeta_{11} 
 - 8975463552 \zeta_{5} \zeta_{11} 
 - 22529041920 \zeta_{7} \zeta_{11} 
 - \frac{1437993422496}{175} \zeta_{13} 
 \no\\&& \textstyle
 + \frac{1504385419392}{35} \zeta_{3} \zeta_{13} 
- 30324602880 \zeta_{5} \zeta_{13} 
- \frac{151130039581392}{875} \zeta_{15} - 
 41375093760 \zeta_{3} \zeta_{15} 
\no\\&& \textstyle
 - \frac{196484147423712}{275} \zeta_{17} 
 + 309361358592 \zeta_{19} 
 - 1729880064 Z_{11}^{(2)} 
 - \frac{1620393984}{5} \zeta_{3} Z_{11}^{(2)} 
 \no\\&& \textstyle
 - 131383296 \zeta_{5} Z_{11}^{(2)} 
 +\frac{138107420928}{175} Z_{13}^{(2)} 
 + \frac{3543865344}{35} \zeta_{3} Z_{13}^{(2)} 
 - \frac{5716780416}{7} Z_{13}^{(3)} 
 \no\\&& \textstyle
 - \frac{674832384}{7} \zeta_{3} Z_{13}^{(3)} 
 + \frac{48227088384}{175} Z_{15}^{(2)} 
 + \frac{3581880576}{25} Z_{15}^{(3)} 
 + 754974720 Z_{15}^{(4)} 
 \no\\&& \textstyle
 - \frac{854924544}{11} Z_{17}^{(2)} 
+ \frac{4963244544}{55} Z_{17}^{(3)} 
+ \frac{818159616}{275} Z_{17}^{(4)} 
+  \frac{175363688448}{1925} Z_{17}^{(5)}\,.
\ee}

\subsection{The first exceptional solution}
The $\su(2)$ sector contains so-called {\it exceptional} solutions, for which the central Q-function is
\be \label{qexc}
\dQ_{2,2} = u^3 + \frac{u}{4} = \left(u-\iotwo\right)u\left(u+\iotwo\right)\,.
\ee
They appear for three-magnon states with even length, i.e. $M=3$ and $L=4,6,8,...$, corresponding to oscillator numbers $\ns=[0,0|L-3,L-4,2,1|0,0]$. Due the position of the Bethe roots \eqref{qexc}, these solutions require extra care when solving the asymptotic Bethe equations, and similarly when solving the TBA equations, which was done up to six loops in \cite{Arutyunov:2012tx}. However, from the point of view of the QSC this state requires no special care.

The anomalous dimension of these multiplets are special in the sense that the $\zeta$-values stemming from the dressing factor are delayed compared to other results. In the QSC, it is not immediately transparent why this delay occurs. 

We here provide the 11-loop result for the first exceptional solution with oscillator numbers $\ns=[0,0|3,2,2,1|0,0]$:

{\footnotesize\be \label{exres}
\!\!\!\gamma &\!\!\!\!=\!\!\!\!& 12g^2-36g^4+252g^6-2484g^8+g^{10}\left( 28188-288\zeta_3 \right) \\
&&+g^{12}\left(-339012+7776\zeta_3+12096\zeta_5-18144\zeta_9\right) \no \\
&&+g^{14}\left( 4214268 - 39744 \zeta_3 - 181440 \zeta_5 + 57024 \zeta_3^2 -260064 \zeta_7  \right. \no\\
&&\quad\quad\left. -34560 \zeta_3\zeta_5 -60480 \zeta_9 - 8640 \zeta_5^2 -96768 \zeta_3\zeta_7 + 665280 \zeta_{11}  \right)\no \\
&&\no
+{g}^{16}\bigg(\!\! -53785620
-820800 \zeta_3  
-699840  \zeta_3^2  
-82944 \zeta_3^3  
+1664064   \zeta_5 
-1510272  \zeta_3  \zeta_5 
\\ &&\quad\quad \no
-290304  \zeta_3^2   \zeta_5 
+ 250560  \zeta_5^2
+ 4257792  \zeta_7
+628992  \zeta_3  \zeta_7 
+ 1451520  \zeta_5  \zeta_7 
+ 4711968 \zeta_9 
\\ &&\quad\quad\no
+ 2903040  \zeta_3  \zeta_9
+{\textstyle\frac{11144736}{5}} \zeta_{11}
-16061760  \zeta_{13}
-{\textstyle\frac{124416}{5}} {\color{black}Z}_{11}^{(2)}\bigg)
 \\ && \no
+{g}^{18}   \bigg(
702413532 
+25507872  \zeta_3
-2282688  \zeta_3^2
-1119744  \zeta_3^3
-248832  \zeta_3^4
-502848  \zeta_5
\\ &&\quad\quad\no
+25653888  \zeta_3  \zeta_5 
+3836160  \zeta_3^2   \zeta_5
+ 5987520  \zeta_5^2  
+ 6635520  \zeta_3  \zeta_5^2 
- 45170784  \zeta_7 
\\ && \quad\quad\no
+ 22037184  \zeta_3  \zeta_7 
+ 6676992  \zeta_3^2  \zeta_7 
- 5766336  \zeta_5  \zeta_7 
- 16027200  \zeta_7^2  
- 75035808  \zeta_9  
\\ &&  \quad\quad\no
+ 10018944  \zeta_3  \zeta_9 
- 38361600  \zeta_5  \zeta_9 
-79511328 \zeta_{11} 
-58848768  \zeta_3  \zeta_{11}  
- {\textstyle\frac{273255552}{5}}  \zeta_{13} 
\\ &&  \quad\quad\no
+ 324324000 \zeta_{15}
+ 311040 {\color{black}Z}_{11}^{(2)}
- {\textstyle\frac{601344}{5}}  {\color{black}Z}_{13}^{(2)} 
+ 145152  {\color{black}Z}_{13}^{(3)}\bigg)
\\ && +g^{20}\bigg(\!\! \no
-\!9354033252 
\!-\! 461062368 \zeta_3 
 \!+\! 198500544 \zeta_3^2 
 \!-\! 2778624 \zeta_3^3 
 \!+\! 2239488 \zeta_3^4 
 \!-\! 348634368 \zeta_5 
 \\ &&  \quad\quad\no
 - 201128832 \zeta_3 \zeta_5 
 + 50865408 \zeta_3^2 \zeta_5 
 + 14681088 \zeta_3^3 \zeta_5 
 - 187012800 \zeta_5^2 
 - 51010560 \zeta_3 \zeta_5^2 
 \\ &&  \quad\quad\no
 - {\textstyle\frac{1310563584}{35}} \zeta_5^3 
 + 343359648 \zeta_7 
 - 351993600 \zeta_3 \zeta_7 
 - 56909952 \zeta_3^2 \zeta_7 
 - 147334464 \zeta_5 \zeta_7 
 \\ &&  \quad\quad\no
 -  221543424 \zeta_3 \zeta_5 \zeta_7 
 + 29465856 \zeta_7^2 
 + 911464704 \zeta_9 
 - 312035328 \zeta_3 \zeta_9 
 - 107619840 \zeta_3^2 \zeta_9 
 \\ &&  \quad\quad\no
 + 75755520 \zeta_5 \zeta_9 
 + 633225600 \zeta_7 \zeta_9 
 + {\textstyle\frac{5846706576}{5}} \zeta_{11} 
 +{\textstyle \frac{740306304}{5}} \zeta_3 \zeta_{11} 
 + 745303680 \zeta_5 \zeta_{11} 
  \\ &&  \quad\quad\no
 + {\textstyle\frac{226451356776}{175}} \zeta_{13} 
 + 1017080064 \zeta_3 \zeta_{13} 
 + {\textstyle\frac{96109333632}{175}} \zeta_{15} 
 - 5951088000 \zeta_{17}
 + {\textstyle\frac{41036544}{5}} Z_{11}^{(2)}
  \\ &&  \quad\quad\no
 + {\textstyle\frac{746496}{5}} Z_{11}^{(2)} \zeta_3
+ {\textstyle\frac{406415232}{175}} Z_{13}^{(2)}  
- {\textstyle\frac{18719424}{7}} Z_{13}^{(3)} 
+ {\textstyle\frac{30710016}{35}} Z_{15}^{(2)} 
- {\textstyle\frac{850176}{5}} Z_{15}^{(3)} 
\bigg) \no
\\ && +g^{22}\bigg(\!\! \no
126583953660 
+ 7084039680 \zeta_{3} 
- 4527778176 \zeta_{3}^2 
+ 453828096 \zeta_{3}^3 
+ 30606336 \zeta_{3}^4 
 \\ &&  \quad\quad\no \textstyle
+ \frac{17915904}{5} \zeta_{3}^5
+ 9380375424 \zeta_{5} 
- 791420544 \zeta_{3} \zeta_{5} 
- 585999360 \zeta_{3}^2 \zeta_{5} 
- 98952192 \zeta_{3}^3 \zeta_{5} 
 \\ &&  \quad\quad\no \textstyle
+ 2203583616 \zeta_{5}^2 
- 573557760 \zeta_{3} \zeta_{5}^2 
- 277447680 \zeta_{3}^2 \zeta_{5}^2 
+ \frac{46625393664}{175} \zeta_{5}^3
- 11355552 \zeta_{7} 
 \\ &&  \quad\quad\no \textstyle
+ 3606814656 \zeta_{3} \zeta_{7} 
- 626621184 \zeta_{3}^2 \zeta_{7} 
- 171777024 \zeta_{3}^3 \zeta_{7} 
+ 4166028288 \zeta_{5} \zeta_{7} 
 \\ &&  \quad\quad\no \textstyle
+ 1438926336 \zeta_{3} \zeta_{5} \zeta_{7} 
+ \frac{88526829312}{55} \zeta_{5}^2 \zeta_{7} 
+ 1029018816 \zeta_{7}^2 
+ 1626863616 \zeta_{3} \zeta_{7}^2 
 \\ &&  \quad\quad\no \textstyle
- 9350341920 \zeta_{9}
+ 5037050880 \zeta_{3} \zeta_{9} 
+ 659653632 \zeta_{3}^2 \zeta_{9} 
+ 1786786560 \zeta_{5} \zeta_{9} 
 \\ &&  \quad\quad\no \textstyle 
+ 3198735360 \zeta_{3} \zeta_{5} \zeta_{9}
- 732139776 \zeta_{7} \zeta_{9} 
- 5322454272 \zeta_{9}^2 
- \frac{73306877904}{5} \zeta_{11}
 \\ &&  \quad\quad\no \textstyle 
+ \frac{21136609152}{5} \zeta_{3} \zeta_{11} 
+ 1512276480 \zeta_{3}^2 \zeta_{11} 
- 1039611456 \zeta_{5} \zeta_{11} 
- 11044598400 \zeta_{7} \zeta_{11} 
 \\ &&  \quad\quad\no \textstyle 
- \frac{432304433856}{25} \zeta_{13}
- \frac{379475114112}{175} \zeta_{3} \zeta_{13} 
- 12640852224 \zeta_{5} \zeta_{13} 
- \frac{5354699627232}{875} \zeta_{15}
 \\ &&  \quad\quad\no \textstyle 
- 16210022400 \zeta_{3} \zeta_{15} 
- \frac{2411128067328}{55} \zeta_{17}
+ 103120452864 \zeta_{19} 
- \frac{1103279616}{5} Z_{11}^{(2)} 
 \\ &&  \quad\quad\no \textstyle 
- \frac{77386752}{5} \zeta_{3} Z_{11}^{(2)} 
- 2985984 \zeta_{5} Z_{11}^{(2)} 
- \frac{175675392}{25} Z_{13}^{(2)} 
- \frac{102518784}{175} \zeta_{3} Z_{13}^{(2)} 
- 10813824 Z_{13}^{(3)} 
 \\ &&  \quad\quad\no \textstyle 
+ \frac{6842880}{7} \zeta_{3} Z_{13}^{(3)} 
- \frac{3975457536}{175} Z_{15}^{(2)} 
+ \frac{103274496}{25} Z_{15}^{(3)} 
+ 28311552 Z_{15}^{(4)} 
 \\ &&  \quad\quad\no \textstyle 
- \frac{13271040}{11} Z_{17}^{(2)} 
+ \frac{31822848}{11} Z_{17}^{(3)} 
+ \frac{13008384}{55} Z_{17}^{(4)} 
+ \frac{1033938432}{385} Z_{17}^{(5)} \bigg) + \CO(g^{24})\,.
\ee}


\subsection{Single-valued MZVs} \label{sec:svmzv}

In this appendix, we list the single-valued multiple zeta-values \cite{Schnetz:2016fhy,SchnetzPrivate} that appear in the anomalous dimensions up to 11 loops.

\be\label{SVMZV}
Z_{11}^{(2)}&=&-\zeta_{3,5,3}+\zeta_3\,\zeta_{3,5}\\
Z_{13}^{(2)}&=&-\zeta_{5,3,5}+11\,\zeta_5\,\zeta_{3,5}+5\,\zeta_5\,\zeta_8\no\\
Z_{13}^{(3)}&=&-\zeta_{3,7,3}+\zeta_3\,\zeta_{3,7}+12\,\zeta_5\,\zeta_{3,5}+6\,\zeta_5\,\zeta_8\no\\
Z_{15}^{(2)}&=&\textstyle\zeta_{3,7,5}-\zeta_5\,\zeta_{3,7}-3\,\zeta_5\,\zeta_{10}+21\,\zeta_9\,\zeta_6+\frac{175}{2}\,\zeta_{11}\,\zeta_4+\frac{637}{2}\,\zeta_{13}\,\zeta_2\no\\
Z_{15}^{(3)}&=&-\zeta_{3,9,3}+\zeta_3\,\zeta_{3,9}+12\,\zeta_5\,\zeta_{3,7}+30\,\zeta_7\,\zeta_{3,5}+6\,\zeta_5\,\zeta_{10}+15\,\zeta_7\,\zeta_8\no\\
Z_{15}^{(4)}&=& \textstyle\frac{3266937}{28028000}\,\zeta_2^6\,\zeta_3
+ \frac{243}{12320} \,\zeta_{2}^3 \,\zeta_{3}^3 
- \frac{27}{7040} \,\zeta_{3}^5 
 - \frac{15351701}{18972800} \,\zeta_{2}^5 \,\zeta_{5} 
 - \frac{1863}{14080} \,\zeta_{2}^2 \,\zeta_{3}^2 \,\zeta_{5} 
 \no\\ &&\textstyle
 - \frac{81}{176} \,\zeta_{2} \,\zeta_{3} \,\zeta_{5}^2 
 + \frac{138257}{61440} \,\zeta_{5}^3 
 + \frac{593993}{2464000} \,\zeta_{2}^4 \,\zeta_{7} 
 - \frac{1053}{704} \,\zeta_{2} \,\zeta_{3}^2 \,\zeta_{7} 
 + \frac{30061}{22528} \,\zeta_{3} \,\zeta_{5} \,\zeta_{7} 
 \no\\ &&\textstyle
 + \frac{44646739}{5644800} \,\zeta_{2}^3 \,\zeta_{9} 
 + \frac{68355}{11264} \,\zeta_{3}^2 \,\zeta_{9} 
 - \frac{3905317}{118272} \,\zeta_{2}^2 \,\zeta_{11} 
 - \frac{31663273}{225280} \,\zeta_{2} \,\zeta_{13} 
  \no\\ &&\textstyle
 + \frac{20651486329}{70963200} \,\zeta_{15} 
 - \frac{81}{880} \,\zeta_{2}^2 \,\zeta_{3} \,\zeta_{3, 5} 
 + \frac{1701}{2816} \,\zeta_{2} \,\zeta_{5} \,\zeta_{3, 5}
 - \frac{8551}{10240} \,\zeta_{7} \,\zeta_{3, 5}
  \no\\ &&\textstyle
 - \frac{243}{704} \,\zeta_{2} \,\zeta_{3} \,\zeta_{3, 7} 
 + \frac{463187}{2365440} \,\zeta_{5} \,\zeta_{3, 7}
 + \frac{261}{1408} \,\zeta_{3} \,\zeta_{3, 9}
 + \frac{81}{352} \,\zeta_{2}^2 \,\zeta_{3, 3, 5} 
 + \frac{81}{704} \,\zeta_{2} \,\zeta_{3, 3, 7} 
 \no\\ &&\textstyle
 - \frac{16663}{202752} \,\zeta_{3, 3, 9}
 - \frac{567}{3520} \,\zeta_{2} \,\zeta_{3, 5, 5} 
 - \frac{150481}{1182720} \,\zeta_{5, 3, 7}
 + \frac{81}{1408} \,\zeta_{3} \,\zeta_{1, 1, 4, 6} 
 + \frac{81}{1408} \,\zeta_{1, 1, 3, 4, 6}
 \no\\
Z_{17}^{(2)}&=&\zeta_{3,3,11}+6\,\zeta_5\,\zeta_{3,9}+15\,\zeta_7\,\zeta_{3,7}+28\,\zeta_9\,\zeta_{3,5}+3\,\zeta_7\,\zeta_{10}+28\,\zeta_9\,\zeta_8
\no\\ &&
+104\,\zeta_{11}\,\zeta_6+396\,\zeta_{13}\,\zeta_4+1449\,\zeta_{15}\,\zeta_2
\no\\
Z_{17}^{(3)}&=&\textstyle\zeta_{3,5,9}+6\,\zeta_5\,\zeta_{3,9}+15\,\zeta_7\,\zeta_{3,7}+69\,\zeta_9\,\zeta_{3,5}+25\,\zeta_9\,\zeta_8+60\,\zeta_{11}\,\zeta_6
\no\\ &&\textstyle
+\frac{1221}{2}\,\zeta_{13}\,\zeta_4+3165\,\zeta_{15}\,\zeta_2\no\\
Z_{17}^{(4)}&=&\textstyle\zeta_{5,3,9}+\frac{10}{3}\,\zeta_5\,\zeta_{3,9}-27\,\zeta_9\,\zeta_{3,5}-\frac{121285}{2073}\,\zeta_5\,\zeta_{12}-\frac{15}{2}\,\zeta_7\,\zeta_{10}
\no\\ &&\textstyle
-44\,\zeta_9\,\zeta_8-\frac{45}{2}\,\zeta_{11}\,\zeta_6+390\,\zeta_{13}\,\zeta_4+2535\,\zeta_{15}\,\zeta_2
\no\\
Z_{17}^{(5)}&=&\textstyle\zeta_{5,5,7}-42\,\zeta_9\,\zeta_{3,5}-7\,\zeta_7\,\zeta_{10}-42\,\zeta_9\,\zeta_8+5\,\zeta_{11}\,\zeta_6+\frac{637}{2}\,\zeta_{13}\,\zeta_4
\no\\ &&\textstyle
+1820\,\zeta_{15}\,\zeta_2\,.
\no
\ee

\newpage

\addcontentsline{toc}{section}{References}

\bibliographystyle{elsarticle-num}
\bibliography{bibliography}
\biboptions{sort&compress}

\end{document}

\begin{figure}[h]
\centering
\begin{picture}(395,300)

\color{black}
\linethickness{0.7mm}

\put(40,200){\line(1,0){20}}
\put(40,180){\line(1,0){40}}
\put(0,160){\line(1,0){80}}
\put(0,140){\line(1,0){80}}
\put(0,120){\line(1,0){80}}
\put(0,100){\line(1,0){80}}
\put(0,80){\line(1,0){80}}
\put(0,60){\line(1,0){40}}
\put(20,40){\line(1,0){20}}

\put(0,60){\line(0,1){100}}
\put(20,40){\line(0,1){120}}
\put(40,40){\line(0,1){160}}
\put(60,80){\line(0,1){120}}
\put(80,80){\line(0,1){100}}

\put(180,140){\line(1,0){20}}
\put(180,120){\line(1,0){40}}
\put(120,100){\line(1,0){100}}
\put(120,80){\line(1,0){100}}
\put(120,60){\line(1,0){100}}
\put(120,40){\line(1,0){100}}
\put(120,20){\line(1,0){100}}
\put(160,0){\line(1,0){20}}

\put(120,20){\line(0,1){80}}
\put(140,20){\line(0,1){80}}
\put(160,0){\line(0,1){100}}
\put(180,0){\line(0,1){140}}
\put(200,20){\line(0,1){120}}
\put(220,20){\line(0,1){100}}

\put(180,300){\line(1,0){20}}
\put(140,280){\line(1,0){100}}
\put(140,260){\line(1,0){100}}
\put(140,240){\line(1,0){100}}
\put(140,220){\line(1,0){100}}
\put(140,200){\line(1,0){100}}
\put(140,180){\line(1,0){40}}
\put(160,160){\line(1,0){20}}

\put(140,180){\line(0,1){100}}
\put(160,160){\line(0,1){120}}
\put(180,160){\line(0,1){140}}
\put(200,200){\line(0,1){100}}
\put(220,200){\line(0,1){80}}
\put(240,200){\line(0,1){80}}

\put(340,180){\line(1,0){20}}
\put(280,160){\line(1,0){120}}
\put(280,140){\line(1,0){120}}
\put(280,120){\line(1,0){120}}
\put(280,100){\line(1,0){120}}
\put(280,80){\line(1,0){120}}
\put(320,60){\line(1,0){20}}

\put(280,80){\line(0,1){80}}
\put(300,80){\line(0,1){80}}
\put(320,60){\line(0,1){100}}
\put(340,60){\line(0,1){120}}
\put(360,80){\line(0,1){100}}
\put(380,80){\line(0,1){80}}
\put(400,80){\line(0,1){80}}

\color{blue}
\footnotesize

\put(80,160){\color{white}\circle*{8}}\put(78,157){0}
\put(60,160){\color{white}\circle*{8}}\put(58,157){1}
\put(40,160){\color{white}\circle*{8}}\put(38,157){3}
\put(20,160){\color{white}\circle*{8}}\put(18,157){5}
\put(0,160){\color{white}\circle*{8}}\put(-2,157){0}

\put(80,140){\color{white}\circle*{8}}\put(78,137){0}
\put(60,140){\color{white}\circle*{8}}\put(58,137){2}
\put(40,140){\color{white}\circle*{8}}\put(38,137){5}
\put(20,140){\color{white}\circle*{8}}\put(18,137){4}
\put(0,140){\color{white}\circle*{8}}\put(-2,137){0}

\put(80,120){\color{white}\circle*{8}}\put(78,117){0}
\put(60,120){\color{white}\circle*{8}}\put(58,117){3}
\put(40,120){\color{white}\circle*{8}}\put(38,117){7}
\put(20,120){\color{white}\circle*{8}}\put(18,117){3}
\put(0,120){\color{white}\circle*{8}}\put(-2,117){0}

\put(80,100){\color{white}\circle*{8}}\put(78,97){0}
\put(60,100){\color{white}\circle*{8}}\put(58,97){4}
\put(40,100){\color{white}\circle*{8}}\put(38,97){5}
\put(20,100){\color{white}\circle*{8}}\put(18,97){2}
\put(0,100){\color{white}\circle*{8}}\put(-2,97){0}

\put(80,80){\color{white}\circle*{8}}\put(78,77){0}
\put(60,80){\color{white}\circle*{8}}\put(58,77){5}
\put(40,80){\color{white}\circle*{8}}\put(38,77){3}
\put(20,80){\color{white}\circle*{8}}\put(18,77){1}
\put(0,80){\color{white}\circle*{8}}\put(-2,77){0}

\put(220,100){\color{white}\circle*{8}}\put(218,97){0}
\put(200,100){\color{white}\circle*{8}}\put(198,97){1}
\put(180,100){\color{white}\circle*{8}}\put(178,97){3}
\put(160,100){\color{white}\circle*{8}}\put(158,97){8}
\put(140,100){\color{white}\circle*{8}}\put(138,97){4}

\put(220,80){\color{white}\circle*{8}}\put(218,77){0}
\put(200,80){\color{white}\circle*{8}}\put(198,77){2}
\put(180,80){\color{white}\circle*{8}}\put(178,77){5}
\put(160,80){\color{white}\circle*{8}}\put(158,77){6}
\put(140,80){\color{white}\circle*{8}}\put(138,77){3}

\put(220,60){\color{white}\circle*{8}}\put(218,57){0}
\put(200,60){\color{white}\circle*{8}}\put(198,57){3}
\put(180,60){\color{white}\circle*{8}}\put(178,57){7}
\put(160,60){\color{white}\circle*{8}}\put(158,57){4}
\put(140,60){\color{white}\circle*{8}}\put(138,57){2}

\put(220,40){\color{white}\circle*{8}}\put(218,37){0}
\put(200,40){\color{white}\circle*{8}}\put(198,37){4}
\put(180,40){\color{white}\circle*{8}}\put(178,37){4}
\put(160,40){\color{white}\circle*{8}}\put(158,37){2}
\put(140,40){\color{white}\circle*{8}}\put(138,37){1}

\put(220,20){\color{white}\circle*{8}}\put(218,17){0}
\put(200,20){\color{white}\circle*{8}}\put(198,17){5}
\put(180,20){\color{white}\circle*{8}}\put(178,17){1}
\put(160,20){\color{white}\circle*{8}}\put(158,17){0}
\put(140,20){\color{white}\circle*{8}}\put(138,17){0}
\end{picture}

\caption{Young diagrams on which one solution to the $\psu(2,2|4)$ Q-system coincide (up to symmetry transformations).}
\label{fig:deg1}
\end{figure}

\begin{picture}(260,90)

\color{gray}
\linethickness{.1mm}

\multiput(15,20)(0,10){5}{\line(1,0){90}}
\put(40,0){\line(1,0){20}}
\put(40,10){\line(1,0){20}}
\put(60,70){\line(1,0){20}}
\put(60,80){\line(1,0){20}}
\put(20,20){\line(0,1){40}}
\put(30,20){\line(0,1){40}}
\put(40,-5){\line(0,1){65}}
\put(50,-5){\line(0,1){65}}
\put(60,-5){\line(0,1){90}}
\put(70,20){\line(0,1){65}}
\put(80,20){\line(0,1){65}}
\put(90,20){\line(0,1){40}}
\put(100,20){\line(0,1){40}}

\multiput(165,20)(0,10){5}{\line(1,0){90}}
\put(190,0){\line(1,0){20}}
\put(190,10){\line(1,0){20}}
\put(210,70){\line(1,0){20}}
\put(210,80){\line(1,0){20}}
\put(170,20){\line(0,1){40}}
\put(180,20){\line(0,1){40}}
\put(190,-5){\line(0,1){65}}
\put(200,-5){\line(0,1){65}}
\put(210,-5){\line(0,1){90}}
\put(220,20){\line(0,1){65}}
\put(230,20){\line(0,1){65}}
\put(240,20){\line(0,1){40}}
\put(250,20){\line(0,1){40}}

\color{black}
\linethickness{0.7mm}

\put(30,20){\line(1,0){60}}
\put(20,30){\line(1,0){70}}
\put(20,40){\line(1,0){60}}
\put(20,50){\line(1,0){60}}
\put(20,60){\line(1,0){60}}
\put(60,70){\line(1,0){20}}
\put(60,80){\line(1,0){10}}
\put(20,30){\line(0,1){30}}
\put(30,20){\line(0,1){40}}
\put(40,20){\line(0,1){40}}
\put(50,20){\line(0,1){40}}
\put(60,20){\line(0,1){60}}
\put(70,20){\line(0,1){60}}
\put(80,20){\line(0,1){50}}
\put(90,20){\line(0,1){10}}

\put(180,20){\line(1,0){70}}
\put(170,30){\line(1,0){80}}
\put(170,40){\line(1,0){70}}
\put(170,50){\line(1,0){70}}
\put(170,60){\line(1,0){70}}
\put(210,70){\line(1,0){10}}
\put(170,30){\line(0,1){30}}
\put(180,20){\line(0,1){40}}
\put(190,20){\line(0,1){40}}
\put(200,20){\line(0,1){40}}
\put(210,20){\line(0,1){50}}
\put(220,20){\line(0,1){50}}
\put(230,20){\line(0,1){40}}
\put(240,20){\line(0,1){40}}
\put(250,20){\line(0,1){10}}

\put(60,40){\circle*{5}}
\put(210,40){\circle*{5}}

\end{picture}